# Master Thesis

Boosting performance:
Gradient Clock Synchronisation
with two-way measured links

Sophie Wenning

April 1, 2025

Examiner: Prof. Dr. Matthias Függer
Co-Examiner: Prof. Dr. Danupon Nanongkai

## Summary

Abstract

This master thesis extends the formal model of the GCS algorithm as presented by (Fan and Lynch 2004, 325), (Lenzen, Locher, and Wattenhofer 2008, 510), and (Függer et al. 2023) to operate under implementation-near assumptions by replacing the one-way measurement paradigm assumed in prior work by the two-way measurement paradigm. With this change of paradigm, we remove many restrictions previously enforced to allow provable performance. Most notability, while maintaining the core behaviour of GCS, we:

1. Lift the requirement for unitary link lengths and thereby create a realistic model for flexible deployment of implementations of GCS in practice.
2. Provide a formal model of frequency sources assumed in prior work.
3. Perform a fine grained distinction between the different components of the algorithm's estimation error and globally reduce its impact by multiple orders of magnitude.
4. Significantly reduce the contribution of the uncertainty to the algorithm's estimation error to be in the range of 10% to 0,1% of the delay per link instead of being in the oder of the delay per link as in prior work and show matching upper bounds on the local and global skew of GCS.

# 1 Introduction

## 1.1 Historical & technical background

With the industrial revolution and the rise of connectivity and communication, the need for a globally consistent notion of time emerged. In other words, the clock synchronisation problem, that is the task of tightly aligning a large number of time references which can be quite far apart from each other, in both their subjective notion time as well as their physical distance from one another.

For example, the clocks all around the country suffered from offsets of up to a quarter of an hour in the late 19th century's Belgium. This became highly problematic with the emergence of national train lines, as a more precise time indication at national level was suddenly required for the sake of efficiency but also for security reasons, e.g, to know when to close a railway crossing gate before a train passes by. A specific Church window in Brussel's cathedral enabled pocket watches and public clocks to align at the precision of a minute instead such as the schedules of the emerging national train network could be synchronised.

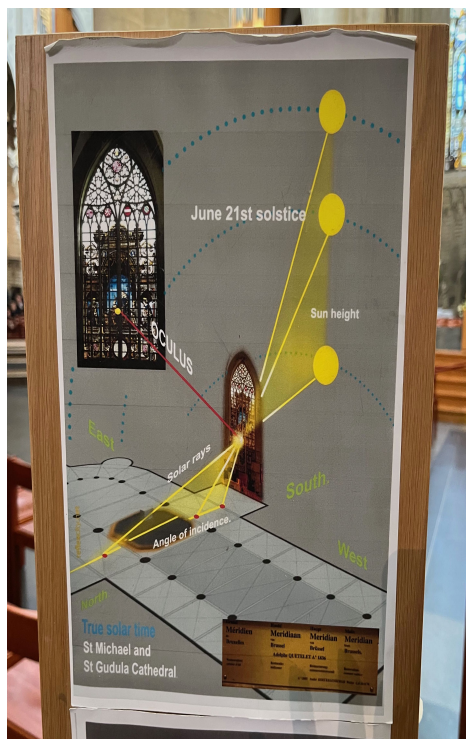

Figure 1: Occulus & trigonometric design. From: private archive.

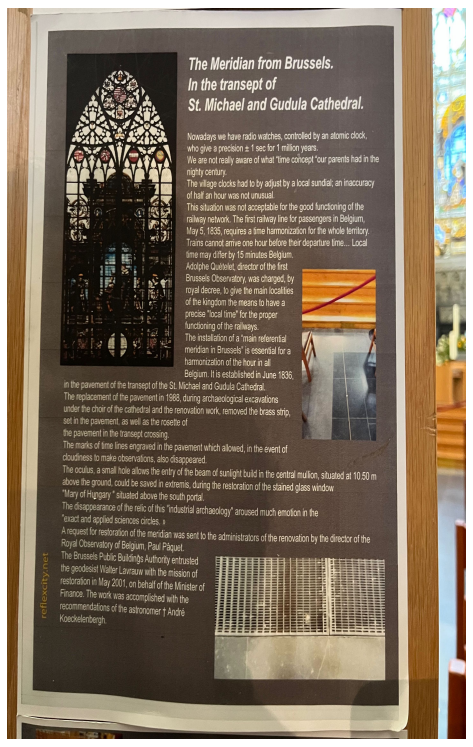

Figure 2: History of creation. From: private archive.

With the always increasing trading speed at stock exchanges, fulled by the emergence of phone lines and later computers throughout the 20th century, the emergence of personal computers and high speed communication networks increased the low-latency requirements [3] at the end of the 20th century, the distributed nature of specific problems pushed forward technical development in both clock building, yielding, amongst others, the invention of atomic clocks, which as of today are the standard references for high accuracy time, as well as various time synchronisation protocols such as NTP [76] and PTP [20] which are the current state of the art protocols for time distribution in computer networks.

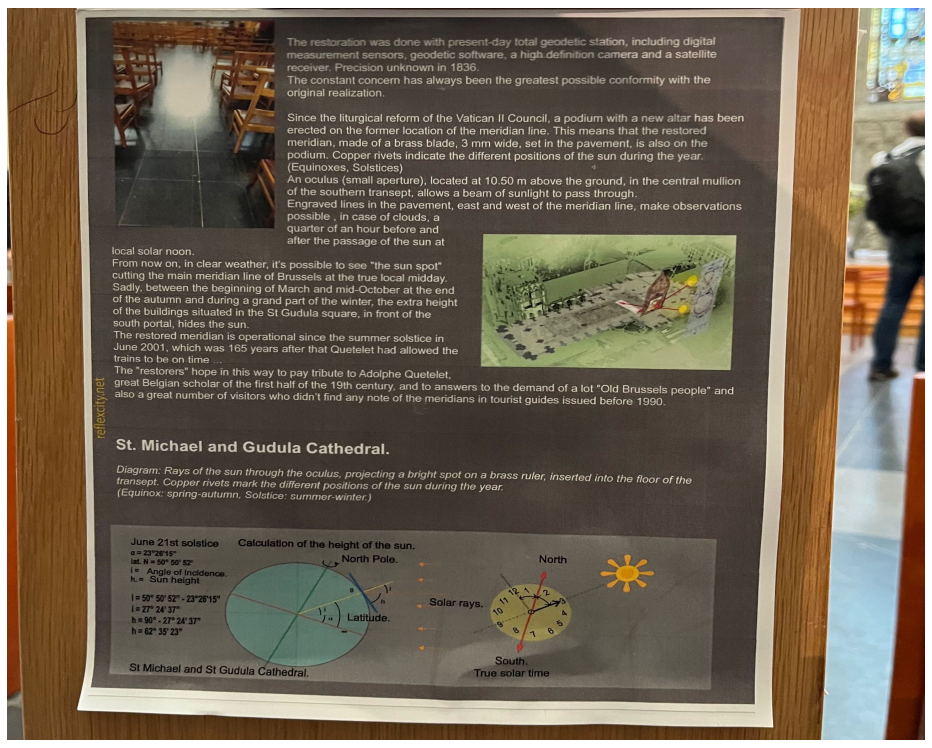

Figure 3: Calculation details based on sun data captured at summer solstice. From: private archive.

Instead of being anchored to a time oracle provided by the trigonometric computations of a church window, today's computer systems are anchored to UTC [51], the global standard time, provided by a clock ensemble [84] of various types of atomic clocks distributed all around the world. With more complex and independently operating distributed systems, such as sensor and cellular networks, wireless time reference transfer became increasingly important. GNSS [9], the satellite system associated to the GPS [78] positioning system, provides a UTC reference signal with nanosecond accuracy [70]. Data center and computational clusters also require higher and higher quality of inter-machine synchronisation to allow for parallelised high speed data processing for AI and distributed data base applications, usually exploiting a high quality reference [2] such as an atomic clock or a GPS access as well as NTP or PTP for signal propagation throughout the network.

## 1.2 Synchronicity is key to performance

The performance of the above applications of time synchronisation to wireless sensor and cellular networks as well as to distributed computations in AI clusters and data storage center is fundamentally tied to the quality of synchronisation [3]. It does not matter in that regard whether that synchronisation has to be local, meaning affecting only the neighbouring nodes involved in the computation process, or global, meaning that the tightness of synchronisation throughout all parties is essential and limiting. This is bound to the nature of the underlying computational problem. Time Division Multiple Access (TDMA) [36], is the standard type of communication protocol in cellular networks and wireless data transmission in sensor networks. As the different parties are only allowed to transmit at specific points in time, the transmissions are partitioned into slots which have to be aligned for all participants in order to prevent interference. The higher the synchronisation quality, the lower the required guard times [53], resulting in shorter dead times and higher data throughput in the network. With their architecture, TDMA based applications usually require low local skew to achieve optimal performance. Computational clusters, in turn, benefit from a low global skew [82]. This is related to the fact that in many cases, global consistency of data access and communication timestamps has to be ensured, requiring that all entities have a perfect notion of synchrony to prevent erroneous accesses and data corruption.

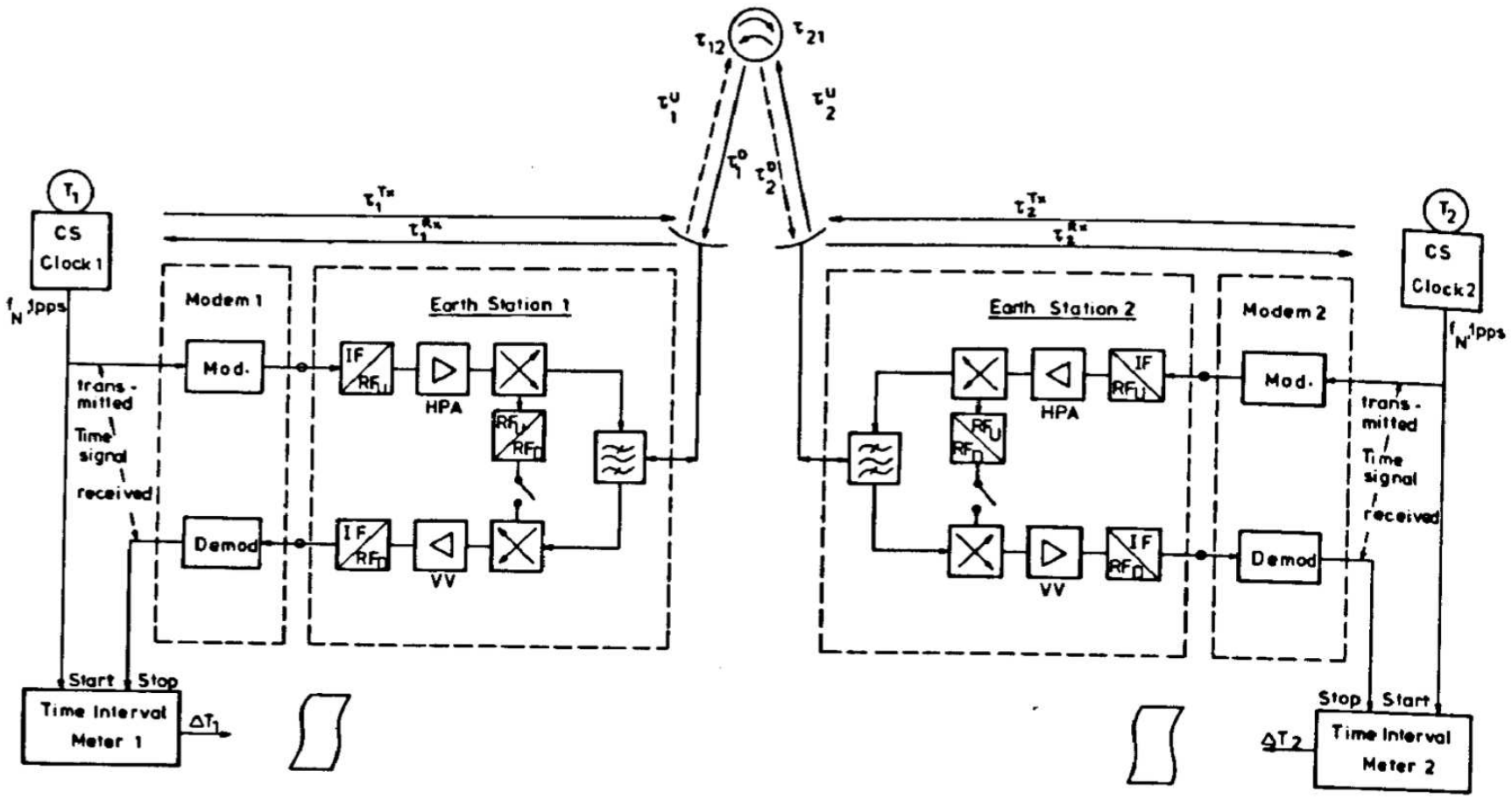

FIG. 1: Two-way time transfer: Basic concept

Figure 4: Two way time transfer over satellite as performed in GPS, the more performant counterpart to GPS common view time transfer [9], as presented in the orginal paper from 1983 [47].

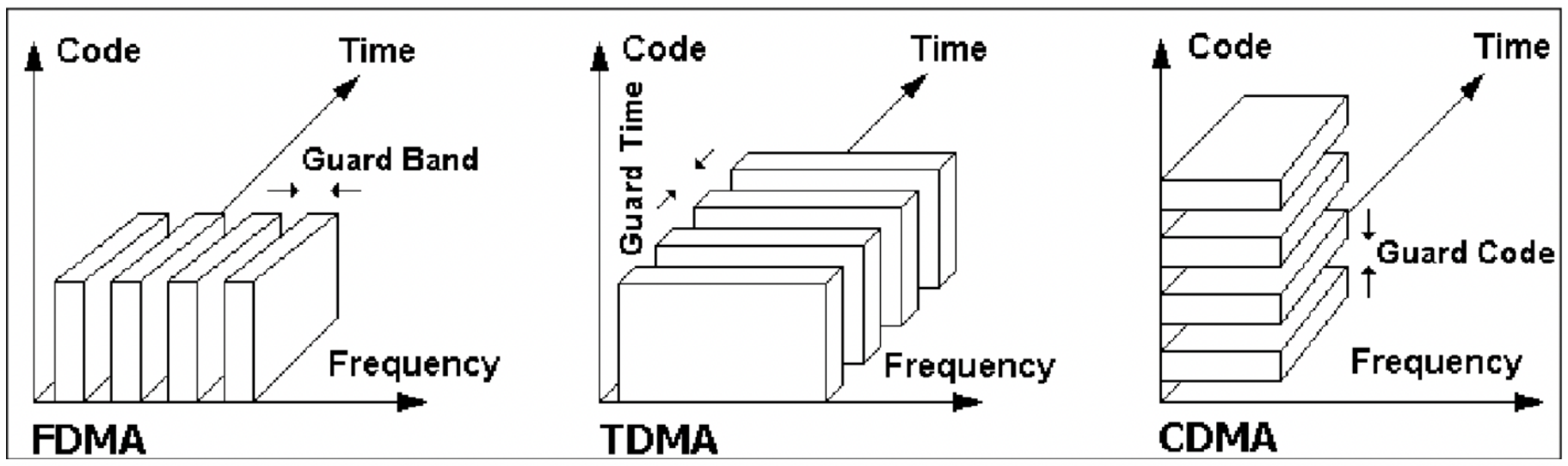

Figure 5: Time Division Multiple Access (TDMA) transmission and its two alternatives, Frequency Division Multiple Access (FDMA) and Code Division Multiple Access (CDMA). From: [62].

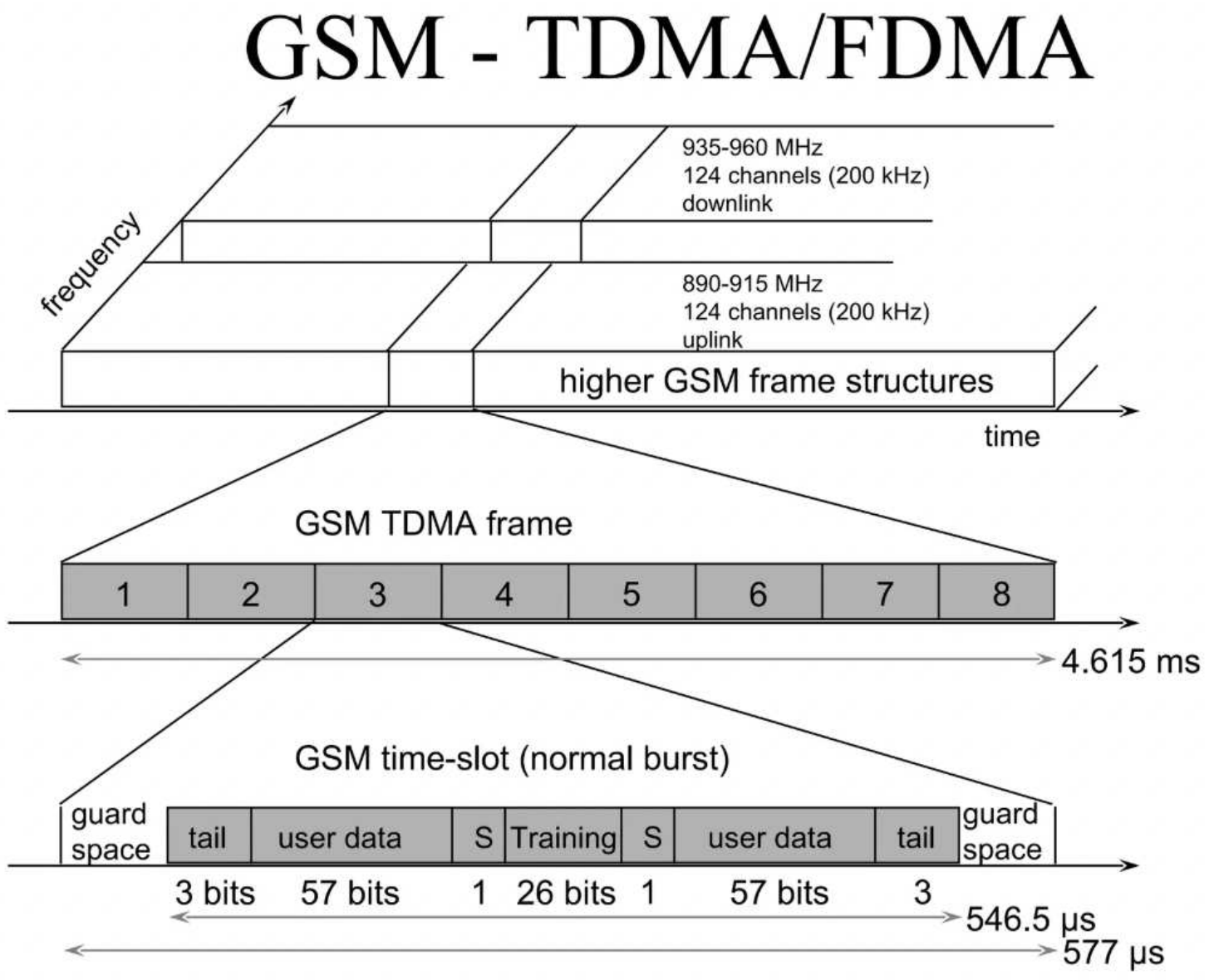

Figure 6: Guard times in a GSM-TDMA transmission frame. From: [24].

## 1.3 Being not so on time - asynchrony induced by noise

In an ideal world, a single time reference per local unit would be sufficient to achieve the above goal of perfect synchrony. However, all local time references - whether they are a high precision, high accuracy reference such as all kinds of atomic clocks or a more common lower quality reference such as relaxation oscillators and quartz oscillators - are affected by noise processes yielding to frequency drift. This drift alters the output of different clocks in a disharmonic way.

While each observable noise type has specific properties that influence the behaviour of a circuit, the details of these effects are considered out of the scope of this thesis. Further information for the interested reader can be extracted from [109]. For our purposes, it is important to note though that specific and important types of noise processes, such as white gaussian noise and $1/f$ noise can be described mathematically, the resulting models cannot be used for prediction.

Hence, it is not possible to directly correct the effects of noise at the source [28]. Not only oscillators but all electronic circuitry suffer from noise [64], adding another error term to the output of any electronic system. This creates the need for synchronisation procedures which can overcome the undesired effects in-duced by these noise processes. For time synchronisation, there are many such algorithms, for example NTP [76] and PTP [20] for centralised synchronisation in large scale networks but also more distributed algorithms which do not aim to map the behaviour of all clocks to a chosen reference clock ensemble. The gradient clock synchronisation method is one of them. While NTP and PTP implement a tree like topology [77] which synchronises all other nodes to a central reference clock ensemble located at the root, the GCS algorithm allows to focus on the immediate neighbourhood of every node in the network, synchronising clocks locally.

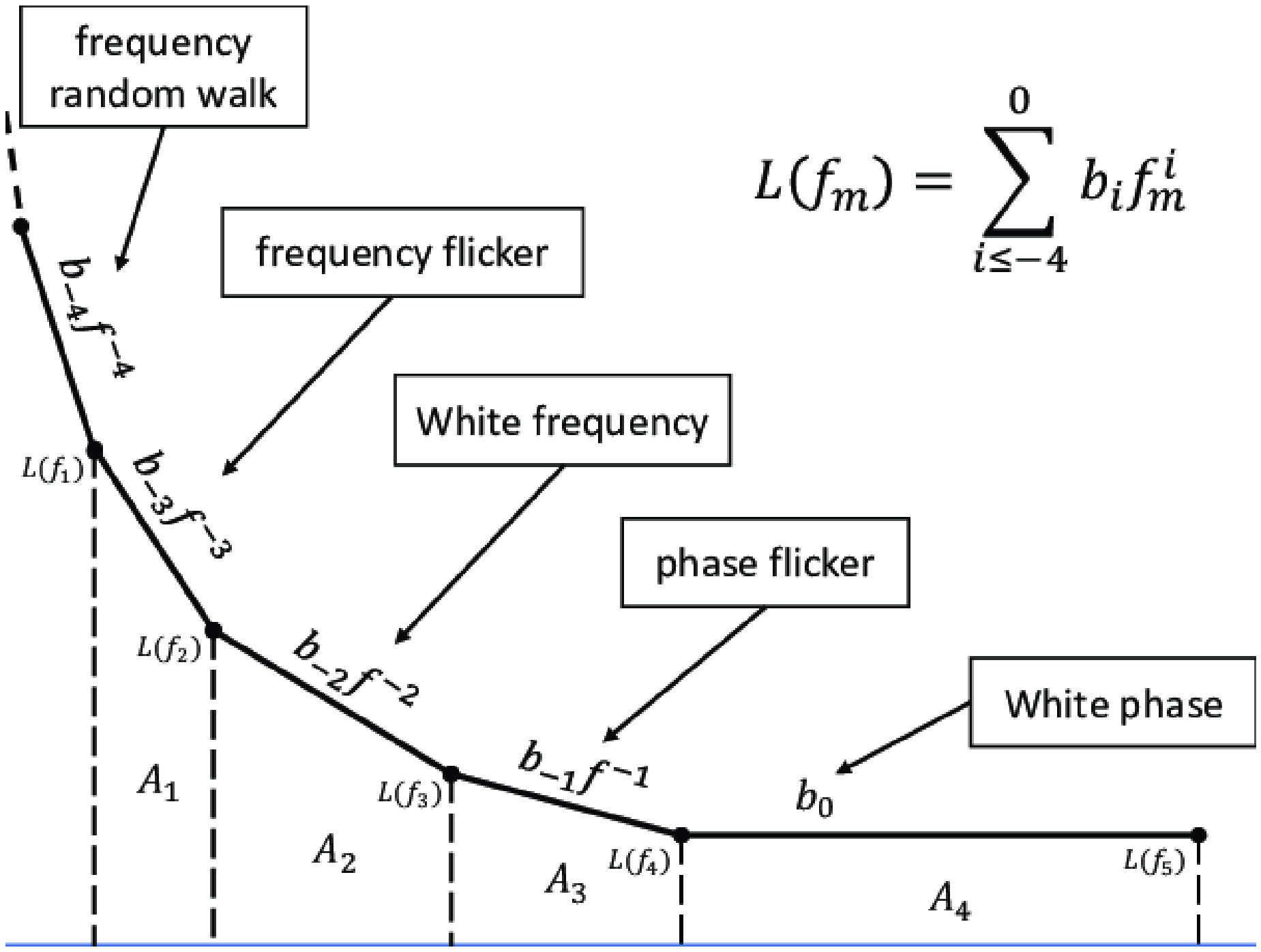

Figure 7: Allan deviation plot of the general structure of phase noise in an oscillator From: [75].

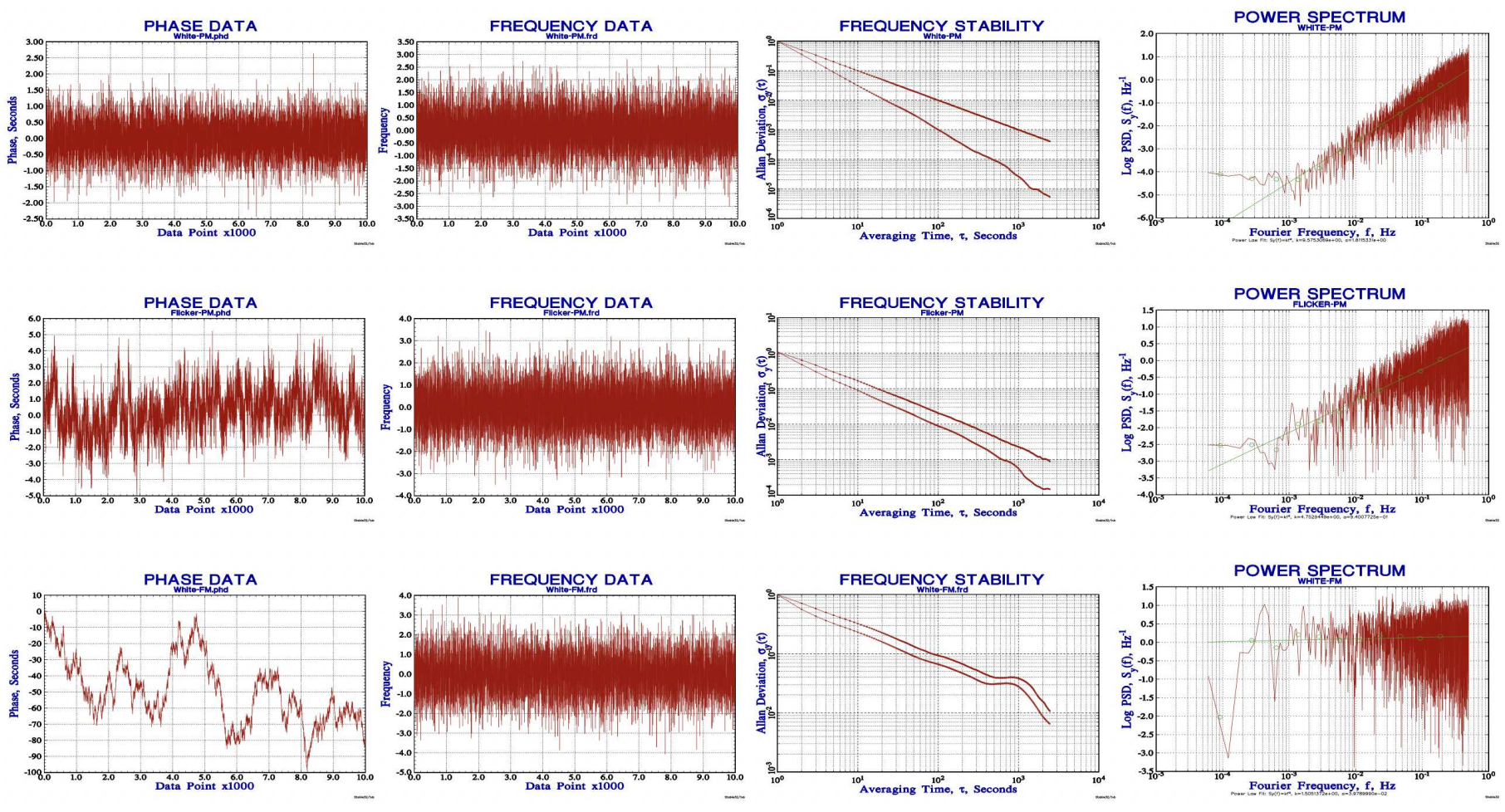

Figure 8: Phase data, frequency data, Allan deviation, and power spectrum plots of white noise (first row) and pink noise (second row) From: [104].

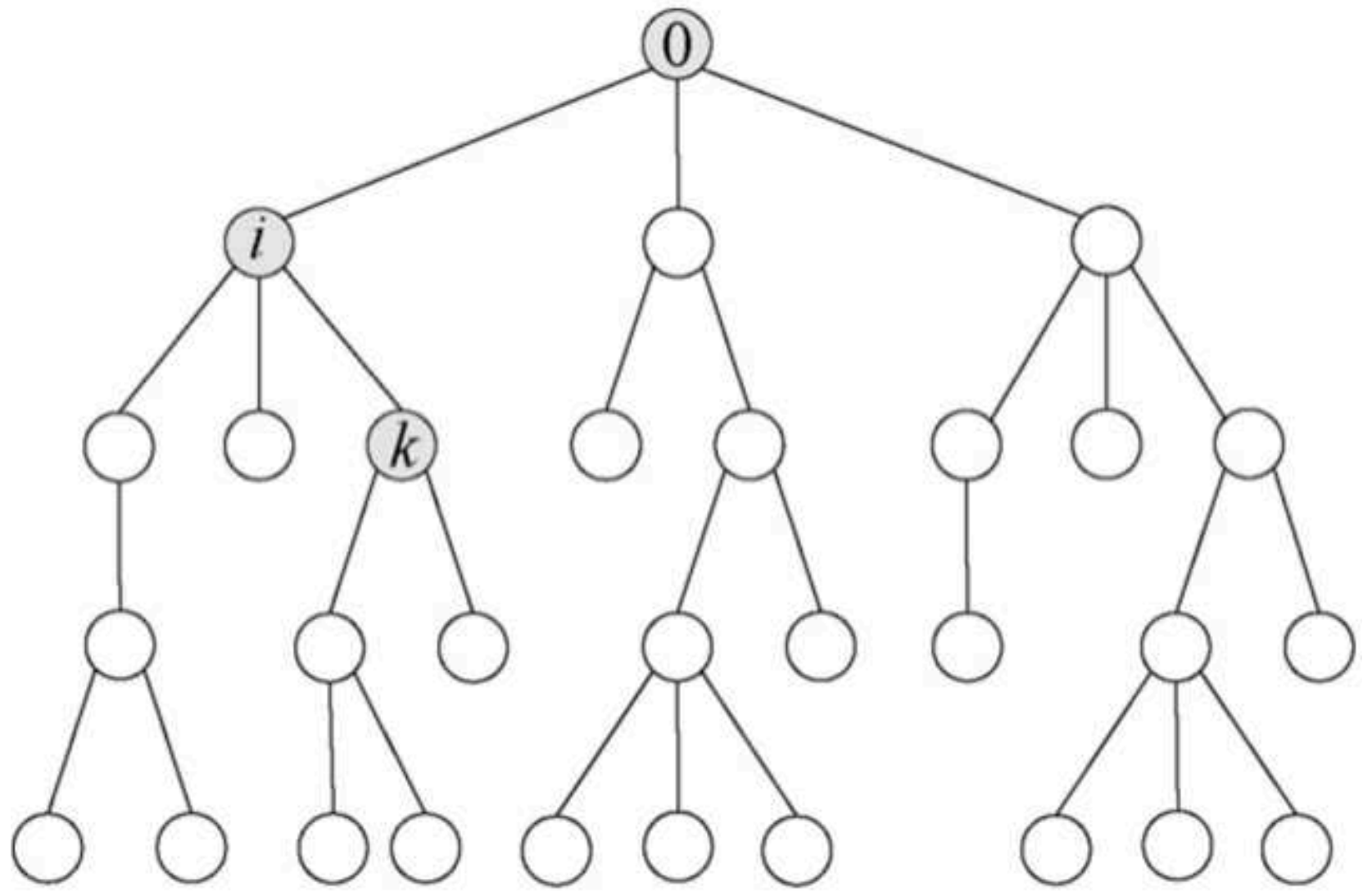

Figure 9: In a tree-like topology, the skew of node k is influenced by all nodes along the path to k. Adapted from: [45].

# 2 Related work

## 2.1 Gradient Clock Synchronisation

This subsection details the origins and core limitations of gradient clock synchronisation.

### 2.1.1 Fan's & Lynch's lower bound

In their 2004 paper, Fan and Lynch [35] establish the gradient clock synchronisation problem by defining the gradient property: the skew perceived by a node grows with to its distance, in terms of path length in the network graph, to the node its estimating the skew of. This property implies that nodes that are close together in distance are also closer synchronised than nodes that are further, up to the diameter of the network, apart from each other. This makes clock synchronisation algorithms based on gradient property highly suitable for decentralised architectures, such as large sensor networks in which a high degree of synchronisation between adjacent neighbours is required but also various kinds of ad-hoc networks, that might benefit from a decentralised boot-up process.

Furthermore, Fan and Lynch prove a matching lower bound on the local skew any algorithm fulfilling this property might have. This lower bound of $\Omega\left(d + \frac{\log D}{\log \log D}\right)$ showcases that the gradient property, beyond binding the local skew to the distance $d$ in terms of propagation delay of a hop in the network, depends on the network's diameter $D$. This implies that large networks might suffer from poor performance, hinting at the approach's bottleneck.

### 2.1.2 A simple GCS algorithm

Based on Fan's and Lynch's groundbreaking work, Lenzen, Locher, and Wattenhofer introduced an algorithm in [66] achieving the conjectured bound of $O(D)$ for the global skew, i.e with a maximum skew within any distance in the network graph equal to the diameter of the network $D$, as predicted by the authors of [35]. This algorithm also achieves a local skew of $\log(D)$. The GCS variant presented in Section 5 and Section 6 of this thesis follows the outline of the algorithm described in Lenzen, Locher, and Wattenhofer's publication as well as on a slightly different variant presented in the unpublished book by Függer et al [39].

## 2.2 One algorithm, many possible use cases

The core GCS algorithm was adapted to more uses cases in further work, of which the following section will present an excerpt. Various types of networks, with both fixed and dynamic topologies were explored. The resulting insights on the network properties of GCS yielded closer investigation of specific scenarios such as fibre wired networks but also application in which communication is based on radio transmission in which could benefit, in terms of performance or robustness, from the distributed synchronisation protocol offered by GCS.

### 2.2.1 GCS in dynamic networks

In Kuhn's, Locher's and, Oshman's 2009 work, the GCS algorithm is adapted to dynamic networks [59]. The resulting algorithm variant is able to deal with non static networks, i.e, allows synchronisation in a network whose set of edges is updated at specific intervals, modelling the dynamic progression of, for example, a wireless network where links between edges are acquired and lost again over time. It is able to deal with distorted delays resulting from lagging messages by filtering out messages arriving outside of an estimated window of correctness. A matching stable bound for the global and the local skew of $O(n)$, where $n$ is the number of nodes present in the network graph, is shown for this procedure. This result is relevant in the context of this thesis as it lays ground for dynamically evolving networks in the context of GCS, of which this thesis will describe a variant of.

### 2.2.2 GCS in wireless networks

Fan, Chakraborty, and Lynch created a variant of the GCS algorithm adapted to operate with high energy performance in wireless networks in their 2004 paper [34]. In [100], Wattenhofer and Sommer present for a GCS-based synchronisation protocol for wireless sensor networks as well as an implementation of a framework running plain GCS as described in [66] for Mica2, a specific type of wireless sensor development platform. This 2009 paper showcases the practical usability of GCS in a real world application, which delivered a decent performance of $4\mu s$ in the evaluation performed by the authors. This paper highlights the practicability of the approach in an actual use case but also underlines some challenges encountered in the implementation, such as issues with the timestamping procedure executed during communication and delay measurements, that will be picked up and discussed in this thesis.

### 2.2.3 GCS in networks on chips

Last but not least, the GCS algorithm was explored in the context of clock distribution on computer chips. Függer, Medina and, Bund proposed PALS, Pleisiosynchronous and Locally Asynchronous Systems, in the paper of the same name [13]. Their work describes a model and variant of GCS allowing to synchronise adjacent, independently operating clock islands on computer chips. The authors also provide a hardware description of the modules required to achieve an implementation of the suggested GCS algorithm as well as simulations on a small grid of such independently clocked blocks. In comparison to classical clock trees, the PALS approach yields an almost stable progression of the skew between adjacent neighbours in clock networks of increasing diameter in comparison to the linear skew growth in the classical approach. This result sets apart some structural advantages of the GCS algorithm when applied to network on chip scenarios, which can benefit from GCS's property of maintaining a tight local skew and a linear global skew even in large scale networks.

# 3 Motivation

## 3.1 Implementing GCS

The GCS algorithm was mostly studied in a theoretical setup so far, yet there are multiple demonstrators that have been build for various use cases such as synchronisation in chips and wireless synchronisation of networks. The related scientific papers mostly approach their respective use case from a highly theoretical perspective, abstracting away many details and specificities that are however relevant at the implementation stage.

### 3.1.1 Theoretical perspective

From a theoretical perspective, a leaner model is preferred as it simplifies both proofs and algorithm design by limiting the amount of parameters as well as the number of interactions between them. This however often results in suppressing important correlations between factors in the model that could be exploited to improve the performance of the algorithm.

The model underlying the basic GCS algorithm presented in [39] provides such a highly abstract description of reality. For example, it is assumed that the physical clocks, regardless of their type, can be modelled as continuous functions. While this is true for quite some frequency references, such as harmonic oscillators, it does not apply to the broad, general case [67]. Moreover, the physical links between nodes taking part in the communication protocol are modelled as having a semi-static, one-sided error. While this simplifies the model, as there is a simple, static bound on how much the quality, in terms of accuracy on the delay length, of a link can vary, the actual link quality of fibre but also wireless links, as it can be observed in live measurements, varies quite a lot over time [31]. This means that the link quality can be much better at many points in time than the worst case assumed in the plain, original model. While this simplification allows provable lower and upper bounds on the skew performance of the algorithm, the resulting bounds assume a worst case scenario which yields much poorer performance than the average case.

From a theoretical perspective, the above approach does not yield any major drawbacks when attempting to translate model and algorithmic pseudocode into an actual implementation. The model of the plain GCS algorithm actually yields a very simple implementation, given sufficiently good estimates on the worst case behaviour of hardware clocks and links. Yet, there are some issues with the model presented in [39] as well as more intricate features of links that can, from an engineering point of view, be exploited to significantly improve the performance in terms of worst case bounds on the skew and that this thesis will attempt to shed light on.

### 3.1.2 Practical considerations and issues

The high degree of abstraction that yielded good results in theory however does not fit two major issues encountered when attempting an implementation of GCS.

The first concerns the clock model. Prior models do not specify any restrictions on the used hardware clocks or their behaviour, yet assume that they will always behave as required. While specific classes of oscillators fulfil these restrictions under certain pre-conditions, the assumption that the clock signal is continuous and the rate differentiable as this restriction is necessary to guarantee convergence of the GCS algorithm. However, this does not hold for all types of frequency references and even for very stable references, not at all times [41]. This is due to the fact that the clock's frequency and phase drifts do not have to be predictable and stable within a pre-specified interval [40].

The second and even more important issue is related to the link model. The one-way measurement scheme deployed in [39] cannot deal with the variable delays and uncertainties mentioned in Section 3.1.1. This means that the current state of the art model cannot be deployed in practice without restrictions as specified in theory. Hence, the core goal of this thesis will be to create a model that can cope with variable delays and uncertainties and thus can be translated into an implementation without further ado.

### 3.1.3 Choice of frequency references types covered

As a conclusion, we aim for a model that suits the theoretical desire for simplicity and allows a straightforward application to practical implementations. This also motivates the choice of the type of oscillators whose properties are represented in our model. In all electronic industries, LC and quartz oscillators are the most common type of frequency references [6]. This is due to their simple circuit design but also to their properties. A simple LC oscillator circuit is composed of an inductor and a capacitor [107]. Various different subtypes are known. One of them is the crystal oscillator, which belongs to the family of the RLC oscillators. It additionally contains a resistor. One example of a very simple crystal oscillator circuit is the Pierce circuit [6].

The most common type of crystal used for crystal oscillators is the quartz crystal. It's cut, that means production quality, has a high impact on the properties of the resulting oscillators [6]. More details will be elaborated in Section 4.

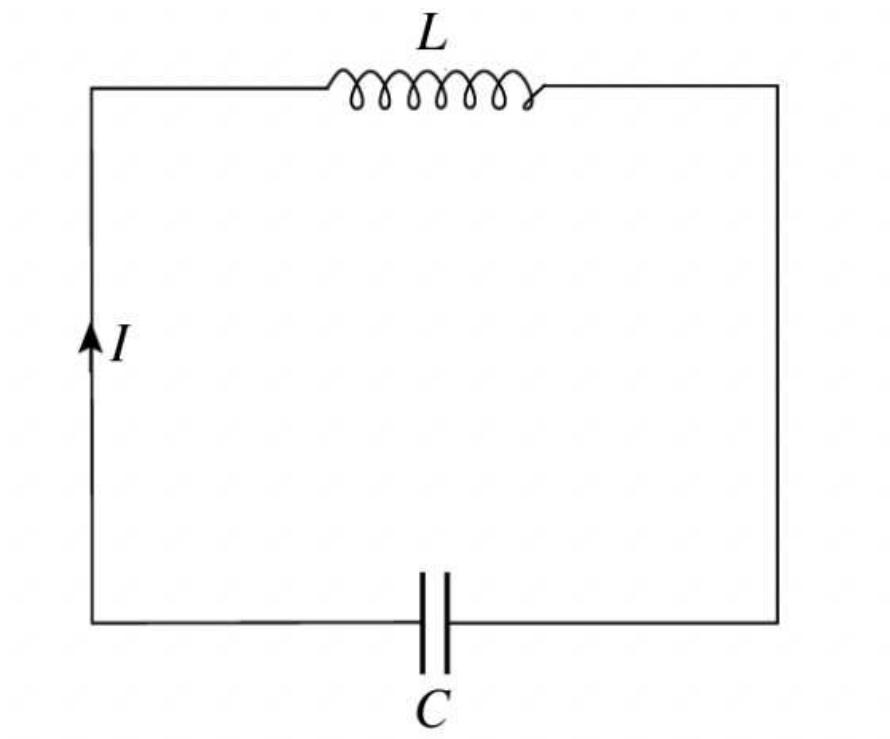

Figure 10: Schematics of a simple LC circuit. From: [107]

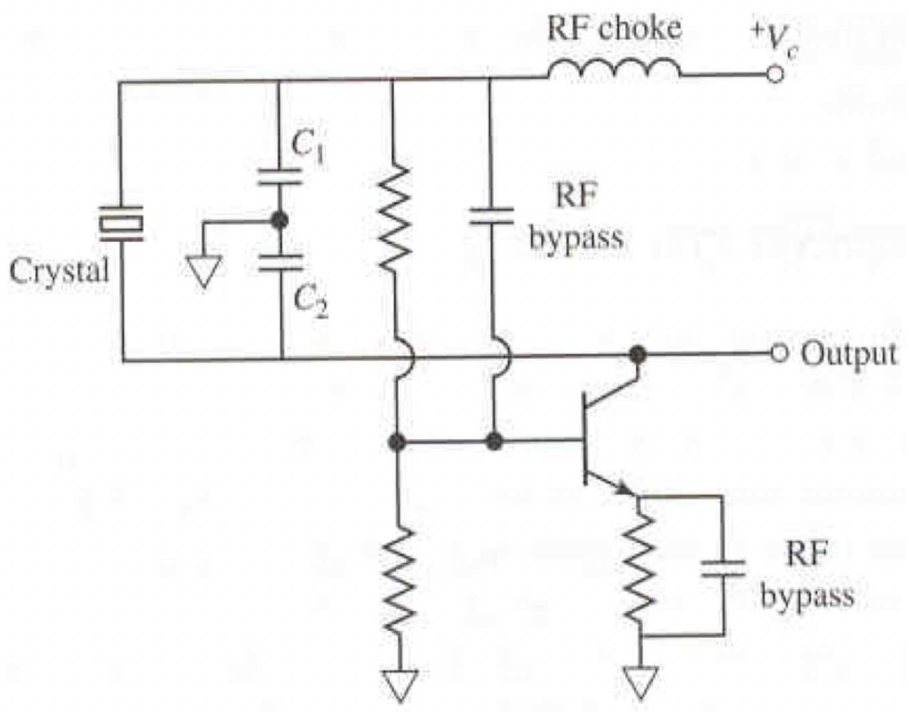

Figure 11: Schematics of an advanced Pierce circuit. From: [102]

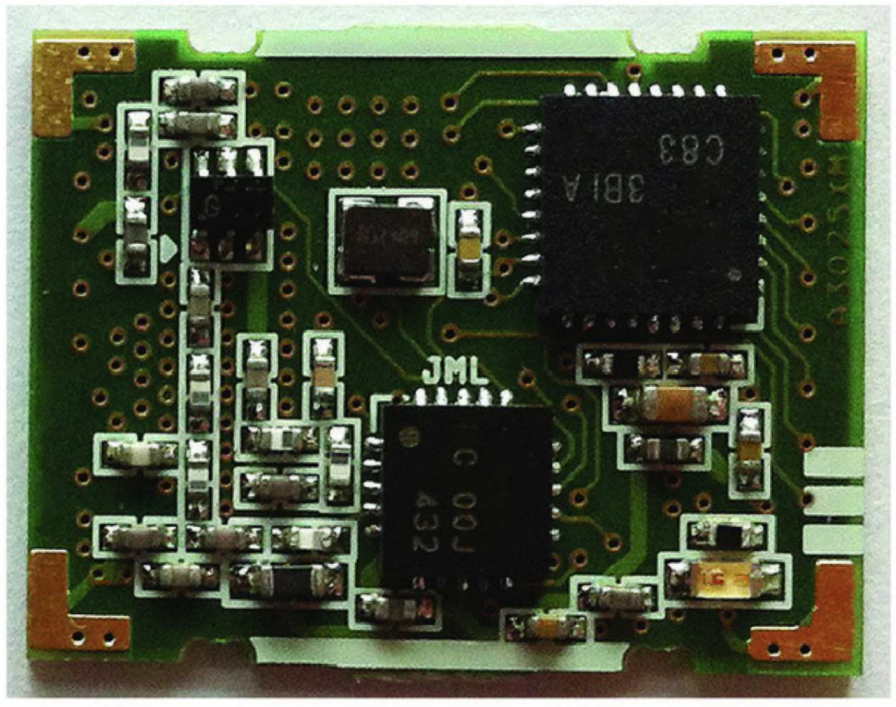

Figure 12: Components of a digitally controlled temperature-compensated crystal oscillator (DCTCXO). From: [57]

Both the LC oscillator and the crystal oscillator are harmonic oscillators [6]. This reflects in their core properties:

1. Stable frequency: LC and crystal oscillators are known to have rather stable frequencies under common environmental conditions [6]. Typically, the frequency of operation of an AT cut, the most common cut, has a frequency constant of 1.661 MHz·mm [94].

2. Band-limitation: additionally, both are band-limited [4]. This means that their rate can only change at a finite speed and that ultimately, their frequency of operation is bounded. The quality factor Q of the oscillator, which corresponds to the natural propensity of damping of the crystal defines this rate. Quartz oscillators have typically a high quality factor [61], implying that their rate can only change slowly and places the possible variations of the frequency of operation within small interval, which implies a high stability.

These two central properties, namely their high stability as well as their bounded frequency, make their behaviour well-defined for our modelling purposes. Details on how we will exploit these two properties in our model will be given in Section 4.4 and Section 4.5.

## 3.2 Filling the gap between theory and practice

### 3.2.1 Core challenges

The core challenge around building a concise yet highly performant model fulfilling this goal is choosing what parameters one models but also what scope the parameter covers. To this effect, the parameters chosen in related work will be reevaluated with the goal of exploiting features of implementation level details, mainly being:

1. Properties of common physical oscillators used as reference clocks

2. Methods used for physical time/delay measurements

3. Properties of physical time/delay measurements

The resulting model will be fitted to the three major use case of graph based synchronisation as performed by the GCS algorithm. As the literature in Section 2.2 suggests, the three core use cases we will study are:

1. Wired networks

2. Wireless networks

3. Networks on chips

With the above, we can define and elaborate the core goals of this thesis.

### 3.2.2 Making realistic assumptions on the network's and clock's behaviour

As a first part of reworking the model underlying the GCS analysis, we will have a closer look at the properties of harmonic oscillators as presented in Section 3.1.3 and build a formal abstraction modelling their behaviour in terms of possible rate variations, predictability and bounds on the clock drift as well as how a correction of the clock rate can be modelled mathematically in Section 4.3.

This re-modelling is done with the purpose of clarifying the impact of the properties hardware oscillators on the performance of the algorithm. More precisely:

1. The properties of the rate and its variations, with the purpose of formally describing that harmonic oscillators can indeed be modelled as differentiable functions.
2. The behaviour of hardware clocks upon rate corrections, with the purpose of showing the limitations of the above assumption, yet proving that it is sufficiently good for practical usage in implementations
3. Delivering a simple clock model that can be used for engineering purposes during any implementation of GCS in any type of system, wether it is a network on chip or a large-scale wired or wireless network.

### 3.2.3 Modelling measurements true to reality

As a second part of this thesis, we will adapt the state of the art model to mirror the properties of measurements as they are performed in the field. As we will see in Section 4.6, the structure of the typical measurement scheme used in practice differs quite a lot from the measurement scheme presumed in prior work. Consequently, we will adapt the model to mirror the behaviour of the two-way measurements as done in practice instead of the theoretical one-way measurements described in related work.

With this fundamental modification of the model, we aim to provide:

1. A realistic link model which mirrors the behaviour of links as they might be observed in practice, ensuring predictability of an implementation's behaviour at the theoretical, i.e, planning stage.
2. A description of potential encountered pitfalls in the process of parameter optimisation, allowing better understandability of the underlying processes and phenomena.
3. An improved model of uncertainties which distinguishes between predictable and non predictable error components, resulting in a highly improved performance in terms of accumulated errors.

4. A roadmap from theory to implementation, clarifying both the theoretical and the engineering background behind method and setup deployed.

5. Overall, an implementation ready framework allowing to fine tune network parameters based on the requirements of the three big use cases: networks on chip, wired, and wireless synchronisation networks.

### 3.2.4 Summary

In summary, this thesis aims to:

1. Build a model that properly mirrors the practical behaviour of most harmonic oscillators.

2. Build a model that fits all sub-use-cases of the three big use cases of GCS in practice, namely wired and wireless synchronisation of any network and synchronisation on chips.

3. Assert the behaviour of clocks, paths and path measurements as they are performed on hardware properly.

4. Not alter the core functionality of GCS as we would like to keep the same guarantees on the algorithm as previous models.

5. Improve the performance of the algorithm in term of lower and upper bounds on the uncertainty and the local and global skew.

# 4 Model

## 4.1 What do we need to model?

### 4.1.1 Gathering requirements

In Section 3, we covered the motivation and resulting goals that would be desirable to achieve. Based on this list, one can formulate properties and performance markers the model and the algorithm should achieve for the oscillators for harmonic and relaxation oscillators:

1. The model should realistically mirror the typical behaviour of the harmonic oscillators in Section 3.1.3. We limit the scope of the model in this work to such oscillators.

2. The model should include computations, in particular time offset measurements, as they happen in real time systems.

3. The model should distinguish between short and long term components of the error.

4. The model should reduce the contribution of the system's uncertainty on the error to the skew visible between adjacent neighbours.

5. As we aim to maintain all properties of the GCS algorithm, the outline of proofs should not fundamentally change.

6. The model should fit the three flagship use cases of GCS namely, wired and wireless synchronisation of computing networks and synchronisation on chips as described in Section 3.2.1.

This list of requirements will be used in Section 7, concluding this thesis, to evaluate the results of the model and algorithm developed in the next two sections of this thesis.

### 4.1.2 High level model

The first step towards a solution that achieves the goals mentioned in the motivation is a description of which aspects of the underlying problem one has to model. The choice of the right level of abstraction as well as the right abstractions are key to deliver an algorithmic model that combines simplicity and guarantees performance. In this work, we will not start from scratch but mainly reuse the model assumptions suggested in prior work, see Section 2. In prior work as well as in this thesis, the computational model of GCS aims at mirroring the following core elements of a real time system:

1. Clocks: are the core of what we aim to describe. Their highly complex behaviour needs to be abstracted to a decent, workable level without losing track of their most important properties. We will more clearly justify why the chosen abstraction fits all important properties.
2. Communication links: are an important part of the algorithmic framework and have a considerable impact on the overall performance. This is why we will pay close attention to them and refine the properties we model about them compared to previous models, in which they were represented as symmetric, uniform, and normalised in delay and uncertainty.
3. Computations: are mostly covered by what one calls "the algorithm", more precisely a description of the algorithmic procedure underlying GCS. Unlike previous models, we will distinguish more clearly between computations related to skew measurements and computations related to skew correction.

With this overview, we will delve into more details.

## 4.2 Modeling computations and message passing

As a first step, the computational model needs to be determined. To fit our purpose of usage, this part of the model needs to account for the side effects of clock skew and message delivery delays.

### 4.2.1 Choosing the right degree of abstraction & asynchrony

We start with choosing a computation and communication model. Three different approaches are known from previous work [39, Chapter 6 and 7]: Synchronous Message Passing (SMP), Asynchronous Message Passing (AMP) and Timed Message Passing (TMP). Their respective properties are summarised below:

SMP [39, Chapter 6.1 and 6.2] offers delay free communication in a fully synchronous setup. Computations and communication is performed in rounds: at the beginning of each round, all entities exchange messages that are delivered instantaneously i.e in zero time. After this step, computations are performed in a "lock step" manner: all computation steps are instantaneous. This computational paradigm allows for very simple and clean algorithms, as all asymmetries in communication and computations are abstracted away. However, it does not provide a realistic setup in general and particularly for our use case and cannot deal with any asynchronous communication.

AMP [39, Chapter 6.3 and 6.4] offers a framework for fully asynchronous communication. Messages are delivered at an arbitrary point in time. The algorithmic procedure needs to be able to deal with unbounded message delays. While this offers a highly flexible model of computation as message transfer is event based meaning that the reception of a message triggers the execution of a new set of instructions. Consequently, the time complexity of an algorithm is determined by the maximum message delivery delay, that can be arbitrarily large, depending on what restrictions on environmental effects are assumed. This yields comparatively poor efficiency and especially, as this model doesn't account for a local clock or notion of time, an inability to measure time intervals which is an undesired property given our aim to reach a computational efficient procedure for synchronisation. Beyond this, the assumption of unbounded message delay is highly impractical in our setup as it doesn't allow detection of communication failures. This inadvertently limits the potential to include fault tolerance already at the level of the chosen computational model.

TMP [39, Chapter 7] offers an intermediate solution: all messages need to be delivered within a bounded delay of known size. Furthermore, the computation time is also bounded. An additional constrain is introduced though: each computational unit is required to have its own hardware clock, that is a local time reference, providing a local notion of time to computations. Note that this notion of time is relative i.e each hardware clock describes a notion of time which is absolute locally but relative compared to others references in a connected network. This offers a compromise in terms of performance and simplicity of performance analysis as well as the possibility to detect crash faults by implementing timeouts. These features make TMP the most versatile and adapted choice for the problem as mentioned in prior work [66]. This assumption still holds in our use case. However, in TMP, one has to deal with uncertainties related to the exact time of message delivery i.e uncertainty about the exact time of arrival as the model doesn't enforce a strict notion of synchrony, but a relaxed one since the local clocks suffer from bounded clock drift and an uncertainty is specified on the message delay. This relaxed notion allows not only to model the drift but also also various kinds of measurement errors which, in practice, have an impact on the performance.

#### 4.2.2 Towards realistic assumptions - TMP with a twist

As we determined, TMP offers the most adapted features and therefore is kept as computational paradigm. This section is devoted to fine tuning a rather broad basic definition lend from [39, Chapter 7], which we described in the section above.

Definition 1. Computational model
The computation time per instruction is clearly limited by an upper bound. Furthermore, within the message passing protocol, computations are assumed instantaneous. Finally, computational units have a local notion of time in form of a local clock which has zero query cost and zero query delay.

The assumption that computations are instantaneous might seem unrealistic in an implementation-near setup as they inherently require time in practice, nevertheless, we abstract this property away in our model of computation. This has multiple reasons. The most important one is the can of worms opened by asymmetries in the computational flow: as we cannot assume that time elapses equally fast at all computational units, the same computation might have different durations at different units, even if their computational cost is unitary in theory. Accounting for this would induce a high degree of freedom in terms of asymmetries between start and end of a computational step at different units. Since all units also exchange messages and message are sent on an event basis, the set of possible variations, more precisely the variance of the computational delay, has to be accounted for. This would require a large amount of parameters, which in turn, would lead to an overly complex analysis. We hence choose to introduce computational delays only at specific, critical points such as the path delay and offset measurements, required to estimate the communication delay in the message passing model. This also motivates the choice of a bounded message passing model, where the upper bound on the message delivery delay includes an upper bound on a computational step.

Definition 2. Message passing model
Message delivery is always guaranteed and message delays are bounded.

For message passing, the query cost and delay to fetch a timestamp from the hardware clock upon sending or arrival of a message are assumed to be zero. This discrimination is rooted in the desire to clearly separate the computation time required by the measurement protocol from the runtime of the GCS procedure. It has to be noted that one could easily argue that fetching the hardware clock in the time stamping procedure can be accounted for in the computation time required by the measurement procedure. Yet this would add in the unnecessary constrain of having to specify a variable delay that in practice depends on many factors. For the sake of simplicity, we hereby abstract away this delay and model it as uncertainty by arguing that, at implementation level, it is unmeasurable and would be approximated by a guess and counted towards an additional error terms in the upper bound on the message delivery delay. Finally, the above assumption that messages are always delivered within the bound exclude faults from our model of computation. Note that this constrain can be relaxed or even lifted, yet we will not cover this as its many implications are considered out of scope of this thesis.

So far, we followed the outline of the basic TMP model. Now, we will define the difference between the classic model of TMP presented in related work [39, Chapter 7] assuming one-way measurements and our model, in which a two-way measurement is performed. In this work, an abstract TMP message can encapsulate the core elements a two-way clock measurement as executed on physical systems in the following way:

Definition 3. Messaging protocol
Messages in our computational model mirror physical two-way measurements:

1. Request: Party A sends its own clock value to party B and requests the clock value from party B.
2. Reply: Party A can then reconstruct the message delivery delay as well as an estimate on the clock value of party B by computing the offset between the respective sending and arrival timestamps transmitted in the payload of the Reply message.

The above abstract messaging protocol describes the high level protocol underlying a two-way path and clock offset measurement as implemented in physical systems. Two parties A and B are connected by a link which is characterised by a message delivery delay [29], that roughly corresponds to the travel time delay, as well as two metrics for uncertainty. The first type of uncertainty in this model is the uncertainty on the message delivery time, that is the uncertainty about the exact time of arrival caused by some unascertainable delay induced by the difference between arrival and processing in terms of time stamping of a message.

Beyond this first type of uncertainty, another type has to be specified as inherent property of physical links: the uncertainty on the path asymmetry [25, 114], that is the difference in delay between forward and backward path in the two way measurement induced by noise processes. The systemic dependencies of these parameters yield very desirable properties that, as in physical implementations of two-way measurements, allow deducting the message delivery delay and obtain clean measurements of the clock skew, that is the offset, between party A and B. A detailed description of the properties of measurement and how they can be used to reduce the global uncertainty in the model are topics covered in Section 5.

## 4.3 Modeling the communication network

Given the abstraction definition of computational units and links described above, we can specify the network topology and parameters in more details.

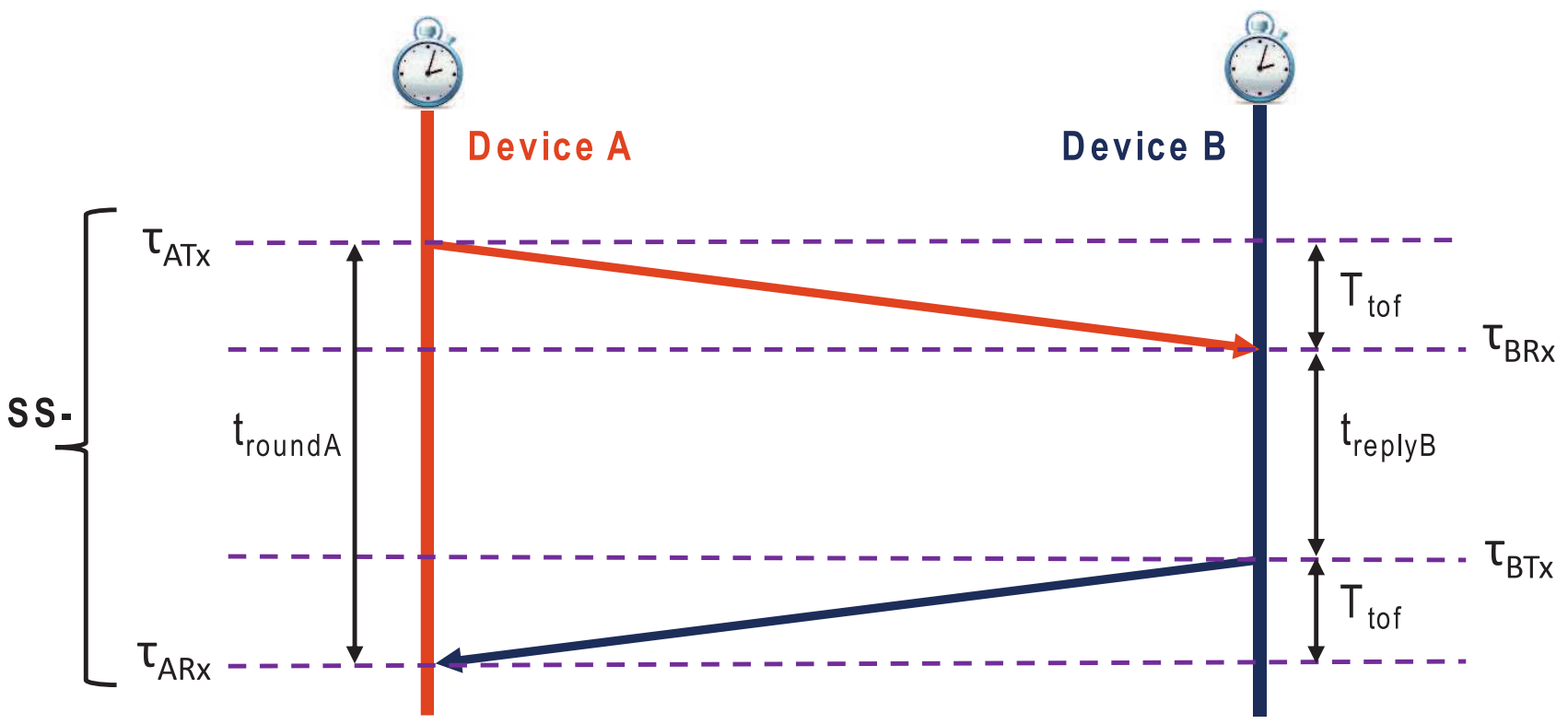

Figure 13: The single-sided two-way measurement paradigm is the state of the art approach for ranging measurements by the IEEE Standard 802.11az [71, 50]. Adapted by hiding double-sided protocol not required in the context of GCS from: [71]

### 4.3.1 Any connected graph can be a network

By definition, our network is described by the graph spanned between computational units, nodes are computational nodes and edges are communication links between units [66]. We do not apply further restriction to the network for all practical use cases which in the scope of this thesis beyond requiring that each communication unit/node has at least one link/edge to another unit as well as that edges in the graph can only exists between neighboured nodes, meaning that communication can only take place within short distances covered by a single edge whose distance cannot surpass the length of two consecutive edges in its local neighbourhood. Compared to prior work, we require one more constrain on the graph: all edges of the network graph need to be bidirectional. Formally, we define the network graph $G$ as:

Definition 4. Network
The communication network is represented by a connected graph $G = (V, E)$ with bidirectional edges. We will use $n$ to denote the number of vertices, which we call nodes, and $m$ to denote the number of bidirectional edges in $G$. Furthermore, $D$ denotes the hop-diameter of $G$, that is the shortest unweighted path of maximum length between any two nodes.

This follows from the desire to be able to model the asymmetry that is present between the forward and backward path on a physical link without requiring two separate edges for the forward and backward path as this would yield the need to fundamentally change how the algorithm handles messages from neighbours and complicate the analysis. This issue can be circumvented by simply encapsulating the delay asymmetry, that is the difference in delay length between forward and backward path as a parameter of the edge. This difference will be called asymmetry factor. This results in a very flexible topology that can be used to model many applications: the same model can cover a network of wired but also wireless stations with various distances between nodes yet also offers the possibility to model the clock network in a processor chip.

### 4.3.2 Fitting our three top-level use cases

We next discuss the properties of the network in concrete applications, especially in the context of the three flagship use cases highlighted in Section 3.2.1 of this thesis.

Wired networks: are the simplest and most straightforward application of this model as the delays in typical wired networks are rather stable, easing up the parameter choices. Typically, we expect complex but regular topologies and low to medium edge degree per node [46]. The resulting network graph is usually rather large but is, comparatively to the number of nodes, rather sparse. This is related to the fact that minimal delays are the critical element in most wired applications. For example, the typical network topology in a data center, where each computational unit will not exceed 10 to 20 fibre connections from and to other units yet the spanned virtual network might have a hypercube or grid topology [63, 113]. These complex topologies are chosen to ensure low latency by allowing for minimal wire lengths and hence minimal delays in peer to peer communication while maintaining highly flexible communication paths. This also allows to keep the asymmetry factor within a rather small interval and keep its variance over time rather small too. This is achieved for example in the White Rabbit protocol developed at CERN. White Rabbit allows for symmetric, delay stable fibre links between two end points [96]. The symmetrisation is achieved by transmitting two signals with different wavelengths over the same fibre and performing some pattern matching on the signals to recover and deduct the asymmetry [92]. The White Rabbit protocol performs well on short distance, i.e for links of length of up to 10 to 20 km, where it achieves an asymmetry of as little as 100 picoseconds between forward and backward path [96]. For larger distances, different protocols are required as amplifiers become necessary to compensate the loss of signal strength [98]. Amplifiers induce a high degree of asymmetry due to being unidirectional in most cases [25], requiring different fibres/wires for forward and backward path. This effect becomes increasingly prominent over larger distances as the noise level grow with the link length [99]. This is an issue in classical fibre networks, as used by telecom providers for example, where links can be several 100km or 1000s of km long [17]. Long fibre links can usually only be used unidirectionally [25, 27]. Hence, this application usually requires using two distinct wires for forward and backward path, resulting in an asymmetry of a few percent between the two [110]. It has to be noted that buffers used for compensation of delays also have a significant and often neglected impact on the performance of such links [99].

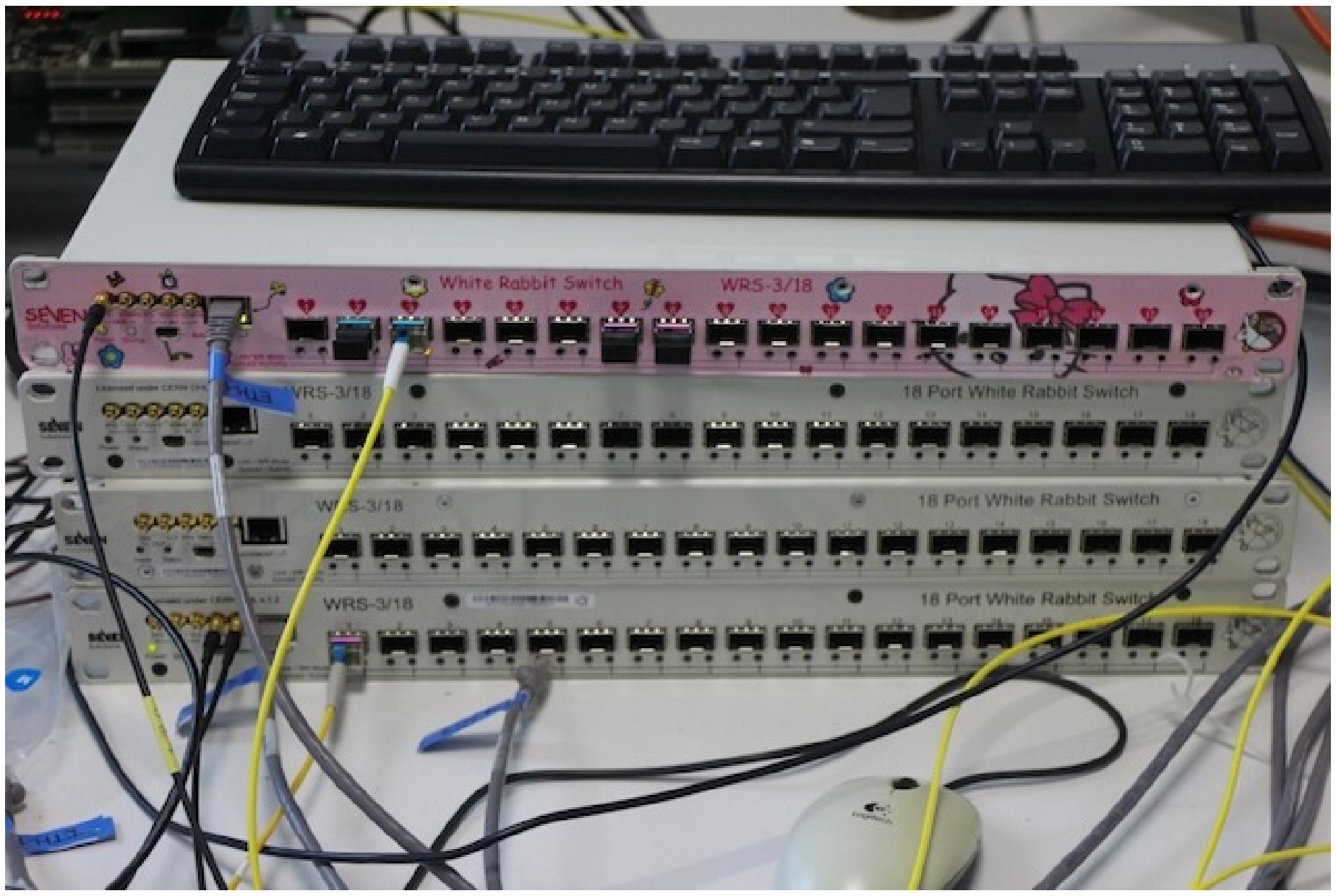

Figure 14: Set of four Seven Solution White Rabbit switches used in an experiment. From: private archive.

Wireless networks highly benefit from our altered model of links as it lifts the constraint of constant delays imposed in prior models. Wireless networks can become very large too, although their topologies are to be expected less complex than in the wired case [56]. However, the maximum edge degree of nodes can be much higher as wireless links are much cheaper to set up than wired connections to peers [56]. In fact, it is only limited by the slots available for scheduling which in turn are bounded by the channel quality available in Multiple Input Multiple Output (MIMO)[95] processing before running into interference issues [43]. As such wireless links operate in a noisier environment prone to various radio effects such damping, reflection, refraction, and most importantly multi-path effects caused by various environmental factors [38], such as temperature and humidity, the signal quality on the links may vary a lot and this very fast. As it is usually not possible to transmit and receive a signal at the same time [62], this results in the asymmetry factor to have a much higher variance over time as the measurement is taken against the local frequency which is not perfectly stable and suffers from an unpredictable frequency drift [40, 41]. In particular, the delay might vary very quickly due to the mentioned environmental factors changing rapidly in between transmission on forward and backward path. From a point of parameter choices, this requires choosing a large upper bound on the delay in the original model whereas in our model the included two-way measurement strategy allows for continuous, adaptive recalibration and consequently smaller bounds and higher performance.

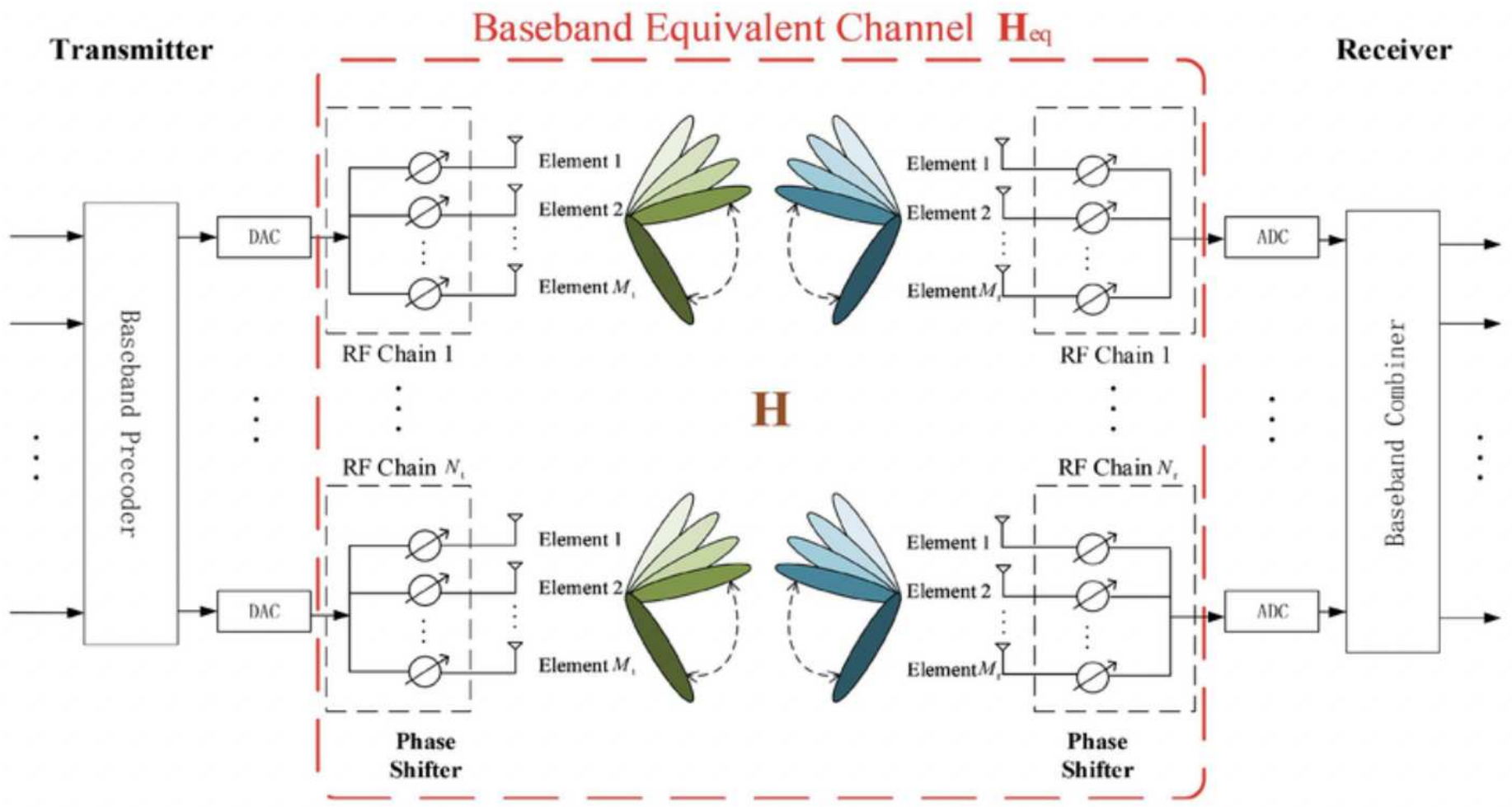

Figure 15: MIMO with beamforming. The antenna array can communicate with multiple devices over different, orthogonal meaning decoupled channels. From: [16].

Networks on chip are a special case of wired networks, where wires are considered to be very short in comparison to the classical wired use case mentioned above. In networks on chip, forward and backward path will always be split into two distinct wires as wires are unidirectional [105] in the transmit direction to prevent shunt faults [10]. However, one transmit wire can have multiple receivers, forming a 1:N relation [105]. Either way, this results in split forward and backward paths in all cases. Furthermore, chips are especially noisy environments due to the density and closeness of components [81, 21]. This induces various effects such as cross-talking [85], where trough induction, parts of the signal on one wire are also transmitted on another, referred to as the moving the edge effect [8], where electrical effects induce the pulse to happen at a later point in time than expected, potentially also altering its shape and spurious pulses [115], where electrical effects in the detection electronics induce a pulse

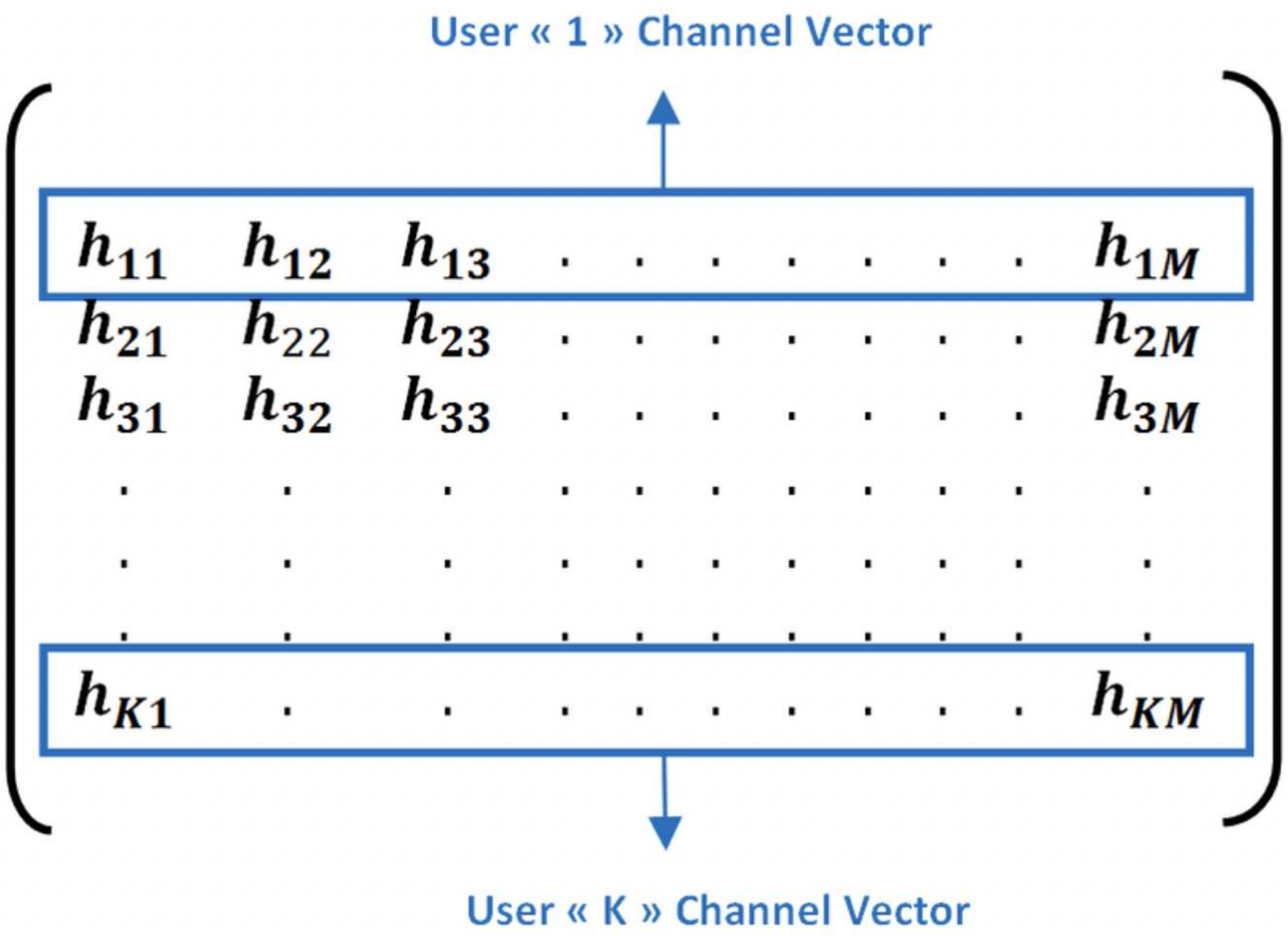

Figure 16: MIMO channel structure. Each Channel corresponds a vector in the channel matrix. From: [69].

that would otherwise not be present.

#### 4.3.3 Two notions of diameter: weighted and hop distance

As mentioned above, we differentiate between two notions of distance in the network graph. The first is the classical hop distance between two nodes, which reflects their distance in terms of edges in a path. To this effect we define the notion of edge length:

Definition 5. Edge length
The edge length $L(e)$ of an edge $e$ in $G$ is a positive real number modelling a notion of distance between the nodes $v$ and $w$ it connects.

With this we can formally define the notion of a shortest hop-weighted path, where $\mathcal{P}_{v,w}$ denotes the set of all paths from $v$ to $w$:

Definition 6. Shortest hop-weighted path
A shortest hop weighted path $h(v, w)$ between two nodes $v$ and $w$ corresponds to the path with the least number of edges from $v$ to $w$. Formally:

$$h(v,w) = \arg \min_{p(v,w)\in\mathcal{P}_{v,w}} \left\{ \sum_{e\in p(v,w)} 1 \right\}$$

With the two definitions above we can formalise the notion of the length of a shortest hop-weighted path:

Definition 7. Length of the shortest hop-weighted path
The length of the shortest hop-weighted path $H(v,w)$ between two nodes $v$ and $w$ is the number of edges in $h(v,w)$. Formally, we have:

$$H(v,w) = \sum_{e\in h(v,w)} 1$$

This metric is a distance metric as in an unweighted setting, each edge always has a unitary, strictly positive length of 1.

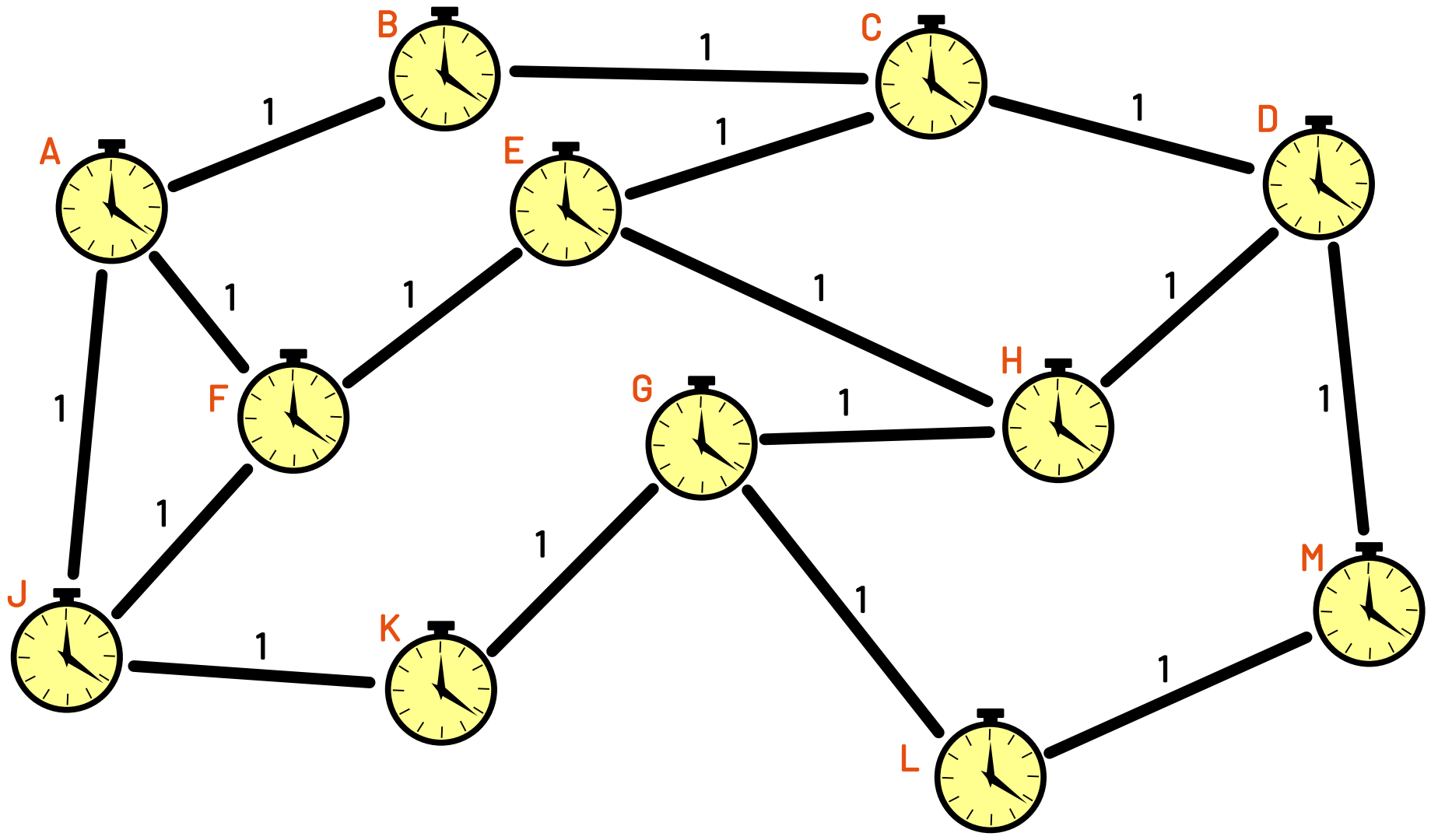

Figure 17: Example graph with labelled hop-distances. Original graphic.

We derive a notion of weighted distance based on this first definition. For an unweighted hop-distance of $H$ for a path from $v$ to $w$, we define a corresponding weighted distance on $G$. This notion will be used by the GCS algorithm on the pruned version of $G$, which weights all edges with the error term $\kappa$ resulting from asymmetry quotient and delay estimation error:

Definition 8. Shortest $\kappa$-weighted path
The shortest $\kappa$-weighted path $P(v,w)$ between two nodes $v$ and $w$ corresponds to the shortest weighted path between $v$ and $w$ for which the sum of all edge weights $\kappa_e = L(e)$ along the path is minimal. Formally:

$$P(v, w) = \arg \min_{p(v,w) \in \mathcal{P}_{v,w}} \left\{ \sum_{e \in p(v,w)} \kappa_e \right\}$$

Note that this only defines a distance function if all coefficients $\kappa_e$ are strictly positive or negative paths and cycles are excluded. As we will see in Section 5.5, the first holds true.

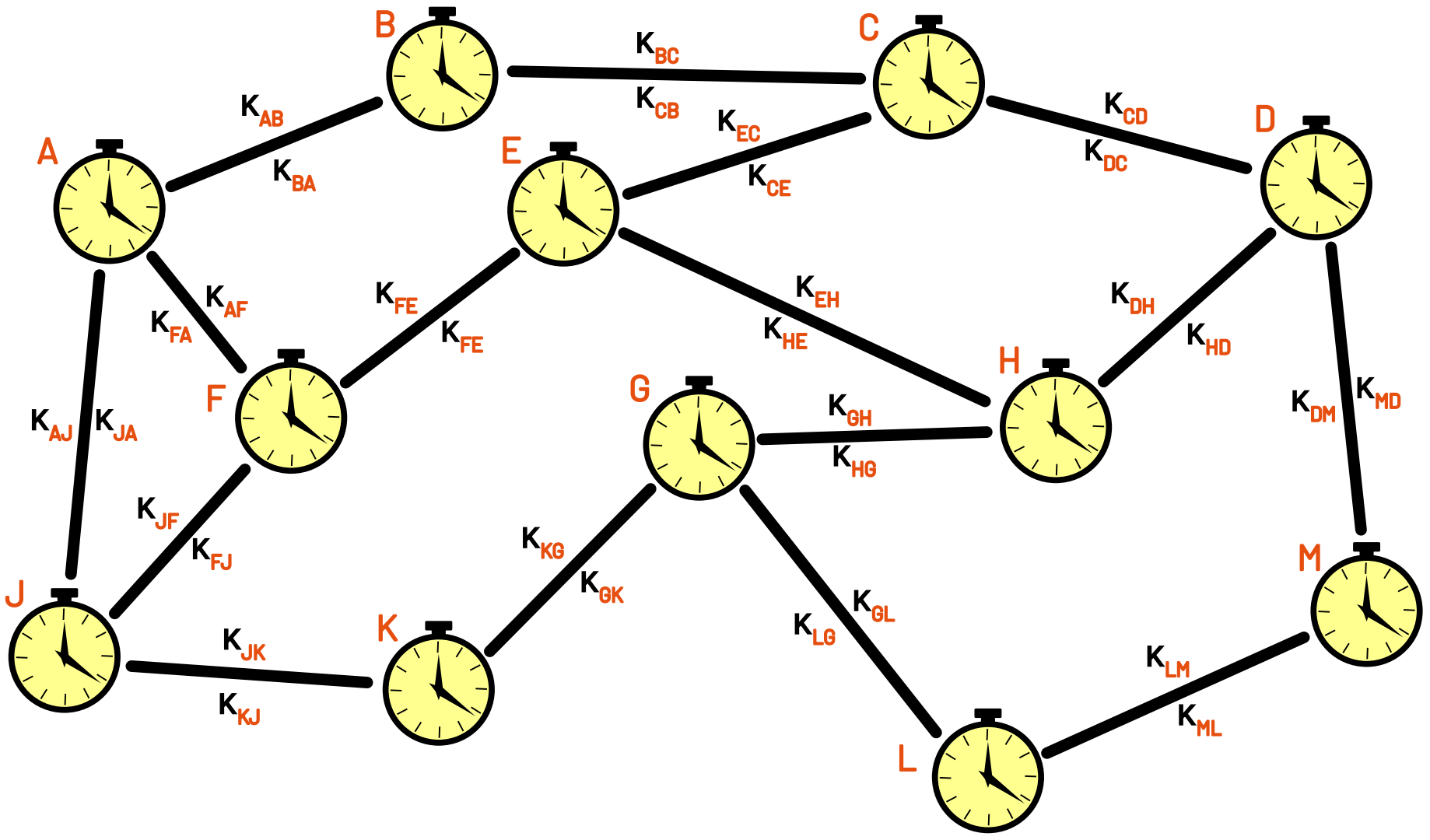

Figure 18: Example graph with labelled weighted distances. Note that, since we assume an asymmetry between forward and backward path between two nodes, the corresponding edge specific weights might differ by the asymmetry factors we will define in Section 4.6.4. Original graphic.

## 4.4 Modeling clock behaviour

### 4.4.1 Model of physical hardware clocks

As mentioned in Section 3.1.3, we only consider harmonic oscillators as time references and aim at defining a highly abstract but still realistic model of the time references in our system. It is impractical to produce a one fits all model for all types of references, as their properties widely differ with the considered type of reference. We thus will attempt building an abstraction of the clocks that are most commonly used in the use cases we consider.

In practice, quartz oscillators are by far the most common type of clocks integrated in systems and chips susceptible to operate our algorithm, followed by LC oscillators [6]. All harmonic oscillators have the central property of being band limited [4] as defined below.

Definition 9. Band limited signal [4]
A signal is band limited if its frequency spectrum is zero outside a predefined and finite frequency range. In practical applications, signals whose spectrums are magnitudes lower or are decaying quickly outside the predefined band spectrum are considered band limited.

This property originates from electronic harmonic oscillators being built from a resonant circuit, which has an inherent limited bandwidth or, equivalently, large quality factor, resulting in the oscillator's output being band limited [6], which filters out specific frequencies of the signal such as the processed signal corresponds to a single peak with fast decaying edges. The bandwidth of this signal is inverse proportional to the quality factor of the oscillator [7]. The higher this quality factor is, the lower the bandwidth of the signal and the stiffer, in the mathematical sense of the term, the differential equations describing the behaviour of the oscillator [4]. The stiffer the equations of the differential description of the system, the more inertia it has. This directly translates into its reaction time: the higher the inertia of the system, the longer it takes the system to respond to an impulse, meaning that its reaction time grows with the inertia, limiting the bandwidth of the reaction time. Relaxation oscillators and ring oscillators cycle trough states at a specific speed [44]. In fact, each step is associated with a notion of progress in terms of speed. This directly implies that their reaction time is bounded and they are, by definition, band limited as well [4]. For any band limited signal that is additionally energy limited, as it is the case with the reaction time of the harmonic oscillators we consider, Bernstein's inequality holds:

Theorem 1. Bernstein's inequality [4]
Given that a signal $s(t)$, band limited to $x$ Hz, can be written as:

$$s(t) = \int_{-x}^{x} e^{2i\pi ft} g(f)\ df$$

for $t \in \mathbb{R}$ and some integrable function $g$, meaning that $g \in L^1$, then:

$$\left| \frac{ds(t)}{dt} \right| \leq 4\pi x \ \sup_{\tau \in \mathbb{R}} |s(t)|$$

for $t \in \mathbb{R}$.

This theorem implies, that if a signal is band limited, its derivative is defined and bounded at all points, as the supremum of the original signal is bounded.

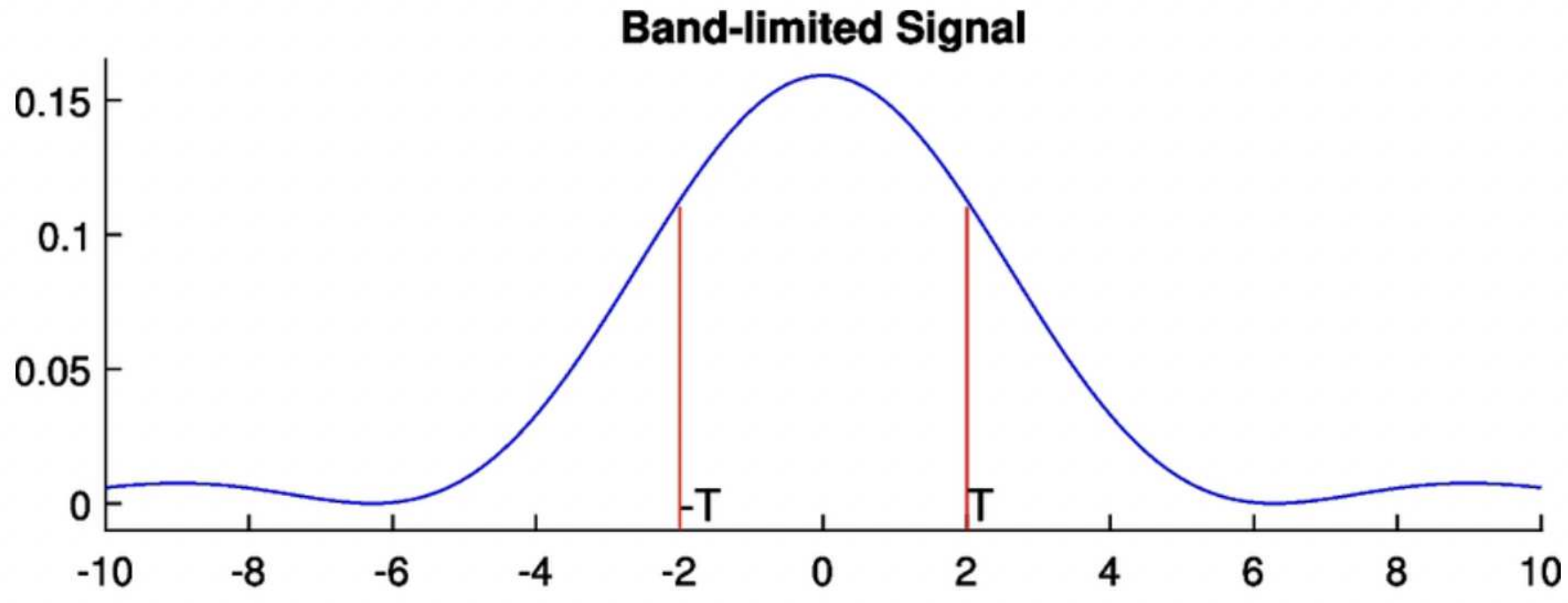

Figure 19: Example of a [-T,T] band-limited signal from: [18]

We thus can model the physical hardware clock present at each node in the network of monotonously growing, differentiable function:

Definition 10. Hardware Clock
Each node $v \in V$ is equipped with a hardware clock $H_v$ which can be modelled as a strictly monotonously growing, differentiable function:

$$H_v : (\mathbb{R}_{\geq 0} \rightarrow \mathbb{R}_{\geq 0}; t \longmapsto H_v(t))$$

It advances at rate $\mathcal{R}_{H_v}(t) = \frac{dH_v(t)}{dt}$ with $\mathcal{R}_{H_v}(t) \in [1, \vartheta]$ when $\vartheta \geq 1$.

The rate of the hardware clock maps the real time to the local notion of time at a node: the rate describes the instantaneous frequency of a clock. It's defined as the derivative of $H_v$, meaning that it models the instantaneous variations in the frequency of $H_v$ at any point in time, at least in theory. In practice, we cannot measure the rate instantaneously as it progresses. The reaction time of the electronics in any case is non zero, meaning that we cannot measure without delays [29] as the frequency is measured as the time interval between two zero crossings of the signal, thus requiring finite time for each measurement. Furthermore, as visible in the definition above, the rate of a clock is not strictly invariant but lives within a bounded range. The rate variations are induced by the so called clock drift, which describes the deviation from the expected oscillation frequency and forms a metric for the stability of a clock, which mirrors the frequency deviation due to various noise and ageing effects [89].

#### 4.4.2 Modelling clock drift

The deviations of clocks is generally described in terms of their clock drift. However, depending on the field, this term has a different meaning. For electrical engineers and physicists it measures the evolution of the frequency offset[12]. In turn, for the theoretical computer science and time synchronisation community it measures the evolution of the phase offset over time [111]. The basic notion of drift is the frequency drift [12], which monitors the evolution of the frequency error [68], that is the deviation to the reference frequency of a clock.

As RC and LC oscillators do not have a stable and well-described drift in literature, we will limit the study of drift in harmonic oscillators to quartz oscillators, for which the observable frequency error originates in the production process [94]. The operating frequency is inverse proportional to the thickness of the quartz cut. For an AT-cut, we have an expected oscillation frequency of 1,6 MHz x mm [94]. In such a crystal, a deviation of 1 ppm translates to a length variation of 160 picometers. This is smaller than than the diameter of a single atomic layer, which averages 1.7 to 6.5 nanometers [97, 49]. As crystals are cut in production and this cut can only be achieved with limited precision which is much less than an atomic layer, deviations of multiple ppm are common.

Furthermore, quartz oscillators are prone to various environmental effects altering their frequency. Beyond temperature and humidity, which have a comparatively small effect, the air pressure has a notable effect on the frequency drift of quartz oscillators [89]. The enclosure in which the crystal is located are prone to air pressure dependent deformations. These deformations result in stress variations, which in turn alter the operation frequency of the quartz. It has to be noted that the worse-case scenario in typical environmental conditions will define the upper bound on the drift. This worse case assumption directly translates to the assumptions we can make about the maximum drift we have to take into account in the model. As a consequence, many upper bounds in the algorithm will be defined by the maximum drift. It is hence advisable to ensure that, if possible, it stays relatively small in any practical implementation. This can be achieved over a procedure called syntonisation [112] which consists in locking the frequency of an oscillator to a reference frequency provider by another reference of higher accuracy. The implications of implementing such a procedure to enhance the stability of the local oscillators is however considered out of scope of this thesis.

In our model, we will consider the phase drift of a clock [42, 11]. As the phase corresponds to the integral of the frequency, the phase clock drift $\mathcal{D}_{H_v}$ of a hardware clock $H_v$, which we will abbreviate as clock drift in the remainder of this thesis, corresponds to the integral of the frequency error. At all times $t_1, t_2 \in \mathbb{R}_{\geq 0}$ with $t_2 > t_1$, the phase clock drift can be computed with the following formula:

$$\forall t_1, t_2 \in \mathbb{R}_{\geq 0} : \mathcal{D}_{H_v}(t) = \int_{t_1}^{t_2} \mathcal{R}(t) - 1 \, dt$$

The frequency drift is defined as the deviation from the normalised clock rate of the hardware clock rate which is equal to 1 in the equation above. With this, we can formalise our definition of clock drift used in our computational model as:

Definition 11. Hardware Clock drift
At node $v$, the clock drift $\mathcal{D}$ of its hardware clock $H_v$ corresponds to the difference between elapsed real time and elapsed local time at node $v$. More formally, for two real times $t_2 > t_1 \in \mathbb{R}$ the clock drift can be modelled as Lipschitz condition:

$$\forall t_2 > t_1 : (t_2 - t_1) \leq H_v(t_2) - H_v(t_1) \leq \vartheta(t_2 - t_1)$$

This means that in our model of computation, the drift is expressed in terms of variations of the clock rate, which lives in range $[1, \vartheta]$, where $\vartheta$ is called maximum clock drift. The maximum clock drift defines an upper bound on the phase error that can result from various effects resulting in noise shifting the operation frequency of our oscillators.

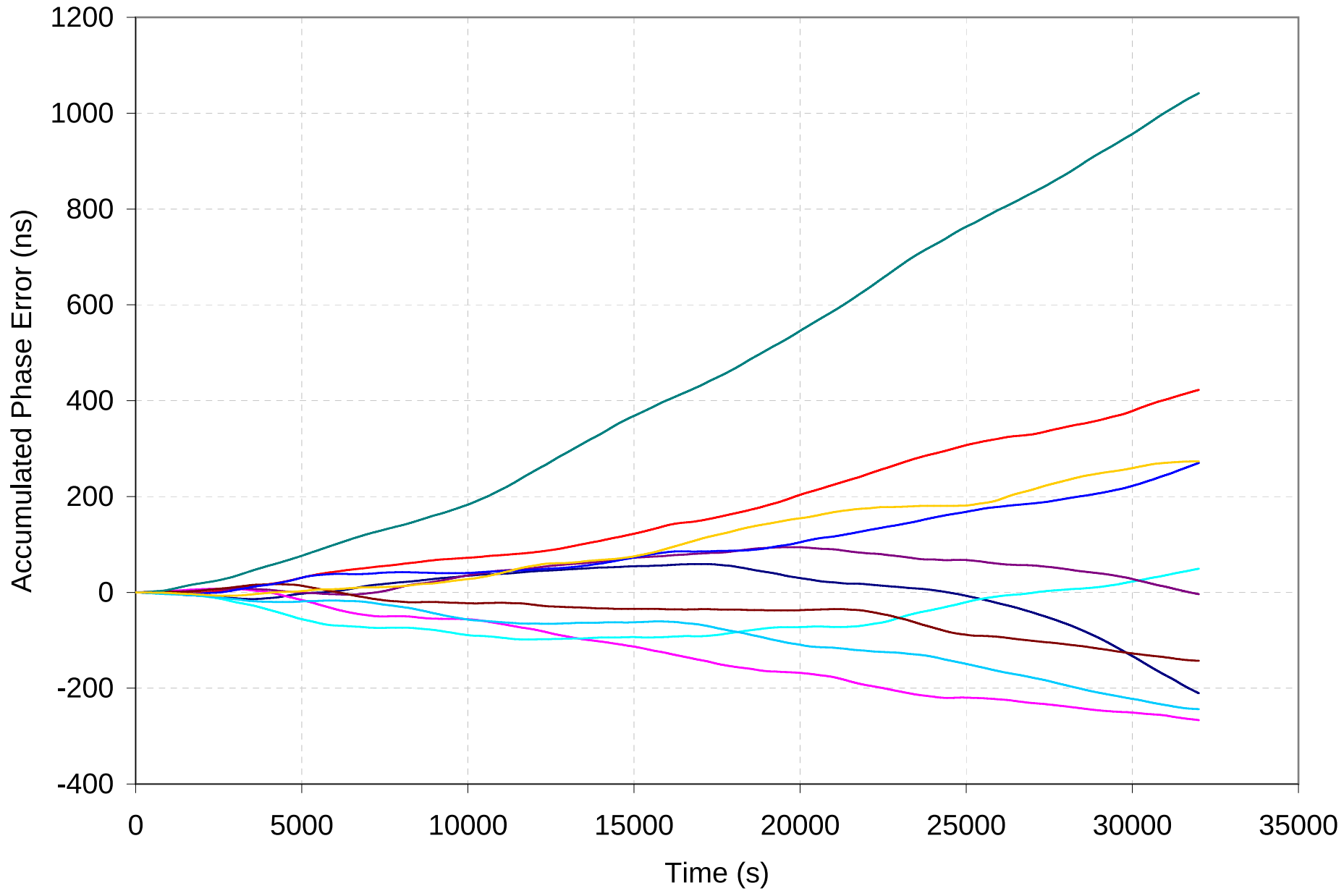

Figure 20: Example plot of accumulated phase error, i.e, phase drift in function of time for different OCXOs. From: [108]

### 4.4.3 Modelling clock corrections

The deviations resulting from the presence of clock drift have undesired side effects in practice, which we would like to correct in a running system. The GCS algorithm allows to determine when and by how much we have to correct our clocks to counterbalance the skew induced by clock drift. Prior work however doesn't mention how these corrections can be applied in practice.

Quarz oscillators usually feature a tuning input, which allows to adjust the operation frequency once the oscillator is integrated to the system it is supposed to clock. Many quartz oscillators are equipped with an electronic frequency control (EFC) port [54, 44] to this effect. This port allows to change the reference voltage applied to the quartz. As the operation frequency of an oscillators is in quasi-linear relation to the applied supply voltage, this port allows to tune the resonance frequency of the crystal. Another way to change the operation frequency is to tune the temperature of the crystal, as this also changes its oscillation frequency [89]. However, both voltage and temperature based tuning affect the stability of the oscillator in a negative way as the system successively undergoes a series of unstable intermediary states at rather high speed until it reaches and stabilises at the desired frequency [73]. To get rid of this undesired effect, the frequency of an oscillator can be corrected with the help of a so called phase or frequency stepper [101, 73]: it allows to slowly push the phase or frequency of an oscillator in small steps, reducing the impact of the change on the stability of the crystal's oscillations.

Finally, one might ask how to determine the voltage (or temperature) that has to be applied to get the crystal to a certain frequency in practice. As quartz oscillators have, as mentioned, a quasi linear voltage to frequency curve [90], it is possible to simply compute the required supply voltage required to reach the desired oscillation frequency.

In our concrete case, the GCS algorithm computes the phase correction factor $\mu$ that has to be applied to the clocks to counterbalance the built-up skew. In our model, the rate of the hardware clocks is only altered at discrete intervals. Formally, we define the correction function, mapping the decision of the GCS algorithm to apply a correction of $\mu$ or no correction at all as:

Definition 12. Skew correction
The local correction function $\mathcal{C}_v : \mathbb{R}_{\geq 0} \rightarrow \{0, \mu\}$ at node $v$ is a piece-wise continuous function and models the rate change to be induced by applying the correction factor $\mu$ to the hardware clock $H_v$ by adjusting its rate. Since $H_v$ can progress at rates up to $\vartheta$ at any time $t \in \mathbb{R}$, we impose that $\mu > \vartheta$ as we require $\mu > \vartheta$ as we cannot tell upfront in the messaging protocol whether we run at rate $\vartheta$ or our neighbour does.

For the simple sake of conceptual separation of concerns, we define the logical clock at a node as counterpart to the hardware clock to which the above

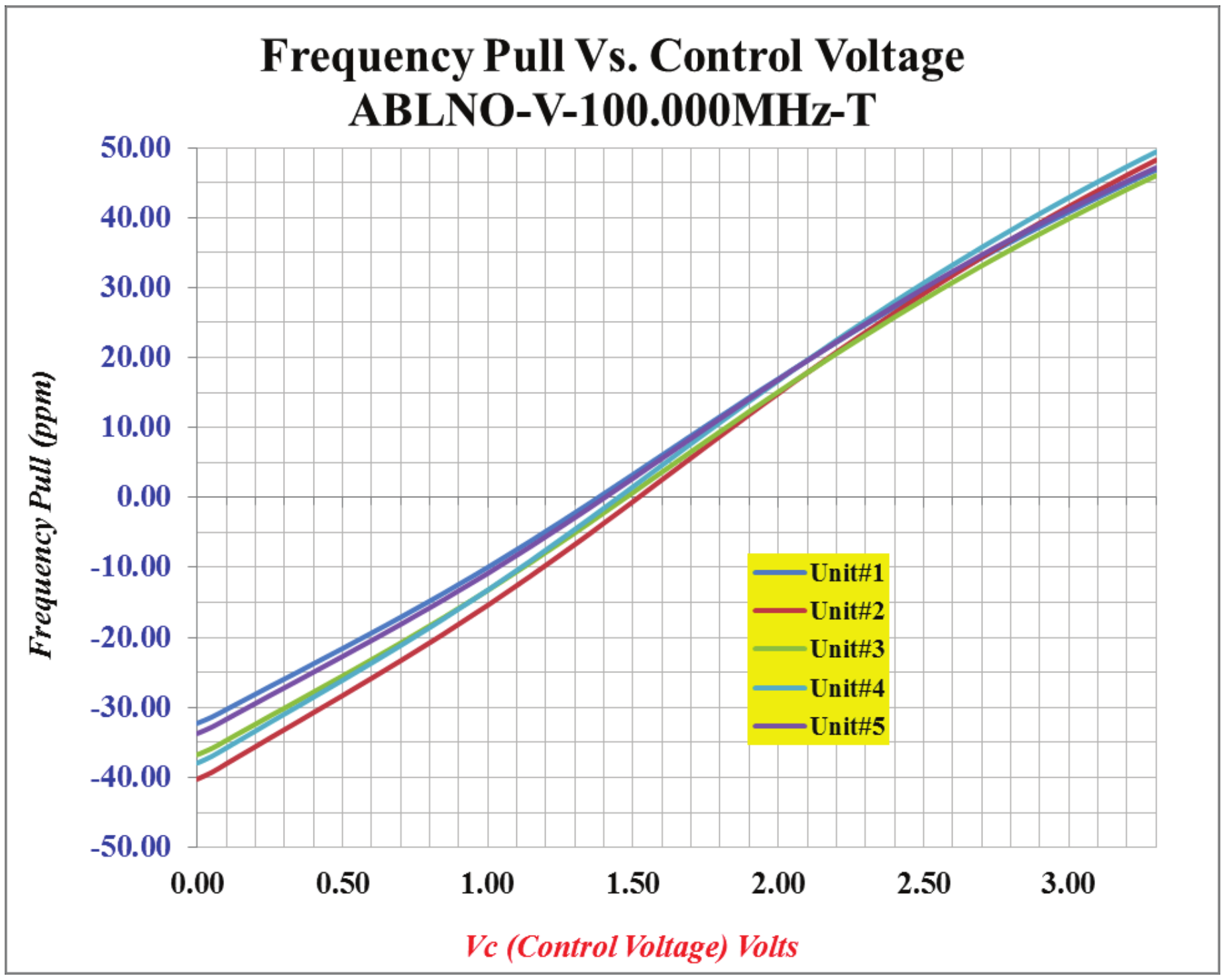

Figure 21: Quasi-linear curve of frequency evolution in tuneable oscillator based on applied control voltage. From: [90]

correction function has been applied. Formally, the logical clock of a node is defined as:

Definition 13. Logical Clock
Each node $v \in V$ is equipped with a logical clock $L_v$ that is derived from its hardware clock $H_v$. More formally, $L_v : (\mathbb{R}_{\geq 0} \to \mathbb{R}_{\geq 0}; t \longmapsto L_v(t))$ is a continuous, differentiable function defined as:

$$L_v(t) = H_v(t) + \int_0^t \mathcal{C}_v(\tau)\, d\tau$$

where $C_v$ is the local correction function, meaning that the rate of the logical clocks mirrors the rate of the corresponding hardware clocks with corrections applied. The rate $R_{L_v} = \frac{dL_v}{dt}$ is $R_{L_v} \in [1, (1+\mu)\vartheta]$.

As the hardware clocks can be modelled as continuous, differentiable functions and the changes made by the piece-wise continuous and thereby locally integrable correction function $\mathcal{C}$ are applied within discrete intervals, the func-tion representing the corresponding logical clock can be approximated by a smooth function by convolving with a mollifier [33].

### 4.4.4 Is the clock model realistic?

Given that the core goal of this thesis is to design a model that upholds realistic assumptions that also hold in practice, one might ask whether our clock model is a good abstraction of the physical behaviour of harmonic oscillators. We assume that our hardware clocks can be modelled as differentiable functions and show this by applying Bernstein's inequality [4]. This indeed accurately but abstractly models the behaviour of oscillators: as mentioned in Section 4.4.1, we assume that any hardware clock of the types listed is band limited, mirroring the fact that harmonic oscillators are not able to change frequency instantaneously.

By the same argument as above, one can justify the choice of assuming that the logical clocks, speed up by a factor up to $\mu$ at discrete intervals, also can be assumed to be differentiable functions. When using an EFC, as described in Section 4.4.3, to tune the frequency of an oscillator, the voltage also undergoes a slow change, increasing the frequency at a certain speed rate and stepping trough intermediate levels [101, 44]. Hence, the oscillator cannot instantaneously change its frequency but moves trough the frequency spectrum at a certain rate before stabilising at the target frequency [6]. This again implies band limitation and hence differentiability.

We mentioned in Section 4.4.3 that frequency and phase tuning can be achieved instantaneously trough a frequency stepper. This implies that the standard argument to show that the frequency is differentiable as the reaction time of the system is limited and thus the resulting signal is band-limited does not hold here. However, we still can assume that the clock function is differentiable after having been altered by a frequency stepper. As frequency measurements are inherently bound to a delay, i.e, cannot be achieved instantaneously, we simply can measure the system's state before or after the frequency jump, and interpolating the clock function in between those two points. This is possible as the Mollifier theorem [33] holds.

Finally, we have a look at the impact of environmental effects on the differentiability of oscillators. Low production quality can cause so called spurious frequency jumps [58, 93] in quartz oscillators, these are random, large jumps in the operation frequency of the crystal. Assuming proper production of the used oscillators, these spurious jumps are very rare. On average, one can expect frequency jumps to occur in only 20% of OCXOs [65] but we choose to ignore their possible presence in our model as it does not affect the differentiability of the corresponding clock function. This is due to the fact that oscillators are still bandlimited during jumps, i.e, the jump does not occur instantaneously but is spread over multiple stages. In practice, frequency jumps can be detected by monitoring the operation frequency of the quartz at periodic intervals [65].

## 4.5 Modeling skews throughout the network

The clock skew, that is the clock offsets between two logical clocks located at different nodes, results from the clock drift mentioned above, shifting the operation frequency of clocks, resulting in them progressing at different speeds. We consider two notions of clock skew in our model, the local and the global skew of the network. The local skew describes the maximum skew between two nodes, the global skew the maximum skew between any two nodes in the graph. Which of these two metrics is more important depends on the practical use case, more specifically the scope of communication and where it takes effect, i.e, whether the communication is focused on peer to peer transmissions or information has to cycle throughout the entire network, determines which of the metric is more relevant for system performance. Note that we define these metrics on the logical clock, that account for the corrections effectuated by the algorithm, as we aim to study the skew of the system also after corrections to the clock rate of that logical clock have been applied.

### 4.5.1 Local skew: clock drift between neighboured nodes

In our network, the local skew describes the clock skew between adjacent neighbours in the network. More precisely, we are looking for the local neighbour, that is another node in unweighted hop distance 1, that has the largest clock offset, positive or negative, to our own clock. Formally, it can be defined as:

Definition 14. Local Skew
The local skew $\mathcal{L}$ corresponds the maximum time offset between the logical clocks of any node $v$ and his neighbour $w$ at time $t \in \mathbb{R}$, that formally is the difference:

$$\mathcal{L}(t) = \max_{(v,w)\in E} \{L_v(t) - L_w(t)\}$$

The local skew plays an important role in all the big use cases we focus on in this thesis. With this it becomes clear that having a near optimal bound on the local skew is a desirable property for the GCS algorithm. Let's take a closer look at the relevance of the local skew in wireless, wired, and on-chip networks:

Wireless networks have many applications in which the local skew is dominant. Due to the inherent property of radio waves to only be able to propagate over limited distances, the communication can only happen rather locally as the propagation of the time signal is bound to short distances, too. For local communication and coordination, the local skew dominates the latency. Low local latency is crucial in many different applications. One prominent example are cellular networks, in which neighboured base stations need to be synchronised in the nanosecond to microsecond range [1, 22, Chapter 5.1.8] to ensure smooth schedule and interference free communication between base stations and devices. Since base stations transmit at very high power, especially compared to devices, base stations and devices cannot transmit at the same time. If they do, destructive interference between stations or stations and devices stalls any data transfer [22, Chapter 21.1.2, Chapter 21.2]. Beyond guaranteeing safety of communication, the local skew also has a significant impact on the performance of cellular networks: the lower the local skew between devices, the more devices can be scheduled in very short slots, enhancing the throughput of the network [22, Chapter 7.1, Chapter 7.2, Chapter 5.1.8]. The 5G standard for example requires that the skew between base stations and base stations and devices does not surpass as little as 10 nanoseconds [1].

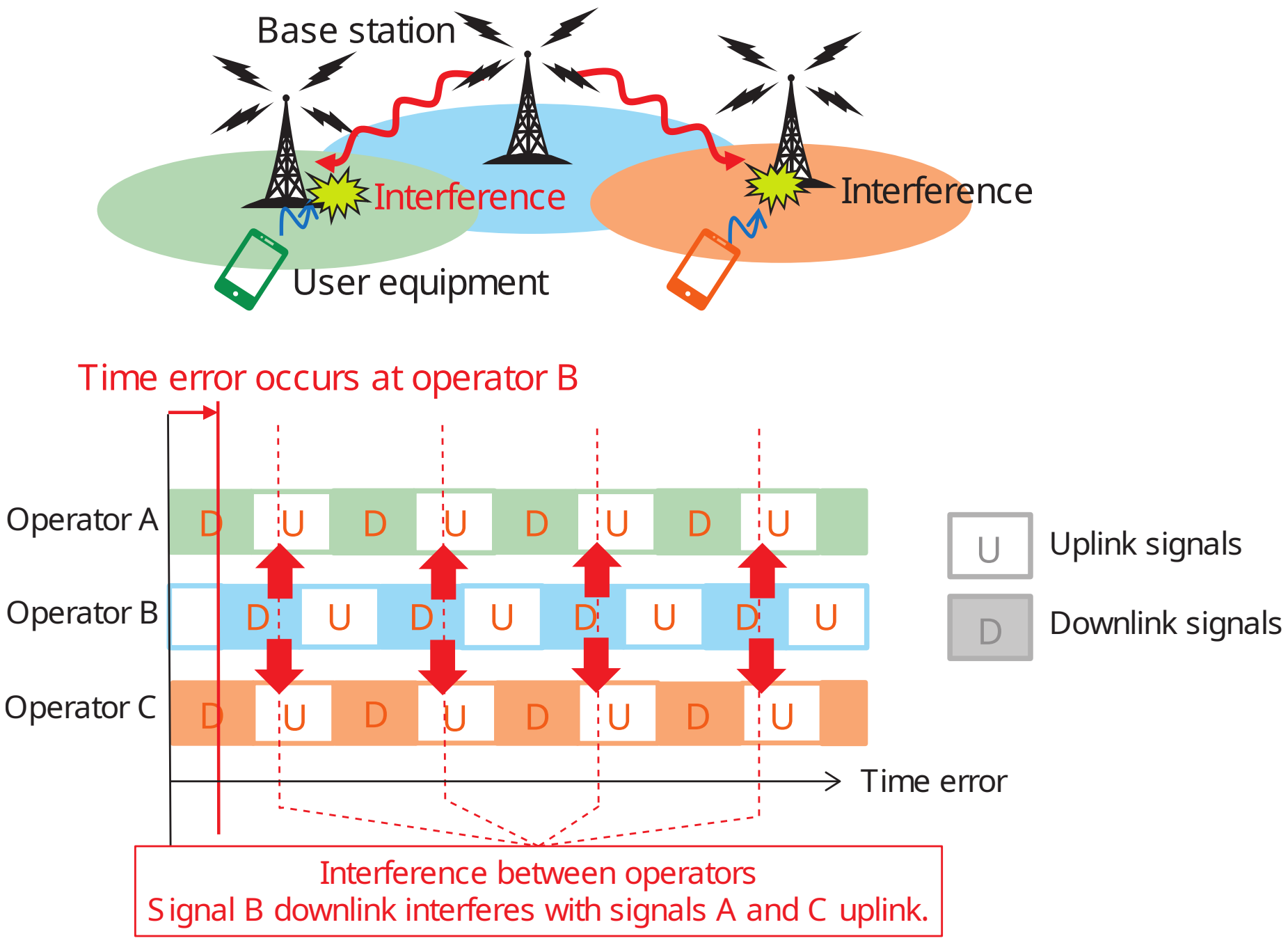

Figure 22: Interference between mis-scheduled uplink and downlink transmission due to lack of accurate synchronisation in a 5G Network. From: [22].

Wired networks also have many application in which communication takes place in an inherently local scope. A prominent example are distributed compu-tational clusters, in which communication again mostly happens between adjacent peers [113]. As the communication latency has a significant impact on the global performance in terms of run time, data is mostly transmitted over fibre links [63]. As the transmit delay is the central limiting element in this scenario, the paths between machines should be kept as short as possible while allowing to reach as many other end points as possible to guarantee flexibility of communication. To this effect, as mentioned in Section 4.3.2, hypercube topologies [63] but also grid-like topologies [113] are often deployed as they allow minimal path lengths while maintaining high in- and output degrees per machine while being regular. The regular structure of the communication network resulting from these topologies are also important in this use case to limit the variations in delay between paths. This allows for an inherent synchronisation by communication, the algorithmic procedure or protocol is executed in a lock-stepped manner, clocked in an implicit way by arriving messages or in an explicit way by a time synchronisation algorithm or protocol.

Networks on chip benefit from a low local skew only in some applications. In a network on chip, the communication network replaces the data bus of classical chips. The network is build using different topologies, i.e, 2D and 3D grids, toruses or general meshes [30]. Data is often propagated only from node to node, i.e, communication is local. Thus, such an architecture would benefit from tight local synchronisation while global synchronisation does not matter as much [26].

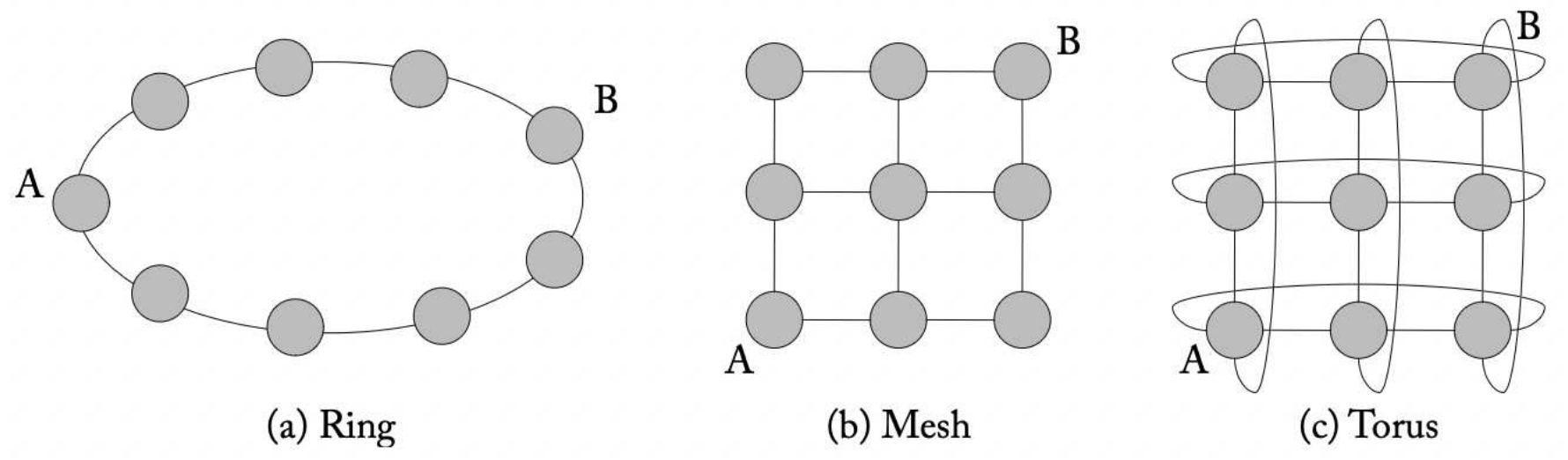

Figure 23: Example of simple Network on Chip topologies. From: [30].

### 4.5.2 Global skew: a network-wide notion of clock drift

Conversely, the global skew corresponds to a network-wide notion of clock skew: we consider the largest clock skew between any two nodes in the network graph, no matter their unweighted hop distance. Formally, this can be modelled as:

Definition 15. Global Skew
The global skew $\mathcal{G}$ is the supremum on the maximum time difference between the logical clock values of any two nodes at any time $t$, being at a distance of up to $D$ from each other. Formally, at a node $v$ we have:

$$\mathcal{G}(t) = \max_{v,w \in V} \{L_v(t) - L_w(t)\}$$

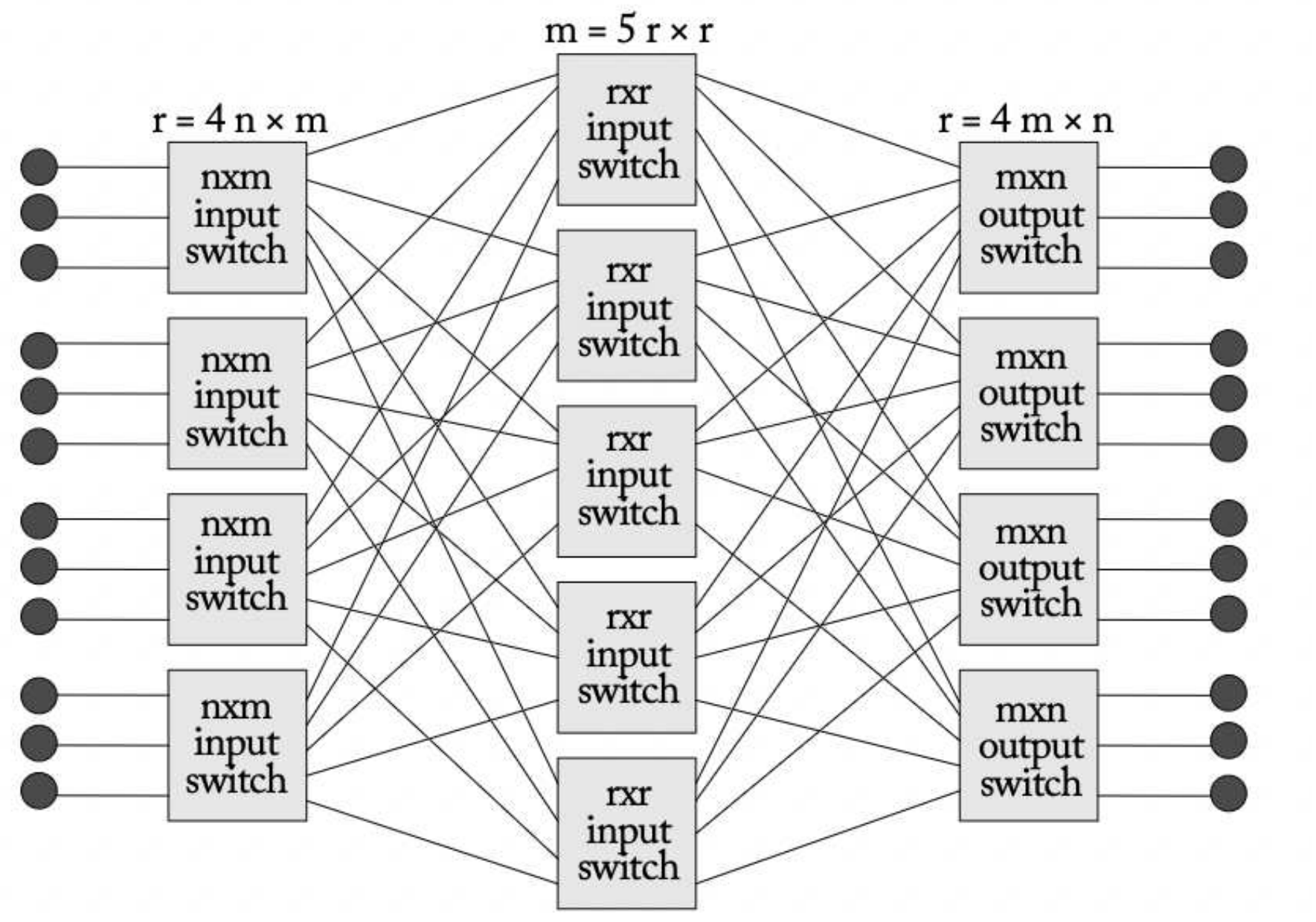

Figure 24: Example of the more complex Clos Network NoC topology which incorporates switches. From: [30].

Wireless networks have little applications in which the global skew is dominant. This is again due to their inherent property of having a limited propagation distance of the time signal, implying that data transmitted over larger distance is transmitted over multiple hops. Consequently, the local skew of the peer to peer transmission dominates the skew and the global latency of communication.

Wired networks are widely used to support communication of financial transaction systems, especially in high frequency trading [83, 37]. The global skew is dominant in this use case as the main purpose of time synchronisation is to achieve global consistency of transactions: for all transactions executed at different times and places, potentially all over a country or even the world, their exact timestamp to a central reference needs to be known. This is important for example in broking [37], where knowing the ordering of transactions is required to check for consistency of transactions and validate a transaction as successful, for example whether a particular stock has been sold or acquired successfully by exactly one other entity. A low global skew and fast data transmission and processing are crucial as transactions are time sensitive: as there are usually many interested buyers for a single stock, buyers need to have almost real time data on the current state and price of the stock to be able to acquire or sell at the desired price, which maps to a specific point in time, and evolves fastly depending on offer and demand. A global accuracy to UTC, with a maximum skew of 50 to 100 milliseconds is required by regulation by both EU [106] and US [5] regulation.

Networks on chip were originally devised to solve the inherent scalability problem of classical bus architectures that require global synchronisation of all participants on the bus [87, 52]. This either requires all participants running on the same clock with tight skew bounds over the whole chip, which complicates clock distribution [103], or long arbitration delays to ensure that all participants agree on the same arbitration result [91] at the end of the process. In the former case, all participants of the bus run in a synchronous fashion. This means that all communication which is required for arbitration must happen within a single clock cycle. Any skew between the clocks of the participants results in the necessity to maintain large guard times within the clock cycle, thus limiting the maximum clock speed. Distributing the clock signal on the whole chip with low skew is a rather difficult task which requires large amounts of power as the signal has to cover large end to end distances. In the latter, asynchronous case, at the end of the arbitration round that is after an arbitration decision is taken, the arbiter has to wait at the minimum for the longest round trip time between arbiter and participant to ensure that no data race between two participants requesting resources on the bus remains unresolved. This means that, as the chip grows in size, the waiting times of the arbiter continuously increases as the communication delays increase, thus severely limiting achievable throughout and latency.

## 4.6 Modeling measurements

This section highlights the impact of the chosen measurement protocol on the performance in terms of skew of both model and algorithm. A simple change of protocol paradigm will allow to greatly reduce the contribution of measurement related parameters on the skew bounds the algorithm can fulfil, ultimately delivering a more performant algorithm for practical applications.

### 4.6.1 One-way versus two-way measurements

The core goal of this thesis is to build a realistic model of links and measurements. To this effect, we alter the modelled measurement paradigm. In previous work, so called one-way measurements are performed to determine the offset between two neighboured nodes since their degree of synchronisation can be determined based on a delay aware measurement. Translated to a message based protocol, all nodes send their clock values to their adjacent neighbours, exchanging exactly one message within the communication protocol. The single message is delivered to the receiver within a message delivery delay of $d$, from which one has to deduct a global value of uncertainty on the exact arrival time of the message present in the system. This approach has many disadvantages, the most important of them being that the exact message delivery delay cannot be determined as the sender misses feedback on the arrival time of the message. As the uncertainty on the arrival time can only be estimated, this results in a comparatively low precision on the message delivery delay. Since the communication delays play an important role in the performance of the algorithm as we will see later in the analysis, the current paradigm limits the skew bounds that can be achieved. Furthermore, the resulting model cannot cope with different paths lengths as the one way paradigm makes it very difficult to account for variations on uncertainty and path lengths, given the high second order uncertainty on the estimation uncertainty.

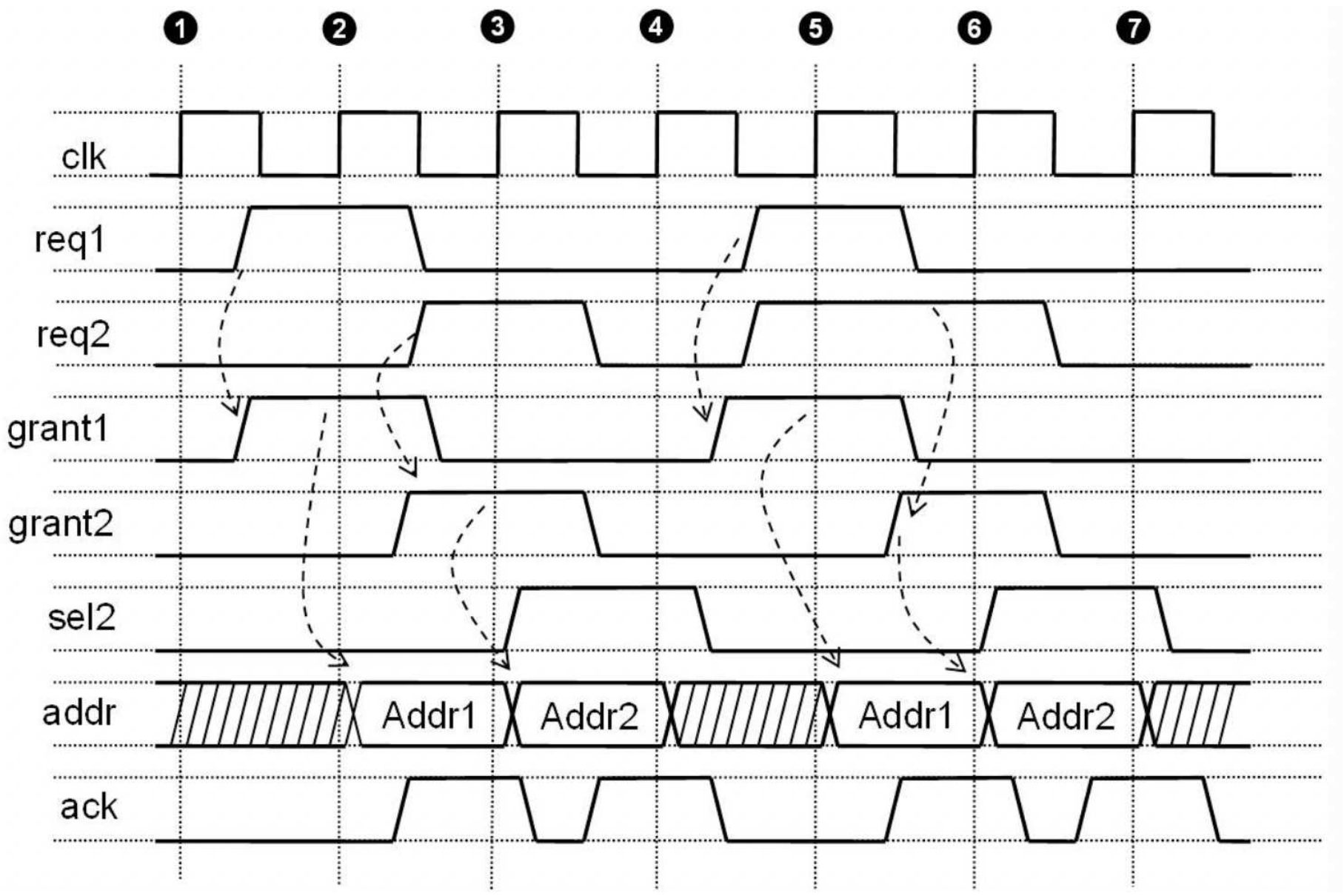

Figure 25: The bus arbiter set grant signals (grant1 and grant2) to requests (req1 and req2) to assign resource slots (addr) to them. From: [86].

In practice, the path delay is prone to many small but also bigger fluctua-tions due to noise and environmental factors. However, the one way measurement paradigm does not adjust well in terms of performance to this issue as it requires very high and potentially even uniform upper bounds on the path deviation that have to be determined beforehand and need to be feed to the algorithm as preset parameters. To cope with this issue most practical implementations requiring some kind of delay or path length measurement and are based on the two-way measurement paradigm. This is preferred in practice as it allows to accurately determine the time elapsed between sending and receiving of a message and hence determine the message delivery delay on the basis of computations instead of estimations. It hence also allows to account for variable path lengths. As already mentioned in Section 4.2.2, the resulting messaging protocol is simple: instead of exchanging only one single message, the node which estimates its neighbour's clock offset sends a timestamped request message to its neighbour. The neighbour responds with a second message, hence the name two way, containing the timestamp of arrival of the request as well as an outgoing timestamp. The reply message is timestamped again upon arrival of the message. Based on this four timestamps and upper bounds on the processing delay as well as the maximum relative drift, the path length can be precisely determined by computing the time elapsed between sending of request and arrival of answer minus the request processing time. This greatly reduces the burden of uncertainty: instead of having to account for both measurement uncertainties, that is the uncertainty on the precision of the measurement and the uncertainty on the path length, we are only left with the first as the second was, up to the measurement uncertainty, exactly determined by during the measurement.

### 4.6.2 What and how exactly do we measure?

With the above, we will delve into the measured parameters. In a two-way measurement, we aim to determine the two following variables: first, the message delivery delay mentioned in Section 4.2.2, corresponding to the path delay in a physical system, and secondly, the clock skew between two nodes, corresponding to the relative clock offset between two logical clocks. In both cases the measurements are made using the local clock as reference. The detailed procedure on how to compute these values for the GCS algorithm is specified as pseudo-code in Section 5.3 and Section 5.4, respectively.

In practice, these two general measurements are implemented in different ways, depending on the requirements of the system. Depending on the type of links the delay measurement is implemented by a different mechanism. For wired fibre links as used in White Rabbit for example, a pattern matching over Digital Dual Mixer Time Difference (DDMTD) [92] is performed on the phase of the optical signal. Each switch has, beyond a local time reference, a source synchronous ethernet reference signal as input. By correlating the phase of the incoming signal with both the local, built-in oscillator and a synchronous ethernet signal for the forward and backward path, the White Rabbit switch is able to precisely, down to picoseconds, determine the phase offset and hence the delay between the incoming signal and the two reference signals.

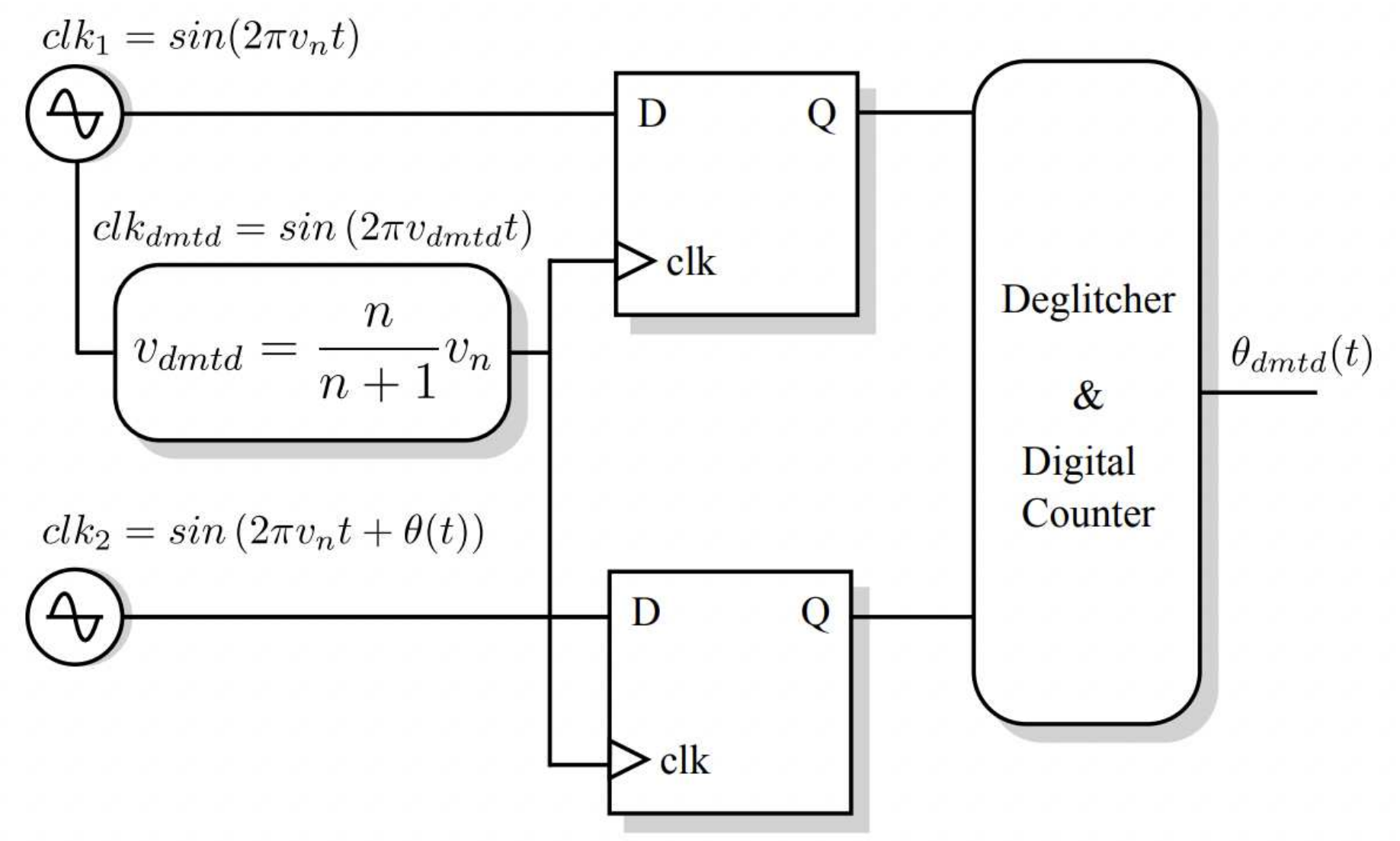

Figure 26: DDMTD circuit with timing reference input signals clk1 and clk2. The common clock signal clkdmtd has an offset of a few Herz to the two respective input signals, allowing the logic implemented by the two D-latches to compute the system's beat signal from which the phase offset between the two input signals can be inferred. From: [79].

A similar mechanism is deployed to determine delays in wireless networks. The direct sequence spread spectrum method modulates the data stream, which has a much larger period, onto a so called spreading code [14, 48]. The spreading code is build of a pseudorandom binary sequence, alternating much faster than the data. This asymmetry is used to demodulate data and spreading code: as the data signal alternates between 0 and 1 much slower than the sequence, it is possible to recenter the data sequence by performing early/late correlation [19, 23]. To this effect the carrier is demodulated from the signal, which is then multiplied by the PRN sequence [23]. Its power spectral density peaks are determined next: if it is perfectly synchronised with the PRN sequence, the peak arrives at the beginning of the sequence [19]. If it is slightly late or early, the peak is slightly off the expected point [19]. This point can be determined and the offset to the sequence corrected to decode the data [23].

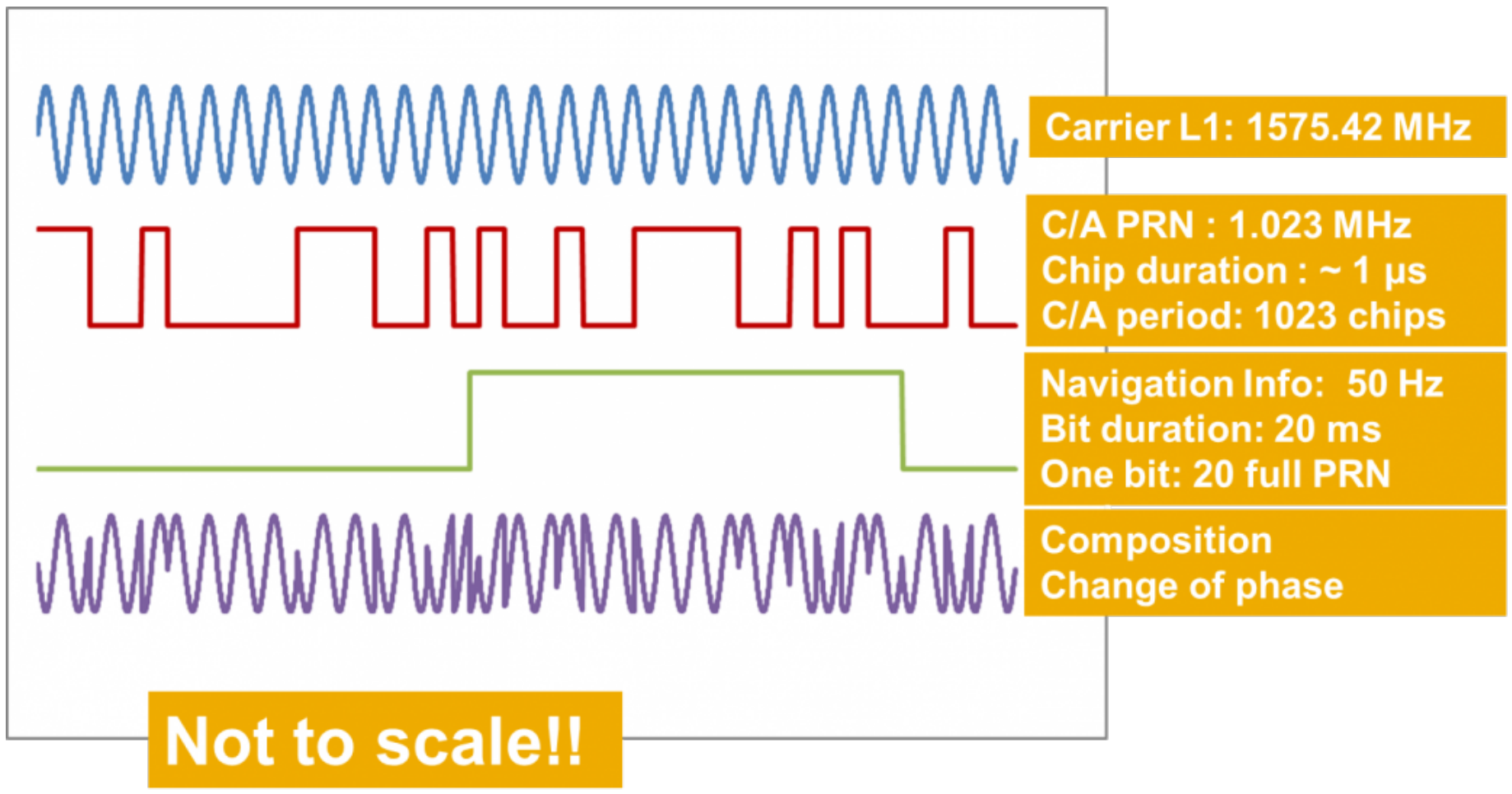

Figure 27: Core structure of GNSS PRN driven signal illustrated with the example of the GPS L1 C/A signal split into its components. From: [32]

### 4.6.3 Modelling message delivery delays

With the above considerations in mind, we will formalise the message delivery delay in the context of two way measurements. First, we will define the message delivery delay, i.e, the time between departure and arrival of a message, is specific and locally bounded for each link in the network. Furthermore, our model assumes there is a global upper bound on the message delivery delay on any message passing trough the network, meaning that each edge of $G$ has a delay of at most $d_{max}$. The actual measured delays can be heterogeneous throughout the network. With this, we formally define the message delivery as:

Definition 16. Message delivery delay
The message delivery delay is the difference in real time between the time $t_1 \in \mathbb{R}$ a message is send from node $v$ and the time $t_2 \in \mathbb{R}$ it is received at node $w$ is bounded by the edge specific message delivery delay $d_{v,w}$ and modelled by the function $M : (\mathbb{R} \times V \times V) \to \mathbb{R}, (t, v, w) \longmapsto M(t, v, w))$ where:

$$M(t_1, v, w) = t_2 - t_1 \leq d_{v,w}$$

where $d_{v,w}$ is an upper bound and $d_{v,w}$ and $d_{w,v}$ are not necessarily symmetric as the path from $v$ to $w$ and the path $w$ to $v$ can vary over time depending on environmental conditions and use case.

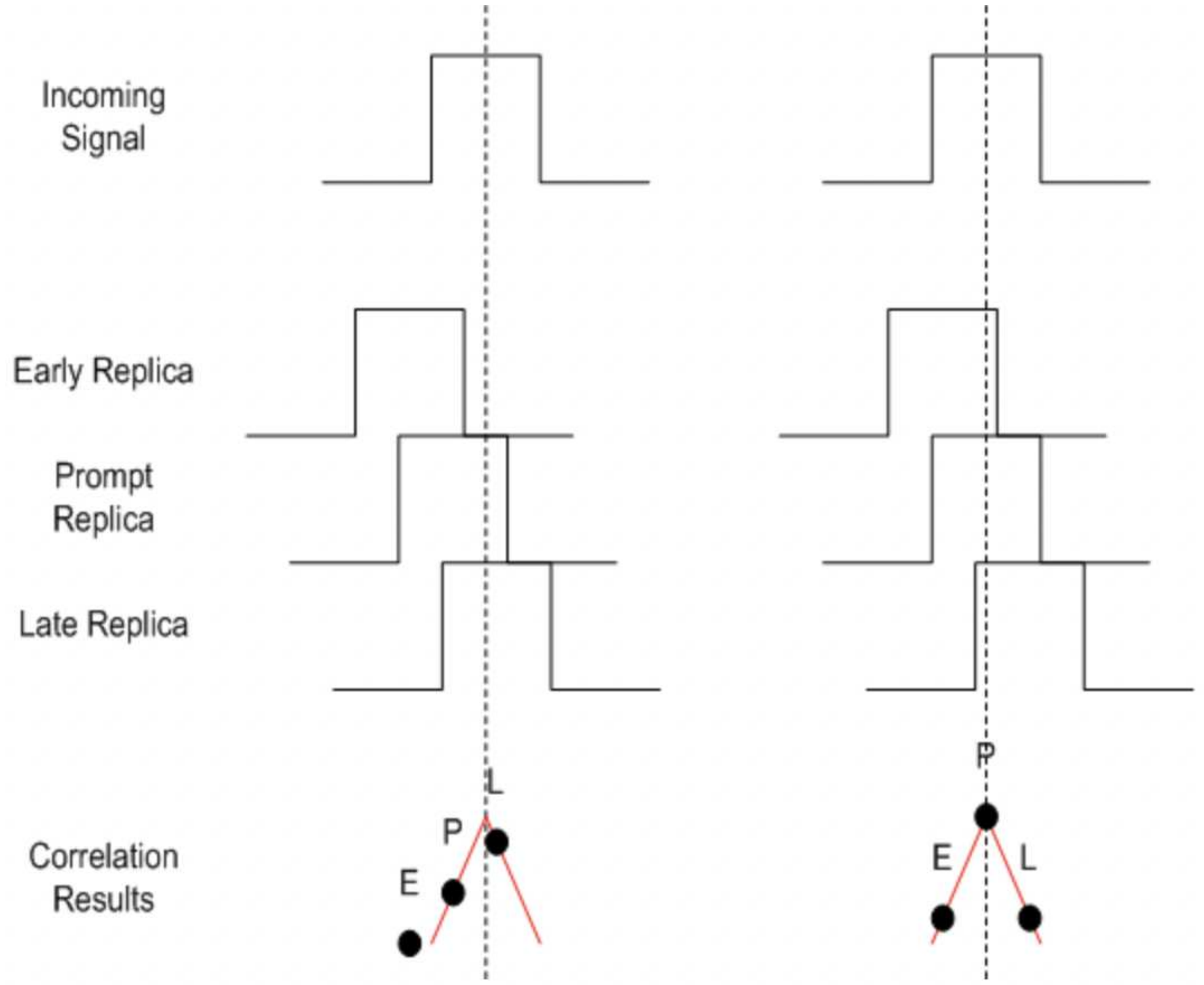

Figure 28: Early/Late correlation on a PRN encoded BPSK signal. The base prompt signal (P) is replicated as an early (E) and a late (L) signal. Depending on whether the timing of the prompt signal aligns (right hand side) or not (left hand side), the signal will be aligned with the center of the prompt replica or not. From: [80].

In our model, we also assume that for all edges and times, we have that:

$$\forall (v, w) \in E : d_{v,w} < d_{max}$$

where $d_{max}$ is the maximum message delay admissible in our network.

With this definition, the message delivery delay in a system based on two-way measurements allows for variable path delay, but only up to a certain, specified degree defined by the upper bounds on each link, $d_{v,w}$. Each such bound is in turn again bounded by the global upper bound $d_{max}$, which specifies the maximum delay on any link. This means that links can have variable delay up to a certain global maximum delay, which is used to determine the worse-case delay of the network. This strict global upper bound is important as it dominates the hop-to-hop delay in the worse-case scenario. Furthermore, we also locally bound the delay of each link within that global upper bound. This bound is required to be allow to bound the asymmetry factor, which is, as the global maximum delay on any link, use case dependent and a central limiting factor to the performance of the algorithm.

### 4.6.4 Modelling measurement uncertainty

As mentioned previously, both one- and two-way measurement have some degree of uncertainty on the exact value of the delay. Previous models group the measurement uncertainty, caused by the precision and accuracy of the equipment with the path length uncertainty, which corresponds to the asymmetry factor in our case. It, however, makes sense to conceptually separate these two notions of uncertainty as their statistical properties are different, yielding different effects and a different impact on the total uncertainty of the system. Formally, we define the measurement uncertainty of an edge as:

Definition 17. Measurement uncertainty
The measurement uncertainty $u_{v,w}$ on each edge $(v, w)$ is the deviation of the path length as well as any possible asymmetries in the message delivery delay:

$$\forall (v, w) \in E : u_{v,w}(t) = \max_{t_1, t_2 \in [t, t+T]} \{|M(t_2, v, w) - M(t_1, v, w)|\}$$

Furthermore, we assume that the measurement uncertainty is bounded by the message delivery delay:

$$\forall t \in \mathbb{R}_{\geq 0} : u_{v,w}(t) < \max\{d_{v,w}; d_{w,v}\} \cdot \epsilon_d + \epsilon_m$$

where $\epsilon_d$ is an upper bound on the maximum path asymmetry and $\epsilon_m$ an upper bound on the measurement uncertainty.

The measurement uncertainty $\epsilon_m$ covers all uncertainties related to the exact time of arrival of a message: due to the inherent timestamp delay mentioned in Section 4.6.5, the exact time of arrival can vary within a small interval [72, 100]. This uncertainty adds up some additional spurious delay to the message delivery delay, yielding a small estimation error.

The path asymmetry factor $\epsilon_d$ formalises the previously mentioned path asymmetry factor. It is usually orders of magnitude smaller than the path length itself. The measurement uncertainty is small in comparison as well in most scenarios, but can be the dominating factor in specific applications such as networks on chip. As we will investigate in more details in the next section, both uncertainties have a static, that means slow changing components, and a stochastic, that means rapidly fluctuating component.

### 4.6.5 Two kinds of uncertainty?

By distinguishing between short and long term fluctuations of the uncertainty, the contribution of the confounding factor can be reduced. In our model, we can consider the path delay variations as static error since the fluctuation in the length of a single path measured in one direction can be fully determined and deducted from the estimation error. The same applies to any slowly evolving error component such as, for example, off-specification delays of device components, slow gradient environmental factors and may more. The contribution of the static, long term error can be deducted in any scenario allowing calibration over some reference signal given that it is visible in the measurement, allowing to precisely compute the bias induced in the measurement [60]. The same applies to semi static errors, which are by nature slowly and predictably evolving. For slow path length variation for example, the bias induced on the measurement can be averaged out, by inverting forward and backward path on two distinct wires for example. Another example are interrupt based timestamps [72]. The interrupt latency in this case is small, although non-zero, inducing a static error offset on the timestamp.

The short term fluctuations of the uncertainty cannot be deducted as they often are unmeasurable parameters given their uncorrelated, independent nature. The short term uncertainty is different for each measurement performed. All types of noise, such as turbulences, jitter [74], voltage[74] and temperature [89] droops count as stochastic unpredictable error. The asymmetry factor also is unmeasurable as it would require a second reference path to determine the length between forward and backward path.

Looking at uncertainties in the context of GCS specifically, we measure the package arrival times to determine the path delay. The measurement of the package arrival time can be implemented in a self calibrating fashion. By sending, for example, a pulse or ping message consecutively over both forward and backward path, the path delay can be determined and deducted. As a result, only the asymmetry factor and the measurement uncertainty remains. As specified above, they are both error terms and hence contribute to the local skew. Although this procedure can be applied to all kinds of wired and wireless use cases in theory, the practical implementation highly depends on the exact setup, deciding on the correlated long term parameters that can be deducted from the measurement. A detailed analysis of the impact on the algorithm's performance will be presented in Section 5.4.2.

## 4.7 A new computational model

### 4.7.1 Summary

This section aims to summarise the changes made to the abstraction as well as the resulting model, comparing it to the previous, highly abstract model from [39] and [66]. For the sake of clarity, the high level computational model can be summarised as:

Definition 18. Computational Model
The computational units have the following properties:

- Each computational unit $v$ consists of a hardware clock $H_v$, modelled by a continuous, differentiable function, which serves as local time reference to any computations undertaken at this unit.
- Additionally, we introduce a logical, also differentiable clock $L_v$, mirroring the behaviour of $H_v$ post the event-based applications corrections of the rate. At each correction, the rate of the logical clock can either left unchanged or speed up by the clock correction factor $\mu$.

The communication in between the computational can be modelled as follows.

Definition 19. Communication Model
The communication network used by the GCS algorithm can be formalised as a directed, connected graph $G = (V, E)$, with $|V| = n$ nodes and where $\forall (v, w) \in E : (w, v) \in E$, satisfying the following properties:

- Each node $v$ of the graph represents one computational unit.
- All $(v, w) \in E$ have an associated edge specific, bounded delay $d_{v,w}$ and an associated edge specific uncertainty $u_{v,w}$ for which it holds that:

  $$\forall (v, w) \in E : d_{v,w} < d_{max}$$

  and

  $$\forall t \in \mathbb{R}_{\geq 0} : u_{v,w}(t) < \max\{d_{v,w}; d_{w,v}\} \cdot \epsilon_d + \epsilon_m$$

  where $\epsilon_d$ corresponds to the asymmetry factor between forward and backward path and $\epsilon_m$ to the measurement uncertainty.
- The asymmetry factor $\epsilon_d$ between forward and backward path is usually magnitudes smaller than the path delay, being $d_{v,w}$ on the forward path and $d_{w,v}$ on the backward path. It holds that:

  $$\max\{d_{v,w}; d_{w,v}\} \ggg \epsilon_d$$

  The same applies to the measurement uncertainty $\epsilon_m$ in most cases.

### 4.7.2 Comparative study

Compared to previous models based on the one-way measurement paradigm as presented in [39] and [66], introducing two way measurements:

- Lifts the constrain of requiring fixed path lengths and delays but allowing to properly determine the path delay and monitor its evolution over time, measurement by measurement.
- Reduces the impact of the uncertainty on the performance of the model, and especially the local skew as they allow to compute and deduct the static path error such as the uncertainty is solely determined by the much smaller short term uncertainty induced by random, unpredictable stochastic errors.

As we will see in Section 5 and Section 6, this significantly, meaning by orders of magnitude, lowers the upper bound on the local skew that can be achieved by the GCS algorithm.

## 4.8 Is the computational model realistic?

### 4.8.1 Scope of the model

The above model can be applied to harmonic oscillators such as LC and quartz oscillators as well as to stable RC, ring and relaxation oscillators in specific cases. It has to be noted that the assumptions also hold for any time reference with a higher stability than quartz oscillators. This means that it can theoretically be applied to a network of high precisions clocks, such as various atomic standards. This doesn't hold true for time references with lower stability than the mentioned harmonic oscillators, as their lower precision and accuracy as well as their less well-defined properties do not allow to assume that the reference can be modelled as differentiable function. This is due to the fact that low quality references are prone to spurious, abrupt frequency and phase jumps [58, 93], that violate in an unrecoverable way the assumption of differentiability.

Providing a model allowing to use GCS with low stability time references is an open and rather difficult problem as a result of their irregular and unpredictable behaviour. This is due to the fact that their frequency widely varies over short time spans. The high and variable drift resulting from that cannot be accounted for in a simple model. We conjecture that it will require adjusting the behaviour of the algorithm such as it counterbalances the high degree of fluctuation.

### 4.8.2 Can the model be applied to implementations?

After considering the scope of the model, one might ask if and how the above abstractions are realistic, i.e, allow a straightforward implementation without requiring a fundamental change of model and abstracted parameters as in previous work. The model specified above provides an almost complete description of what has to be implemented in practice. However, some aspects which are relevant in the implementation process are not included in the communication model.

First of all, timestamps count as separate messages in practice [72]. This is due to the fact that they cannot be produced instantaneously, inducing bias on the measurement value. Solving this issue however is considered simple in implementation. In most cases, the same device path can be used for forward and backward path, allowing to deduce the error as additional part of the path delay.

Spurious frequency jumps are another issue in practice [41]. As mentioned in Section 4.4.4, they are however rare given proper production of the oscillator. Frequency jumps can thus be seen as negligible in most use cases for the type of oscillators covered by our model. However, it has to be noted they are more frequent in radiation intense environments, such as space applications, in which appropriate measures and potentially an adaptation of the model would become necessary.

Finally, the performance of the model presented in this thesis is mostly limited by not accounting for the time dependent variations of the clock drift. The upper bounds on the skew are solely inferred with the help of the global upper bound $\vartheta$ on the drift. Furthermore, it is assumed that all nodes drift at the highest possible rate in all algorithmic conditions, including timeouts. In practice however, the nodes might have much smaller drifts on average, resulting in a much lower skew and much smaller time out values in the average case. Making use of this properties is to be considered non trivial and left for future work.

# 5 GCS Parameters & Algorithm

## 5.1 Solving the clock synchronisation problem

After defining the model of communication and computation, we will focus on the formal definition of synchronising time references with the GCS algorithm.

### 5.1.1 The clock synchronisation problem

The GCS algorithms aims to solve the clock synchronisation problem. Adapting the definition [39, Chapter 7, Def. 7.2], it can defined as follows:

Definition 20. Clock Synchronisation Problem
The clock synchronisation problem requires to bound the global skew:

$$\mathcal{G} = \sup_{t \in \mathbb{R}_{\geq 0}} \mathcal{G}(t)$$

over all executions $\mathcal{E}$, where:

$$\mathcal{G}(t) = \max_{v \in V}\{L_v(t)\} - \min_{w \in V}\{L_w(t)\}$$

with the suprema/maxima taken over all possible executions.

### 5.1.2 Resulting core properties of GCS

The above definition of the clock synchronisation problem does not specify how tight in terms of precision the synchronisation between two nodes can be. Hence, nodes can be tightly or loosely synchronised and are not assigned to different precision levels in terms of their skews. As we would like to tie the quality of synchronisation to some notion of distance, we need to introduce some discrete notions of precision on the estimated skews.

As described in [35], this property of discrete skew levels is a central feature of the GCS algorithm: instead of a tight, boolean notion of synchronisation, it allows to define a synchronisation gradient, where the degree of synchronisation decreases with the hop-distance in the communication network. This distance is determined by the estimation error of a hop, more precisely by the uncertainty on the message delay we defined in Section 4.6.4. This is due to the estimation error of each local measurements being source of the visible the local and global skews.

With this in mind, we can describe the subroutine executing the two-way measurements and the clock estimations for the GCS algorithm in Section 5.2 and Section 5.3. Afterwards we will prune the network graph with the resulting estimation error in Section 5.4 and Section 5.5, before describing the structure the GCS algorithm and the impact of the modifications in Section 5.6 and Section 5.7.

## 5.2 Performing delay and clock offset measurements

### 5.2.1 From practice to theory

In Chapter 4, we discussed the properties as well as the vastness of implementations of two way delay measurements. In order to allow the GCS algorithm to make use of their properties, an abstract summary of the core steps of such a measurement needs to be formalised. First of all, we need to define what parameters are actually benchmarked during the measurement of both message delivery delay and actual offset measurement between clocks.

Path delay measurements allow, as mentioned in Section 4.6.1, to estimate the delay based on the difference between send and arrival times of a message, quantified over timestamps. They correspond to the difference between send time and arrival time of the first message, called Request and send and arrival time of the second message, called Reply, in the protocol defined in Section 4.2.2. The computation time required for processing the first message at the receiver explicitly does not count towards the path delay but is described by an auxiliary parameter.

Given a scenario in which a node $v \in V$ is the sender, initiating the measurement and $w$ is another node, called receiver in our scenario, whose clock offset we aim to determine relatively to the local clock at node $v$, the protocol requires four timestamps:

1. $L_v(t_1)$, timestamp of the departure of the request message at node $v$
2. $L_w(t_2)$, timestamp of the arrival of the request message at node $w$
3. $L_w(t_3)$, timestamp of the departure of the answer message at node $w$
4. $L_v(t_4)$, timestamp of the arrival of the answer message at node $v$

where $t_1, t_2, t_3, t_4 \in \mathbb{R}_{\geq 0}$ are in chronologically ascending order. Hence, we split up the measurement protocol into two subroutines: Algorithm 1 describes the computations executed at node $v$ whilst Algorithm 2 describes the parts of the protocol performed at node $w$.

In the receiver protocol executed at node $w$ and described in Algorithm 2, the node computes the arrival time of this message upon receiving the current clock value of $v$, and at the end of processing appends the outgoing timestamp $L_w(t_3)$ to the answer containing the timestamp received from node $v$ as well as the timestamp of arrival of this message and it sends back to $v$. The sender protocol, executed at node $v$, is a little more involved. Upon being triggered by the GCS algorithm, node $v$ sends its current clock value to $w$ and awaits the response. Upon receiving the answer package within a given time frame, which we will define more precisely in the upcoming Section 5.2.2, Algorithm 1 reports success by computing the timestamp of arrival of the answer and appending it to the timestamps received in the answer package.

Clock offsets are inferred from the timestamps created in the course of the path delay measurement protocol as we will see in Section 5.3.2. The timestamps determined by the `ReqTime()` procedure allow to directly compute a clock offset between the clock values of $v$ and $w$ by estimating the time elapsed between two consecutive timestamps in the respective local timescales.

### 5.2.2 Determining the timeout window

Unlike computations by Definition 1, transmissions are not deemed instantaneous in our model. Hence, the `ReqTime()` procedure needs to be provided with an upper bound on the transmission time to ensure timely and efficient termination of the procedure. This bound mirrors the effect of the message delivery delay on the time elapsing between emitting the answer and receiving the answer at node $v$ in Algorithm 1 but also the processing time required by the receiver to execute all intermediate steps of the protocol in practice, where each computational step requires time. Potential unexpected delays are also included in the processing time $P$, for which we define and use an upper bound $P_{max}$, covering the expected worse case scenario of highest possible additional delays due to computations and unexpected additional delays which we will not quantify in this model for the sake of simplicity. Still specific computational delays have a direct influence on the timeout window: the measurement uncertainty $\epsilon_m$ directly contributes to the response time as the additional delay required to cover the uncertainty between arrival of the message, successful time-stamping and further processing is on the critical path used to quantify the message delivery delay, which we would like to keep as neat as possible for the sake of lowering the estimation error on path delay but also clock offset.

With this we can define the `timeout` condition, implementing the await condition in the while loop of Algorithm 1. As we need to wait for the response to travel to the receiver and back to the sender, the response time is at least twice the maximum message delivery delay $d_{max}$. To this, the maximum expected processing time $P_{max}$ is added to account for all secondary computational effects delaying the response from node $w$ as is the the measurement uncertainty $\epsilon_m$. Last but not least, all of the above factors contributing to the timeout window are multiplied by the maximum clock drift. This is required as both the clock at node $v$ and at node $w$ could drift with rate $\vartheta$ relative to each other.

**Algorithm 1** ReqTime(w) algorithm at node $v$

```
Send L_v(t_1) to w                                   ▷ Send own current clock to neighbour
while ΔL_v(t) < (2d_max + P_max + ε_m) · ϑ do                            ▷ Wait for timeout
    L_v(t_4) ← (L_w(t_2), L_w(t_3))            ▷ t_2 arrival of L_v(t_1) and t_3 outgoing time
end while
return Est(w, L_w(t_2), L_w(t_3), L_v(t_1), L_v(t_4))                  ▷ Estimation 5-tuple
```

**Algorithm 2** ReqTime(w) algorithm at node $w$

```
if received L_v(t_1) from some neighbour v then
    Compute L_w(t_2)                                  ▷ Timestamp of arrival of L_v(t_1)
    Compute L_w(t_3)                                      ▷ Outgoing timestamp at w
    Send Answ(w, L_w(t_2), L_w(t_3), L_v(t_1))                            ▷ Answer to v
end if
```

## 5.3 Computing clock estimates

### 5.3.1 Estimating logical clocks values algorithmically

Given the measurement procedure `ReqTime()` presented in Section 5.2, we can subsequently compute the message delivery delay determining the path length as well as estimating the neighbour's logical clock value based on the relative message delay offset between the two nodes. These computations are implemented by the procedure `ComputeEstimates()`, executed right after the `ReqTime()` procedure, providing the 5-tuple required to determine both delay and offset, has terminated. With this, let's delve deeper into how delay and offset are estimated in Algorithm 3.

The path delay is a system parameter we aim to computed as it is relevant to compute the current estimate of the local skew. To this effect, we first compute $t_v$, describing how much time has elapsed locally between start and end of the measurement at node $v$. We then compute $t_w$, the time required by node $w$ to process the request message and send out the answer message in the `ReqTime()` procedure. Note that the `ReqTime()` procedure itself couldn't determine this time interval to get a better timeout bound as only the arriving answer contains the necessary information needed to estimate the processing time. We consequently compute the message delivery delay to be the difference between the time the measurement took seen from the perspective of node $v$ and the process-ing time required by node $w$ to complete its part in in the measurement protocol.

Averaging the path delay is a measure taken to amortise the impact of uncertainties and asymmetries on the quality of the estimates. It allows to reduce the impact of all uncertainties, especially the path asymmetry as it is dominant in most cases by symmetrising the error. This is helpful as we do not know whether the forward or the backward path is longer since we only measure the path itself, without being able to compare it to some reference measurement. Furthermore, averaging enables us to obtain a real time estimate of the delay, which is, on average, much smaller than the worst case bound.

The offset estimates can be computed directly from the four timestamps determined by `ReqTime()`. The offset corresponds to the difference between the time elapsed in the local notion of time at node $v$ and the local notion of time at node $w$, captured by the respective logical clocks, in between sending and arrival time of the request and answer message seen from the perspective of node $v$. Assuming that $v$ and $w$ are perfectly synchronised, this offset is zero. Based on this, let's derive the formula for the offset deployed in `ComputeEstimates()`. For the forward path we have:

$$L_v(t_2) = L_v(t_1) + (d_{v,w} \pm d_{v,w,asym})$$

For the backward path, we have:

$$L_w(t_4) = L_w(t_3) + (d_{v,w} \mp d_{v,w,asym})$$

where $d_{asym}$ corresponds to the actual asymmetry on the forward and backward path delay respectively. The offset between these times at node $v$ and $w$ correspond to the clock offset per path. We for the request message offset hence have:

$$\begin{aligned}
\tilde{o}_{req} &= L_w(t_2) - L_v(t_2) \\
&= L_w(t_2) - (L_v(t_1) + (d_{v,w} \pm d_{v,w,asym}) \\
&= L_w(t_2) - L_v(t_1) - d_{v,w} \mp d_{v,w,asym}
\end{aligned}$$

For the answer message offset, we have:

$$\tilde{o}_{ans} = L_w(t_4) - L_v(t_4) = L_w(t_3) + d_{w,v} \mp d_{w,v,asym} - L_v(t_4)$$

Combining these two offsets and averaging for the same reason as before, we obtain:

$$\begin{aligned}
\tilde{o} &= \frac{\tilde{o}_{req} + \tilde{o}_{ans}}{2} \\
&= \frac{L_w(t_2) - L_v(t_1) - d_{v,w} \pm d_{v,w,asym} + L_w(t_3) + d_{w,v} \mp d_{w,v,asym} - L_v(t_4)}{2}
\end{aligned}$$

As one might notice, the path delays and the actual asymmetries cancel out, leaving us with an estimation error of the offset and a cleaned up offset of:

$$o = \tilde{o} - \frac{d_{avg} \cdot \epsilon_d}{2} - \epsilon_m$$

The logical clock estimate allows the node $v$ to determine the logical clock value of node $w$ in between two consecutive measurements. The offset provides us with the skew between the two nodes and hence is subtracted from the current clock value at node $v$. Note that the offset can be negative, but that its sign gets accounted for, i.e, a negative offset would be added to the current value of $L_v$ instead of being deducted. As we however keep the same offset for some time, namely a measurement cycle which is bounded but can be longer than the clocks are stable, we need to account for the maximum drift rate between the two nodes. In the worse case, one of the two nodes advances at rate 1 whilst the other has rate $\vartheta$. Given that we can not tell which of the two is the faster one without comparing to a reference measurement, we need to assume the worst case, in which the offset to a faster node is overestimated. This is also required to ensure mutual exclusiveness of the algorithmic conditions and triggers, presented in Section 5.6 and Section 5.7. Furthermore, this property is essential for the convergence as it ensures that the algorithm overcorrects fast running nodes, preventing a race condition in which slow nodes would push all nodes to arbitrarily speed up.

---

Algorithm 3 ComputeEstimates(w) algorithm at node $v$

---

Require: $Est(w, L_w(t_2), L_w(t_3), L_v(t_1), L_v(t_4))$ ▷ Data of neighbour present

  $t_v = (L_v(t_4) - L_v(t_1))$ ▷ Compute relative time at $v$

  $t_w = (L_w(t_3) - L_w(t_2))$ ▷ Compute relative time at $w$

  $d_{avg} = \frac{t_v - t_w}{2}$ ▷ Compute average message delivery delay

  $\tilde{o} = \frac{L_w(t_2) + L_w(t_3) - (L_v(t_1) - L_v(t_4)}{2}$ ▷ Compute offset

  $\tilde{L}^v_w(t) = L_v(t) - \tilde{o} - d_{avg} \cdot (\epsilon_d + \vartheta - 1) - \epsilon_m$ ▷ Estimate clock of $w$

  return $d_{avg}$, $\tilde{L}^v_w(t)$

---

### 5.3.2 Estimating logical clock values formally

As mentioned in the previous section, the logical clock estimates computed at node $w$ at node $v$ will allow the GCS algorithm to determine its modus operandi. To study the impact of the logical estimates on the algorithm, we formalise the behaviour of the above logical clock estimates as:

Definition 21. Logical Clock Estimates [39, Eq. 8.6]
Each node $v$ computes an estimate $\tilde{L}^v_{v,w}$ of the time difference to its neighbour node $w$ over the directed edge $(v, w)$ up to an directed edge specific estimation error $\delta_{v,w}$:

$$L_w(t) - \delta_{v,w} \leq \tilde{L}_w^v(t) \leq L_w(t)$$

The estimator function $\tilde{L}_w^v : (\mathbb{R} \to \mathbb{R}; t \longmapsto \tilde{L}_w^v(t))$ is continuous and differentiable but only for the duration of a computational cycle, see Section 5.5.2.

This definition mirrors the behaviour of the estimates as determined by the `ComputeEstimates()` procedure. We deduct the estimation error $\delta_{v,w}$ from the clock offset assuming the worst case. The resulting lower and upper bound on the estimate are incorporated into the algorithmic triggers, specified in Section 5.7, allowing to decide on whether the correction factor $\mu$ will be applied by GCS or not when accounting for the estimation error.

## 5.4 Estimation error and error-weighted network graph

### 5.4.1 How big is the estimation error?

Given the influence of the estimation error on the quality of measurements and estimates, we aim at quantifying it formally. As a first step, we need to define the scope of the error metric. Since our estimates are ultimately based on physical link measurements, the resulting error is mainly dependent on the length of the measurement, as a result of physical properties of such a measurement. Hence, we define our error metric on the measurement interval, which we assume to have length $T$. In this subsection, we will discuss how, given the mathematical description of the error, one can optimise the length of this interval as well as the number of consecutive, distinct measurements taken to minimise the measurement error.

With this, we will dwell into the structure of the error. Four core factors have an influence on the global estimation error $\delta$:

1. Measurement uncertainty: mainly models the physical noise effects leading to the uncertainty on the measurement quality, such as arrival times and resolution but also upper bound on asymmetry factor of the path between $v$ and $w$, described by the parameter $u_{v,w}$.

2. Link uncertainty: mainly models the distortion effects on the link due to environmental noise leading determining the stability of the time variable message delivery delay modelled by the edge-dependent parameter $d$.

3. Clock drift rate: the clock drift, or more specifically the stability of both clocks, has a direct influence on the quality of the measurements as a higher drift error lowers the quality of the message delivery delay estimations.

4. Side-type of error: depending what type of error, so one or two-sided error is taken as reference in the measurements, one might need to convert this error to fit our theoretical model which, out of historical reasons, operates with one-sided errors.

Given this, we can both specify and bound the estimation error formally. To keep things simple and easy to optimise for performance, a very simple abstraction taking into account only the measurement uncertainty as well as the high level effect of clock drift was chosen. Further details on this decision will be discussed in Section 5.4.2.

Definition 22. Estimation error
Each logical clock estimator has an edge specific estimation error $\delta_{v,w}$ for an edge $(v, w) \in E$. This error mirrors the impact of the message delivery delay, the uncertainty, and the maximum clock drift on the quality of the estimation on any arbitrary measurement interval of length $T$. It is defined as:

$$\forall t \in \mathbb{R}_{\geq 0}, (v, w) \in E : \delta_{v,w}(t) = 2 \cdot (u_{v,w}(t) + \max\{d_{v,w}; d_{w,v}\} \cdot (\vartheta - 1))$$

For a bidirectional edge $e = (v, w)$ the difference of estimation errors per direction is bounded for all times and especially during the measurement intervals $[t, t + T]$ and can be obtained from the bound on $u_{v,w}$:

$$\forall t \in \mathbb{R}_{\geq 0}, (v, w) \in E : \delta_{v,w}(t) < \delta_{v,w,max}$$

where $\delta_{v,w,max} = 2 \cdot (\max\{d_{v,w}; d_{w,v}\} \cdot (\vartheta - 1 + \epsilon_d) + \epsilon_m)$ as an upper bound on the effective estimation error on an edge.

In this definition, the estimation error results from adding up the link specific uncertainty, which also covers the asymmetry factor as well as the measurement uncertainty, and the uncertainty on the distortion of the link length caused by clock drifts relative to each other without there being a possibility to distinguish who's ahead and who's behind. In the worst case, one node advances at rate 1 while the other one proceeds at rate $\vartheta$, resulting in a drift gap and matching distortion of the message delivery delay by a factor of $\vartheta - 1$.

The estimation error can subsequently be bounded by inserting the worst case upper bound on the link specific uncertainty $u_{v,w}$. Note that in both cases, we map our two-sided measurement error as it is given by any physical measurement back to a one sided error by symmetrising and adding up the error.

### 5.4.2 Determining the optimal measurement interval - a task hard to plough trough

As mentioned in the previous section, we aim to optimise the measurement interval parameter $T$ such as the uncertainty $u$ and the message delivery delay $d$ are minimised. On important question one might ask regarding this optimisation is whether an enhanced distinction between different effects causing the uncertainty and the message delivery to grow is helpful in achieving a better solution.

Ultimately, the quality of the estimation is bound to the stability of all the factors mentioned above. This is due to the fact that a stable channel between two nodes allows to have an optimal estimation of the neighbours clock as no noise alters the perception of the neighbours clock frequency and phase. Hence, one might think about about taking a closer look at the components of the globally perceived uncertainty with the goal of lowering its impact on the estimation accuracy. A straightforward strategy could be to distinguish between fast and slow fluctuations, allowing to compensate predictable slowly changing errors. However, as we will see, this distinction doesn't really help lower the overall uncertainty. This is due to the fact that the stability of a link, especially of fibre links which are not already prone very important environmental factors, is almost fully independent of the duration of the measurement interval. This means that measuring for longer, such as to isolate long and short term error components doesn't help with lowering the global uncertainty as the unpredictable, fast fluctuations in the error dominate beyond a certain, fastly reached point, which we will call cut-off point.

To be more precise, the cut-off point, from which measuring for a longer time span doesn't help lowering the uncertainty, corresponds to the point in time at which the link noise surpasses the detection, i.e, the measurement noise. It hence has to be determined for each link individually. For a typical fibre link, having medium length in the range of a few tens of kilometres, this point is reached after a very short amount of measurement time [60]. The exact cut-off point depends on the length and the operating environmental conditions of the fibre link. With this, it becomes apparent that the fibre noise, i.e, the fluctuations induced on the propagation trough the fibre by environmental factors such as vibrations, temperatures fluctuations etc, is the limiting factor. Unsurprisingly, this effect increases with the cable length as the link noise accumulates and increases with the link length. The shorter the fibre, the further in time the cut-off point is located. This means that for short paths, the cut-off point is pushed back as a result of the physical properties of the fibre cable.

Over short distance and for very short measurements, that last less than 1 second, the error bound can be improved by exploiting the short and longterm components of the error [60]. This is due to the fact that the detection noise, which is responsible for the measurement uncertainty, is mostly white gaussian noise due to the nature of the underlying processes where white noise dominates jitter [15]. This is due to the fact that the bandwidth of the white noise component of the error is much larger than the bandwidth of the flicker component, resulting in a higher contribution to the corresponding integral [55] (See Figure 7). More formally, the uncertainty $u_f$ on the remote frequency estimation, i.e, the neighbour's clock thus can be written as:

$$u_f = \frac{\epsilon_m}{T}$$

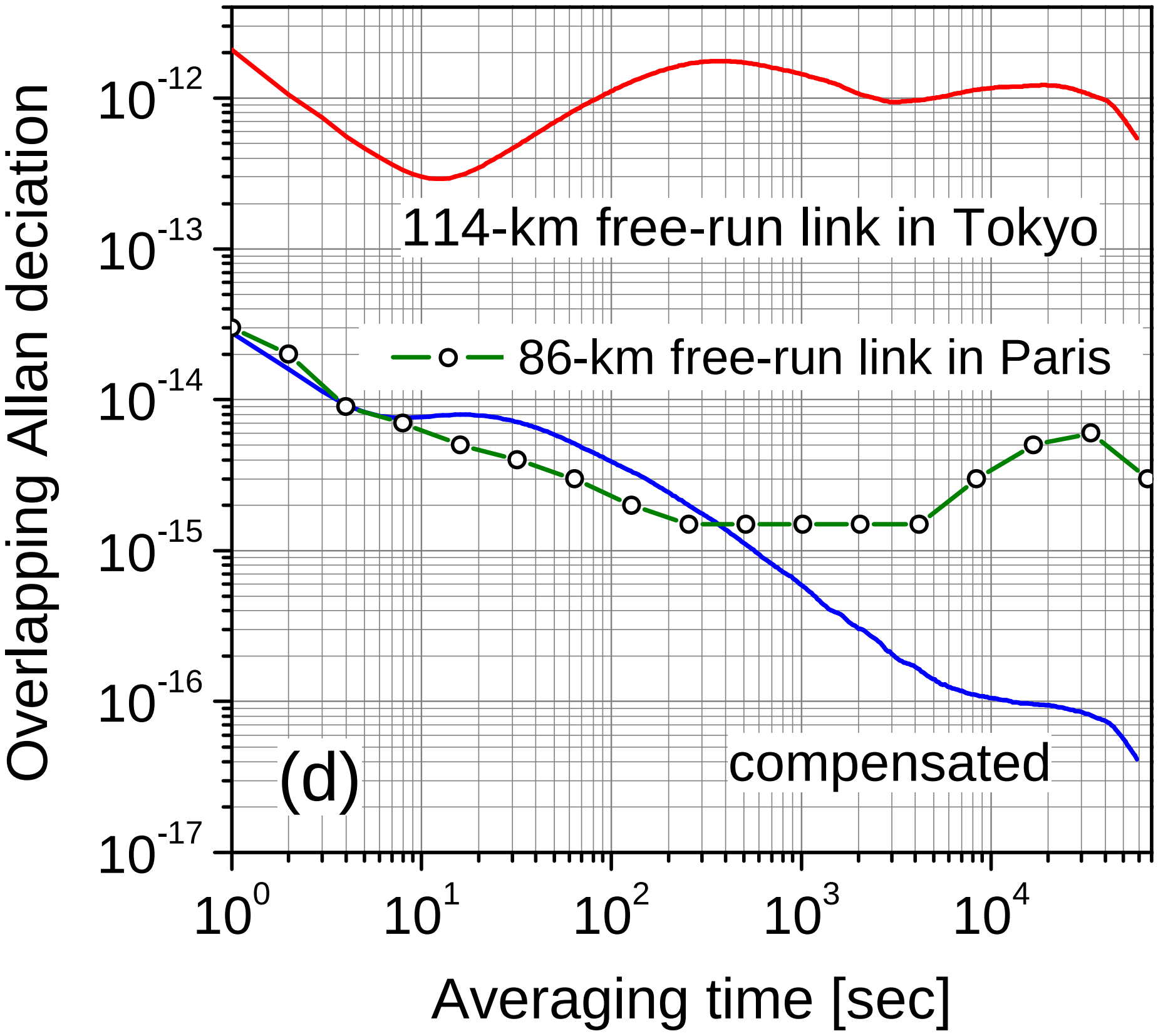

Figure 29: Comparing the free-run Allan Deviation of fibre links in Tokyo and Paris yields the conclusion that measuring for longer periods (x-axis) does not yield a reduction of the measurement uncertainty (y-axis) on the performed one-way measurements. However, performing two-way measurements (blue line) does yield a reduction of the observed global uncertainty. From: [60]

It hence helps to measure for longer as this statistically spreads out the error.

However, it might be even more helpful to average over multiple measurements, allowing to keep the average measurement error below a certain threshold that could guarantee certain bounds to the performance of the algorithm as we have for $N$ consecutive measurements that:

$$u_{f,N} = \frac{\epsilon_m}{\sqrt{N}\,T}$$

Note that the above equations only apply if the measurement intervals are short, i.e, as long as white gaussian noise dominates jitter as correlation effects on start to emerge on jitter in between single measurements.

Two techniques allow to determine the cut off point given the above relations between length of measurement interval and measurement uncertainty:

1. High precision noise measurements: require specific, high precision hardware to create highly accurate model of link and detection noise over repeated measurements.

2. In-situ measurements: allow to estimate the quality of the link by transferring data without specific equipment yet requires rather stable reference clocks at both end nodes to guarantee a decent level of accuracy.

In situ measurements are preferred as they do not require specific equipment, adding overhead, and allow for self-contained calibration of links in our model. To calibrate a network of nodes in situ, all links are measured and a set containing those who are the most average in their behaviour chosen as reference. Out of this set, the average worst case link noise can be determined. An additional safety margin can be applied to this worst case bound, yielding a reliable bound on the link noise for a given measurement interval length. This bound can be used to determine the cut-off point given a matching bound on the measurement error.

While the abstract process might sound simple to operate, variable and hard to identify environmental effects might alter the quality of the reference links or falsify their value as the condition between calibration point and regular operation might vary quite a lot. This makes it very difficult to compute a reliable estimation of the link measurement without costly regular recalibration.

Furthermore, the above description abstracts away the impact of the frequency and phase drift of the local clocks used in the measurements. In practice, low quality oscillators might start to rapidly drift during the reference measurements, yielding the internal clock drift error to dominate the link uncertainty measurement. However, dealing with rapidly drifting clocks in our estimation model is very difficult: the clock drift is to be considered non-linear and unpredictable in its behaviour [40], only allowing for numerical solutions on a case by case basis. As coping with this highly non-trivial issue would open a can of worms, especially as the drift can be triggered by environmental factors such as steep but also slow temperature changes at almost arbitrary points in time, we settle for the assumption that the short term stability of the oscillators considered in our model is sufficiently high on average to guarantee decent measurement quality.

### 5.4.3 Pruning the graph with the edge specific estimation error

The previous two sections presented the cause but also the procedure of determining the estimation error as well as a formal description of it. With this, the practical use of the estimation error as algorithmic parameter can be elaborated: to the GCS algorithm, the estimation error is a metric for estimating the distortion of the neighbour's logical clock value excluding the effects of the clock drift. Each node hence has to know the distortion on the paths to its neighbours to be able to bound the estimation error of the logical clock. To this effect, we prune the graph edges, which correspond to the communication paths, with the estimation error. This yields the following definition:

Definition 23. Edge specific weights
Each edge $e = (v, w) \in E$ in the network graph $G$ is associated with an edge specific weight:

$$\widehat{\kappa}_e(t) = 2 \cdot (\max\{d_{v,w}; d_{w,v}\} \cdot (\vartheta - 1 + \epsilon_d) + \epsilon_m)$$

The edge specific weight parameter $\widehat{\kappa}_e$ fulfils the equation $\widehat{\kappa}_e = \delta_{v,w,max} \geq d_{v,w}$. This is due to the fact that we already have single sided, bidirectional error as in the original definition of the algorithm. Given the pruned graph, we can define weighted distances, that take into account the quality of the taken edges:

Definition 24. X-weighted distance
The $\widehat{\kappa}_e$-weighted distance $\mathrm{dist}^X(v, w)$ between nodes $v$ and $w$ is defined as:

$$\mathrm{dist}^X(v, w) = X \sum_{e \in P(v,w)} \widehat{\kappa}_e$$

where $P(v, w)$ is a shortest weighted path between $v$ and $w$.

This definition contains a placeholder variable, X, which will later be replaced by the level description, a metric used by the algorithmic conditions of GCS to describe allowed amount of skew allowed per discrete level $s$.

## 5.5 Algorithm for gradient clock synchronisation

All definitions in this subsections are adapted from Chapter 8 of the unpublished book by Függer et al. [39].

### 5.5.1 An abstract definition of GCS

To define the GCS algorithm, we first need to define the problem that it is supposed to solve. An intuitive description of the goal of the algorithm is to keep the local skew between two neighboured nodes as small as possible. In a more formal manner, the clock synchronisation problem is defined as:

Definition 25. Clock Synchronisation Problem
The clock synchronisation asks for a solution satisfying that any two adjacent nodes $v$ and $w$ neighboured in $G$ satisfy:

$$\sup_{\forall t \in \mathbb{R}_{\geq 0}} \{L_v(t) - L_w(t)\} \leq \chi$$

where $\chi$ is some arbitrary threshold on the synchronisation quality set as goal on a per use case basis.

The clock synchronisation procedure in turn is a high level description of how the algorithm proceeds to solve the above problem.

Definition 26. Gradient Clock Synchronisation Procedure
The Gradient Clock Synchronisation algorithm solves the clock synchronisation problem, that is to correct the accumulated clock drift between real and local time at node $v$ by:

1. Estimate the clock skew between its own clock and the clock of adjacent neighbours in the network, performed by the procedures `ReqTime(_)` and `ComputeEstimates(_)` .
2. Based on this estimation of the skew, deciding whether or not to apply a corrective factor $\mu$ to its hardware clock value, performed by the core routine `GCS(_)`.

### 5.5.2 The GCS Algorithm

Algorithm 4 describes the GCS algorithm as it is operating at some node $v$. In fact, all nodes in the network run the same algorithmic procedure. In more details, the GCS algorithm proceeds as follows:

Step 1 - Wake-up: at the beginning of each computational cycle, that last a measurement interval of duration $T$ plus the clock stabilisation time $T_{stab}$, so after $T + T_{stab}$ local time has elapsed, the algorithm initiate a new measurement and correction round. We call this a computational cycle. To this effect, it first stops applying the correction factor from the previous measurement interval. Whilst the correction factor could theoretically be maintained in the upcoming measurements and estimates, accounting its effect is rather verbosy and would add an additional error factor to the estimates. We hence decided to skip this part and assume that clocks are free running, implying that they can drift with rate at most $\vartheta$ starting from this point.

Step 2 - Performing measurements: after completing initialisation and setting the clock as free running, the `ReqTime(_)` procedure performs the message delivery delay measurements that will be used to estimate the neighbour's logical clock value, as specified in Section 5.3.1, Algorithm 1 and Algorithm 2. This procedure is repeated for all neighboured nodes in the set $N(v)$ of node $v$.

Step 3 - Compute Clock Estimates: with the data gathered by the `ReqTime(_)` procedure, the `ComputeEstimates(_)` procedure, specified in Section 5.2.1, algorithm 3, can compute clock estimations valid for the remaining measurement interval, and this for each node that is in the neighbourhood of node $v$.

Step 4 - Evaluate triggers and set correction factor: with the estimates computed by the `ComputeEstimates(_)` for all neighboured nodes, we can evaluate the triggers, presented in Section 5.7. Based on wether the neighboured nodes fulfil the slow of the fast trigger from the point of view of node $v$, the algorithm either decides to apply a corrective factor $\mu$ to the logical clock of $v$ or not. If the node gets placed in slow mode, this means that the clock of node $v$ is ahead of its neighbours, it will continue proceeding at the current clock rate. If, however, the fast trigger holds, meaning that node $v$ is behind its neighbours time-wise, the rate of the logical clock is speed up by factor $\mu$. For the sake of completeness, if none of the two conditions apply, the node proceeds at its own rate by default.

**Algorithm 4** GCS algorithm at node $v$ [39, adapted from Alg. 7]

```
if L_v(t) mod (T + T_stab) = 0 then
    Set dC_v/dt = 0                                  ▷ Stop correcting clock
    ∀w ∈ N(v): ReqTime(w)                            ▷ Perform measurements
    ∀w ∈ N(v): ComputeEstimates(w)                   ▷ Compute clock estimates
    if ST then
        Set dC_v/dt = 0             ▷ Slow mode, L_v progresses at own rate
    else if FT then
        Set dC_v/dt = μ                  ▷ Fast mode, L_v is speed up by μ
    else
        Set dC_v/dt = 0                    ▷ Default, progress at own rate
    end if
end if
```

## 5.6 Slow and fast conditions

All definitions in this subsections are adapted from Chapter 8 of the unpublished book by Függer et al. [39].

### 5.6.1 Purpose of conditions in GCS

To decide wether or not to adapt the rate of the logical clock, the GCS algorithm implements two conditions defining the minimum or maximum gap between the local logical clock and the logical clock of the neighboured nodes. This skew bound depends on a fixed factor, that is the edge weight $\kappa$ indicating the error seen on the edge and a dynamic, adjustable factor, the skew level $s$.

The edge weight $\kappa$ describes the error, i.e distortion coefficient of the clock estimate, seen over the edge between node $v$ and its neighbour $w$. We aim to spread out this error over the path, if it contains more than a single edge, to ensure proper switching between the slow mode, in which the node proceeds at its own rate, and the fast mode, in which it speeds up its logical clock to catch up with its neighbours. The purpose of this spreading is the ability to visualise the cumulative skew over a path, instead of adding the errors of all edges without amortising the error: in fact the skew behaviour of a path, where measurements are taken over longer distances, might differ from the sum of the local one-to-one measurements.

The skew level $s$ in turn describes the degree of freedom we allow on the skew threshold of the condition. The larger the value of $s$, the bigger the allowed skew between nodes, given constant edge weights. The skew levels, defined as multiples of two, even or odd depending on wether the fast or slow condition are considered, also ensures that both conditions will be mutually exclusive i.e cannot hold at the same time, as this would result in undefined behaviour.

### 5.6.2 The slow condition

If a node fulfils the slow condition, it runs faster, in terms of the threshold set in the condition, than its neighbours. More specifically, SC-1 implies that $v$ is $(2s-1)\kappa_e$ faster than at least one node in its neighbourhood but also at the same time SC-2 implies that all other nodes in the neighbourhood are at most $(2s-1)\kappa_e$ ahead $v$. In this case, the algorithm will set it to slow mode i.e it will continue running at its own rate without there being any corrections. More formally, we have:

Definition 27. Slow condition (SC)[39, adapted from Def. 8.14]
A node $v \in V$ fulfils the slow condition at time $t \in \mathbb{R}_{\geq 0}$ for a skew level $s \in \mathbb{N}_{>0}$

if and only if:

$$\text{SC-1} : \exists e = (v, x) \in E : L_v(t) - L_x(t) \geq (2s - 1)\kappa_e$$

and

$$\text{SC-2} : \forall e = (v, y) \in E : L_y(t) - L_v(t) \leq (2s - 1)\kappa_e$$

#### 5.6.3 The fast condition

If a node fulfils the fast condition, it runs slower, in terms of the threshold set in the condition, than its neighbours. More specifically, FC-1 implies that at least one node is $2s\kappa$ ahead of $v$ is but also at the same time SC-2 implies that no other node is more than $2s\kappa$ behind $v$. In this case, the algorithm will set it to fast mode i.e its logical clock will be corrected by speeding up its rate by a factor $\mu$. More formally, we have:

Definition 28. Fast condition (FC)[39, adapted from Def. 8.13]
A node $v \in V$ fulfils the fast condition at time $t \in \mathbb{R}_{\geq 0}$ for a skew level $s \in \mathbb{N}_{>0}$ if and only if:

$$\text{FC-1} : \exists (v, x) \in E : L_x(t) - L_v(t) \geq 2s\kappa_e$$

and

$$\text{FC-2} : \forall (v, y) \in E : L_v(t) - L_y(t) \leq 2s\kappa_e$$

Note that in practice one would need to add an additional threshold parameter to this definition (as well as to the slow condition above) to prevent the nodes from oscillating at infinitely small granularity between slow and fast condition.

## 5.7 Slow and fast triggers

All definitions in this subsections are adapted from Chapter 8 of the unpublished book by Függer et al. [39].

#### 5.7.1 Why do we need triggers additionally?

The slow and fast condition defined in the section above are mutually exclusive but also strict, as they evaluate the actual clock values, making them unable to take into account the uncertainty on the estimates of neighbour's clock values the node actually receives. As the error induces some bias on the decision, i.e might alter when we switch from slow to fast mode and vice-versa, the algorithm cannot make this decision being agnostic of the error. Hence, the triggers are parameterised, i.e change their values at the beginning of each measurement interval, as the error does. We hence implement triggers, which take the neighbour's node logical clock estimate as reference value. As mentioned in Section 5.3.2, we assume a one-sided error in all cases, only one trigger, namely the fast trigger, has to cover the estimation error $\delta$, which varies over time. Since the clock estimator will underestimate the value of a clock node, we need to relax the condition by amount the error can contribute. This results in the following definitions of slow and fast trigger.

### 5.7.2 The slow trigger

The slow trigger behaves analogously to the slow condition, as we consider the error to be one-sided, with the sole exception that we consider the neighbour's clock estimates, containing the estimation error, as reference clock values when computing the skew gap. Formally, we thus have:

Definition 29. Slow trigger (ST)[39, adapted from Def. 8.16]
A node $v \in V$ fulfils the slow trigger at time $t \in \mathbb{R}_{\geq 0}$ for a skew level $s \in \mathbb{N}_{>0}$ if and only if:

$$\text{ST-1} : \exists (v, x) \in E : L_v(t) - \tilde{L}_x(t) \geq (2s - 1)\kappa_e$$

and

$$\text{ST-2} : \forall (v, y) \in E : \tilde{L}_y(t) - L_v(t) \leq (2s - 1)\kappa_e$$

### 5.7.3 The fast trigger

The uncertainty on the decision wether the slow or the fast condition currently apply is covered by the fast trigger. Since the error is assumed one-sided on each edge, we respectively add or subtract the potential estimation error of $\delta$ from the condition to create the trigger. This formally results in:

Definition 30. Fast trigger (FT)[39, adapted from Def. 8.15]
A node $v \in V$ fulfils the fast trigger at time $t \in \mathbb{R}_{\geq 0}$ for a skew level $s \in \mathbb{N}_{>0}$ if and only if:

$$\text{FT-1} : \exists (v, x) \in E : \tilde{L}_x(t) - L_v(t) > 2s\kappa_e - \delta_{v,x}(t)$$

and

$$\text{FT-2} : \forall (v, y) \in E : L_v(t) - \tilde{L}_y(t) < 2s\kappa_e + \delta_{v,y}(t)$$

Since $v$ might underestimate $x$'s clock value by up to $\delta$, it might satisfy FC-1 but doesn't notice this fact due to the offset induced by the error, which the trigger FT-1 counterbalances by subtracting the error from the targeted value. This analogously applies to FC-2 and FT-2.

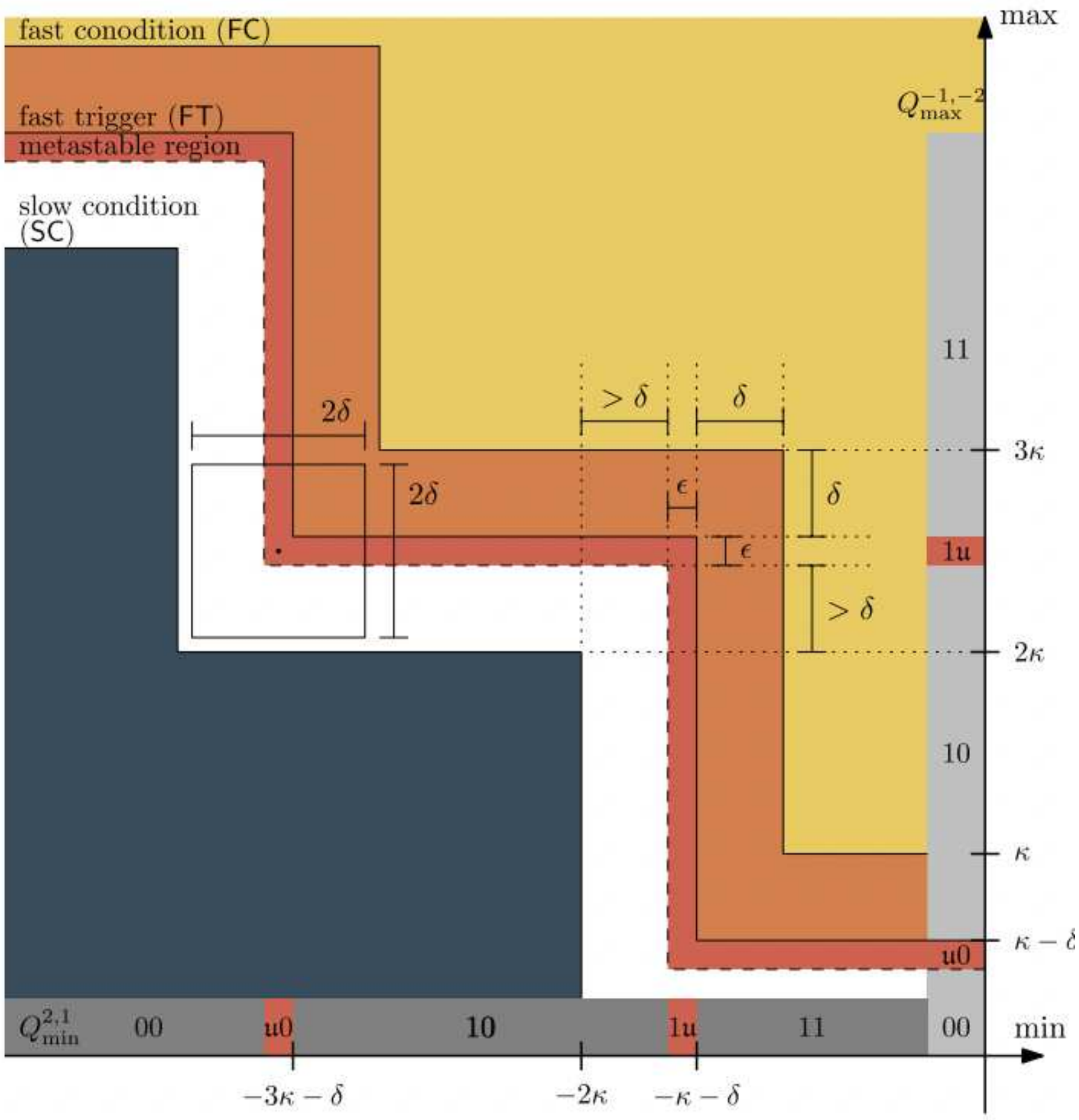

Figure 30: Visualisation of fast and slow triggers and conditions for the GCS variant deployed in PALS. From: [13]

## 5.8 Estimating skew: the potential function

### 5.8.1 What's the purpose of the potential function?

Potential functions are a technique widely used in theoretical computer science and complexity estimation, based on the potential method [88]. In our case, we will create a potential function to estimate the growing skew between any two nodes in the networks by computing, hop by hop, on the path between the two nodes, more precisely of how much passing an edge can add to the skew.

A potential function is a suitable choice here as it fits the frame of adding a certain cost, in our case skew, at each computational step, in our case hop in the graph, translating each hop taken in the graph into a potential of added skew.

At first glance, this might seem a little odd as we cannot categorically bound the skew between any two nodes in the network without predefined restrictions on their behaviour. However, as we have a well defined bound on the maximum clock rate, which translate into a maximum distance in terms of clock values over time, and a bounded error, the potential function can be bounded and used to show certain properties of the algorithm, namely bounds on the local and global skew.

### 5.8.2 Formal definition

More concretely, the potential function describing the skew is a function of the path uncertainty $\kappa$, which expresses the bias seen on the path, a matching skew level $s$, describing the degree of freedom on the skew as well as the difference in the logical clock values between starting and end point of the path. Adapting the definition [39, Chapter 8, Def 8.19] of the unpublished book by Függer et al, formally we have:

Definition 31. Skew potential function
The skew potential function $\Psi_v^s(t)$ describes the maximum skew between a node $v$ and all other nodes in $G$ for a skew level $s \in \mathbb{N}_{>0}$ at time $t \in \mathbb{R}_{\geq 0}$:

$$\Psi_v^s(t) = \max_{w \in V} \left\{ L_w(t) - L_v(t) - (2s-1) \sum_{e \in P(v,w)} \widehat{\kappa}_e(t) \right\}$$

The node that maximises this function, is called leading node and defines the skew level potential function $\Psi^s(t)$:

$$\Psi^s(t) = \max_{w \in V} \{\Psi_w^s(t)\}$$

Note that, whenever $\Psi_v^s$ or $\Psi^s$ are positive, the slow and fast trigger will allow to bound its growth and hence the skew that can build up between nodes on a path, enabling to prove bounds on how fast the skew can grow given a bound on how much nodes in fast mode can lag behind and how much nodes in slow mode can be ahead given the corrections performed by the algorithm. The details of these bounds are topic of Section 6.3 and Section 6.4.

# 6 Analysis

All proofs presented in this section follow the outline of Chapter 8, Sections 8.5.2 & 8.5.3 of the unpublished book by Függer et al [39].

## 6.1 Prerequisites

### 6.1.1 Goal of the analysis

The aim of this chapter is to formally analyse the behaviour of the GCS algorithm described in Section 5.5.2 and derive formal bounds on global and local skew. To this effect, we will prove that the algorithmic conditions and trigger indeed are mutually exclusive and allow to implement the catch-up and wait-up lemmas, describing how fast skew can grow at drifting nodes as well as how fast the nodes lagging behind can catch up with them. The catch-up and wait-up lemma will also enable us to quantify how much skew can be build up and caught up respectively within a specific time frame, yielding an upper bound on the local skew. With this local bound, we will derive a lower bound on the global skew as well as interpret the results of the analysis.

### 6.1.2 Scope of the analysis

To reach the goals specified above, we first need to determine the scope of the analysis, both in terms of how long we observe the behaviour of the algorithm in time as well what exactly parameters we aim to assess and quantify with a metric.

To this effect, we will first discretise the algorithm's notion of time. This needed as by definition, both hardware clocks and logical clocks are continuous functions operating over the (positive) real numbers whereas the algorithmic conditions implementing the two modes of operation of the algorithm, namely fast and slow mode, are either active a disjoint times or switching at discrete points in time defined by start time and the length of a measurement interval. More formally, Algorithm 4, triggers a measurement of duration $T$ at the beginning of each computational cycle, computes the clock estimates and based on the computed estimates corrects the logical clock rate for another $T_{stab}$ time, called stabilisation time. The stabilisation time can be determined by putting in relation and evaluating the different parameters. For a detailed description, we refer to [39, Chapter 12.3].

We thus set the analysis interval $I$ to match one computational cycle of duration $I = [t, t + T + T_{stab}]$, starting at time $t$. Note that this parameter takes into account any drift that can be visible locally at a node and can induce bias. To track the progress of the algorithm over multiple computational cycles, we can index each cycle with natural numbers. We can define a sequence $\mathcal{S}$ of executions with indices in $n \in \mathbb{N}_{\geq 0}$:

$$\mathcal{S} = \{I_0, I_1, ..., I_k\} \text{ for } k \in \mathbb{N}_{\geq 0}$$

where $n$ denotes the index of a single cycle and $I_0$ is the boot-up cycle, which forms as special case which will be described in detail later.

It has to be noted that, because we work over disjoint intervals, many time dependent variables like the delay $d(t)$, the uncertainty $u(t)$, the estimation error $\delta(t)$ and the edge specific weights $\widehat{\kappa}(t)$ are fixed within the interval, i.e, do not vary for the duration of the analysis interval as they are only updated after each new measurement round. The actual analysis starts after these parameters have been determined by the current measurement round that occurred within the analysis interval. We hence will look at parameters $d$, $u$, $\delta$, and $\kappa$ evaluated at time $t + T$:

$$\begin{aligned} d &:= d(t+T) \\ u &:= u(t+T) \\ \delta &:= \delta(t+T) \\ \kappa &:= \widehat{\kappa}(t+T) \end{aligned}$$

These static parameters are valid until the end of the analysis interval, i.e, until time $t + T + T_{stab}$.

During the above analysis time, we aim to determine the average skew over different path lengths in order to study the evolution of the skew as a function of the path length, which will be the central assessment parameter for skew in the graph. This is connected to the gradient property of the GCS algorithm, presented in Section 2.2.1, which inherently relates increasing distances to increasing skew. For the local skew, this means studying any pair of adjacent neighbours; for the global skew path lengths of lengths up to $D$, the diameter of the network.

## 6.2 Theorems, Lemmas, and Proofs

Beyond the known parameters listed in Section 4 and Section 5, the up-following theorems, lemmas and proofs are partially based on specific assumptions which are hidden from the core algorithm. One of them is initialisation. As we will see in Section 6.2, many proofs rely on the assumption that there is some, rather small upper bound on the drift between any two nodes in the graph at boot-up time, i.e, at the beginning of the boot-up cycle.

### 6.2.1 Conditions and triggers

In Section 5.7, we have claimed that slow and fast trigger are mutually exclusive and described the intuition behind this claim. We will now formally prove this statement:

Lemma 1. ST & FT are mutually exclusive [39, adapted from Lemma 8.17]
At all times $t \in \mathbb{R}_{>0}$, no node $v$ in $G$ can simultaneously satisfy their fast and slow trigger.

Proof. Suppose $v$ satisfies FT. This means that $v$ satisfies FT-1 and FT-2 by Definition 30 for some skew level $s > \mathbb{N}_{\geq 0}$.

For any level $s' > \mathbb{N}_{\geq 0}$ with $s' > s$, we have:

$$\begin{aligned} \forall (v,x) \in E : L_v(t) - \tilde{L}_x(t) &< 2s\kappa_e - \delta_{v,x} && \text{by Def. 30} \\ &\leq (2s' - 1)\kappa_e && \text{by } \kappa_e = \delta_{e,max} > \delta_e \end{aligned}$$

This contradicts ST-1 for $s' > s$. If $s' \leq s$ then:

$$\begin{aligned} \exists (v,y) \in E : \tilde{L}_y(t) - L_v(t) &> 2s\kappa_e + \delta_{v,y} && \text{by Def. 30} \\ &\geq (2s' - 1)\kappa_e && \text{by } \kappa_e = \delta_{e,max} > \delta_e \end{aligned}$$

This contradicts ST-2 for $s' \leq s$. We conclude that a node $v$ doesn't fulfil ST if it satisfies FT. The proof that a node cannot satisfy FT if satisfies ST is analogous. □

Furthermore, we have claimed that the algorithmic triggers imply that the corresponding conditions holds whenever the trigger holds. A proof of this statement looks as follows:

Lemma 2. Conditions imply triggers [39, adapted from Lemma 8.18]
If a node $v$ satisfies FC or SC respectively at time $t$, then $v$ also satisfies FT or SC respectively at time $t$.

Proof. Suppose FC holds for $v$ at time $t \in \mathbb{R}_{>0}$, then, by Definition 18, there exists some skew level $s \in \mathbb{N}_{>0}$ such that:

$$\begin{aligned} \exists e = (v,x) \in E : \tilde{L}_x(t) - L_v(t) &\geq L_x(t) - \delta_e - L_v(t) \\ &\geq 2s\kappa_e - \delta_e && \text{by Def. 21} \end{aligned}$$

and

$$\begin{aligned} \forall e = (v,y) \in E : L_v(t) - \tilde{L}_y(t) &\leq L_v(t) - L_y(t) + \delta_e \\ &\leq 2s\kappa_e + \delta_e && \text{by Def. 21} \end{aligned}$$

This implies that FT holds. Analogously, if SC holds, for some $s \in \mathbb{N}_{>0}$ we have:

$$\begin{aligned} \exists e = (v,x) \in E : L_v(t) - \tilde{L}_x(t) &\geq L_v(t) - L_x(t) \\ &\geq (2s-1)\kappa_e \qquad \text{by Def. 21} \end{aligned}$$

and

$$\begin{aligned} \forall e = (v,y) \in E : \tilde{L}_y(t) - L_v(t) &\leq L_y(t) - L_v(t) \\ &\leq (2s-1)\kappa_e \qquad \text{by Def. 21} \end{aligned}$$

implying that ST holds. □

With Lemma 1 and Lemma 2, we have shown that the two triggers implemented by Algorithm 4 are mutually exclusive and, whenever the fast or slow trigger holds, the stronger algorithmic conditions hold too. The algorithmic triggers define, based on the chosen skew level, how much drift can be accumulated before we start correcting the clock but also how much fast running nodes can run ahead of their slower neighbours. By exploiting the bounds set by the triggers as well as a description of the growing skew described by the potential function, we will show that fast running clocks, satisfying the slow trigger wait up in the next section. In the section after that, we will show how the fast trigger enables slow running nodes to catch up with exactly this build up skew.

### 6.2.2 The Leading Lemma: fast running clocks wait up

In Section 5.8.2, Definition 31 provided a description of the potential function. By showing a bound on its growth, we can immediately follow a bound on the local skew at the node considered by the potential function. This result will be exploited in Section 6.2.4 and Section 6.2.5 to define a general bound on local and global skew.

Lemma 3. Leading Lemma [39, adapted from Lemma 8.20]
If a node $w$ is leading at time $t$, it satisfies both slow condition (SC) and slow trigger (ST).

Proof. Let's suppose that our node $w$ is a leading node, that is a node that maximises its potential function, at some time $t$. Let $v$ be the matching neighbour such as $w$ maximises its potential function for $v$:

$$\Psi_v^s(t) = L_w(t) - L_v(t) - \text{dist}^{2s-1}(v,w) > 0$$

That in particular implies:

$$L_w(t) > L_v(t)$$

thus $w \neq v$. If so, then there is a level $s \in \mathbb{N}_{>0}$ for which it holds:

$$\Psi_v^s(t) = L_w(t) - L_v(t) - \text{dist}^{2s-1}(v, w)$$

with $\Psi_v^s(t) > 0$. For any other node $y \in V$, it holds that:

$$L_y(t) - L_v(t) - dist^{2s-1}(y, w) \leq \Psi_v^s(t)$$

Rearranging to determine the distance between $w$ and $y$, we get:

$$dist^{2s-1}(w, v) - dist^{2s-1}(y, v) \leq L_w(t) - L_y(t)$$

$\Psi_v^s$ is strictly monotonously growing in the hop distance, as all edge weights and skew levels are strictly positive. The potential function by definition selects the path $(w, v)$ that has minimum weight.

For any $\{v, y\} \in E$, this translates to $H(v, w) \geq H(y, w) - 1$ in terms of hop-distance, as any path with lower edge weight must be shorter than the path selected by $\Psi_v^s$. This implies for any bidirectional edge $e = (w, y) \in E$:

$$L_y(t) - L_w(t) \leq (2s - 1)\kappa_e$$

This means that SC-2 holds at node $w$. Now, if the path between $x$ and $w$ is exactly one hop shorter than the path between $w$ and $v$, i.e, $H(x, v) = H(w, y) - 1$, then for such an edge $\{v, x\} \in E$, this results in:

$$L_w(t) - L_y(t) \geq (2s - 1)\kappa_e$$

by the same argument as above. Thus SC-1 holds. As we showed in Lemma 2, if a node $w$ satisfies the slow condition, it also satisfies the slow trigger. Showing that SC holds is thus sufficient to prove the claim. □

In Lemma 3, we have shown that leading nodes indeed fulfil the slow condition as it implies that the growth of the potential function is indeed bounded by $\vartheta - \frac{dL_w}{dt}$. This will enable us to show that nodes fast mode can catch up as the nodes in slow mode “wait-up” by not accumulating unquantifiable amounts of skew. The Wait-Up Lemma formalises this statement:

Lemma 4. Wait-Up Lemma [39, adapted from Lemma 8.21]
Suppose that for a node $w \in V$ we have $\Psi_{w,v}^s(t) > 0$ at all times $t \in (t_0, t_1]$, then:

$$\Psi_w^s(t) \leq \Psi_w^s(t_0) - (L_w(t_1) - L_w(t_0)) + \vartheta(t_1 - t_0)$$

Proof. Fix an arbitrary node $w \in V$ and interval $t \in (t_0, t_1]$ and a skew level $s \in \mathbb{N}_{>0}$ such as $\Psi_{w,v}^s(t) > 0$ at all times $t \in (t_0, t_1]$. Then, for another node $v \in V$ and a time $t \in (t_0, t_1]$, we can define:

$$f_v(t) = L_v(t) - (2s - 1) \sum_{P(v,w)} \kappa_e$$

One can observe that:

$$\max_{v \in V}\{f_v(t) - L_w(t)\} = \Psi_w^s(t)$$

For any $v$ for which $f_v(t)$ attains it maximum, i.e, $f_v(t) = L_w(t) + \Psi_w^s(t) > 0$, it implies that $v$ is a leading node by the leading lemma (Lemma 3) as we have that:

$$L_v(t) - L_w(t) - \text{dist}^{2s-1}(v,w) = \Psi_w^s(t) > 0$$

The node $v$ hence is in slow mode by Lemma 2. That, by definition, implies that $v$ is running at a rate of at most $\vartheta$. By applying the the following helper lemma, whose proof of correctness can be looked up in [39, adapted from Lemma 8.33]:

Lemma 5. Bounded growth
For $k \in \mathbb{N}$, let $\mathcal{F} = \{f_i | i \in [k]\}$ where each function $f_i : [t_0, t_1] \to \mathbb{R}$ is differentiable and $[t_0, t_1] \subset \mathbb{R}$. We define the maximum function $F : t \in [t_0, t_1] \to \mathbb{R}$ as $F(t) = \max_{i \in [k]}\{f_i(t)\}$. If $\mathcal{F}$ has the property of $i \in [k]$ and $t \in [t_0, t_1]$ under the condition $f_i(t) = F(t)$ and $\frac{d}{dt} f_i(t) \leq r$ for some $r \in \mathbb{R}$. Then, we have that:

$$\forall t \in [t_0, t_1] : F(t) \leq F(t_0) + r(t - t_0)$$

Applying Lemma 5, it follows for each node $v$ and interval $t \in (t_0, t_1]$ with $r = \vartheta$ and $F(t_0) = 0$ that:

$$\begin{aligned}
\Psi_w^s(t_0) - \Psi_w^s(t_1) &= \max_{v \in V}\{f_v(t_0)\} - \max_{v \in V}\{f_v(t_1)\} - (L_w(t_1) - L_w(t_0)) && \text{add 0} \\
&\leq \vartheta(t_1 - t_0) - (L_w(t_1) - L_w(t_0)) && \text{apply Lemma 5}
\end{aligned}$$

yielding the claim. □

By Lemma 4, the skew potential of a leading node is bounded by its clock value as well as by the maximum clock drift that can accumulate between the two end points of the probing interval. This directly translates into a bound on the potential function itself:

Corollary 1. [39, Corollary 8.22]
$\forall w \in V, s \in \mathbb{N}_{>0} : \Psi_w^s(t_1) \leq \Psi_w^s(t_0) + (\vartheta - 1)(t_1 - t_0)$

Proof. By the wait up lemma, we have that:

$$\Psi_w^s(t_1) \leq \Psi_w^s(t_0) - (L_w(t_1) - L_w(t_0)) + \vartheta(t_1 - t_0)$$

Since the rate of the logical clocks is bound by the rate of the hardware clocks and that we have the constraint that $\frac{dL_w}{dt} \geq \frac{dH_w}{dt}$, we have:

$$\Psi_w^s(t_1) \leq \Psi_w^s(t_0) - (H_w(t_1) - H_w(t_0)) + \vartheta(t_1 - t_0)$$

Since the rate of of the hardware clocks fulfils $\frac{dH_w}{dt} \geq 1$ at all times $t$, we get:

$$\Psi_w^s(t_1) \leq \Psi_w^s(t_0) + (\vartheta - 1)(t_1 - t_0)$$

which proves the claim. □

### 6.2.3 The Trailing Lemma: slow running nodes catch up

In this section, we will prove that the bounds defined in the previous section will allow us to show that nodes in fast mode can indeed catch-up the skew to their neighbours running in slow mode. To this effect, we first define the notion of a trailing node, that is a node that lags behind its neighbours and is supposed to satisfy fast condition and fast trigger respectively.

Definition 32. Trailing Nodes [39, adapted from Def. 8.23]
A node $w \in V$ is considered trailing at time time $t \in \mathbb{N}_{\geq 0}$ if:

$$\exists s \in \mathbb{N}_{>0}, v \in V : L_v(t) - L_w(t) - \mathrm{dist}^{2s}(v,w) \\ = \max_{x \in V} \left\{ L_v(t) - L_x(t) - \mathrm{dist}^{2s}(v,x) \right\} > 0$$

We now formally show that trailing nodes indeed implement slow condition and slow trigger:

Lemma 6. Trailing Lemma [39, adapted from Lemma 8.24]
Suppose a node $w \in V$ is trailing at time $t$, then $w$ satisfies both FC and FT.

Proof. We will prove that a node $w$, fulfilling the definition of a trailing node, also satisfies FC. If so, there is some node $v \in V$ and a matching skew level $s \in \mathbb{N}_{\geq 0}$ satisfying:

$$L_v(t) - L_w(t) - \mathrm{dist}^{2s}(v,w) = \max_{x \in V}\{L_v(t) - L_x(t) - \mathrm{dist}^{2s}(v,x)\} > 0$$

This means that $L_v(t) > L_w(t)$ and $v \neq w$. Hence, as in the Catch-up Lemma, for a $y \in V$ with $H(v,w) \geq H(y,w) - 1$, we have:

$$L_v(t) - L_w(t) - \mathrm{dist}^{2s}(v,w) \geq L_v(t) - L_y(t) - \mathrm{dist}^{2s}(v,y)$$

This implies that for all edges $(w,y) \in E$, the difference of their weighted difference, i.e, their skews, is positive:

$$L_y(t) - L_w(t) - \mathrm{dist}^{2s}(v,w) + \mathrm{dist}^{2s}(v,y)) \geq 0$$

Hence, FC-2 is fulfilled as:

$$\forall (v,y) \in E : L_w(t) - L_y(t) \leq 2s\kappa_e$$

Similarly, we can show that FC-1 holds as here is some edge $(v,x) \in E$, where $v \neq w$ and $v$ and $x$ in hop-distance of $H(v,x) = H(v,w) - 1$, such as:

$$\exists (v,x) \in E : L_x(t) - L_w(t) \geq 2s\kappa_e$$

As conditions imply the triggers by Lemma 2, as FC-1 and FC-2 hold, FT-1 and FT-2 also hold, proving the claim. □

With help of Lemma 6, we can show that trailing nodes can eventually catch up with their faster running neighbours. The time this process requires is described by the potential function $\Psi^{s-1}(t_0)$ and formalised in the Catch-Up Lemma:

Lemma 7. Catch-Up Lemma [39, adapted from Lemma 8.25]
For $s \in \mathbb{N}_{\geq 0}$ and $t_0, t_1 \in \mathbb{R}_{\geq 0}$:

$$t_1 \geq \begin{cases} t_0 + \frac{\mathcal{G}(t_0)}{\mu}, & \text{if s=1} \\ t_0 + \frac{\Psi^{s-1}(t_0)}{\mu}, & \text{otherwise} \end{cases}$$

Then, for any $w \in V$, we have:

$$L_w(t_1) - L_w(t_0) \geq t_1 - t_0 + \Psi_w^s(t_0)$$

Proof. Given a node $w \in V$ and a neighbour $v$ of $w$ such as $\Psi_w^s(t_0) > 0$ at start time $t_0$, we have:

$$\Psi_w^s(t_0) = L_v(t_0) - L_w(t_0) - \text{dist}^{2s-1}(v, w) > 0$$

With this we can define:

$$f_x(t) := L_v(t_0) + (t - t_0) - L_x(t) - \text{dist}^{2s-2}(v, x)$$

for $x \in V$, allowing us to observe that:

$$\begin{aligned} \Psi_w^s(t_0) &= L_v(t_0) - L_w(t_0) - \text{dist}^{2s-1}(v, w) \\ &\leq L_v(t_0) - L_w(t_0) - \text{dist}^{2s-2}(v, w) && \text{dist}^{2s-1}(v, w) > 0 \\ &= L_v(t_0) + (t_0 - t_0) - \text{dist}^{2s-2}(v, w) && \text{add 0} \\ &= f_w(t_0) && \text{Def. of } f_w(t_0) \\ &= f_w(t_0) && \text{Observation 1} \end{aligned}$$

as the distance is a strictly positive value. Within an interval $t \in [t_0, t_1]$ we thus have:

$$\begin{aligned} L_w(t_1) - L_w(t) - (t_1 - t) \geq 0 \geq f_w(t) & \\ &= L_v(t_0) + (t - t_0) - L_w(t) - \text{dist}^{2s-2}(v, w) && \text{Def. of } f_w(t) \\ &= f_w(t_0) + (t - t_0) - (L_w(t) - L_w(t_0)) && \text{Def. of } f_w(t_0) \\ &\geq \Psi_w^s(t_0) + (t - t_0) - (L_w(t) - L_w(t_0)) && \text{by Obs. 1} \end{aligned}$$

if the maximum value of the skew function is $\max_{x \in V}(f_x(t)) \leq 0$.

Now, if $\max_{x \in V}(f_x(t) > 0)$, then for a trailing node $y \in V$, where $\max_{x \in V}\{f_x(t)\} =$

$f_y(t)$, we have:

$$\begin{aligned}\max_{x\in V}(L_v(t) - L_x(t) - \mathrm{dist}^{2s-2}(v,x) &= L_v(t) - L_v(t_0) - (t - t_0) + \max_{x\in V}\{f_x(t)\} && \text{Def. of } f_x(t)\\ &= L_v(t) - L_v(t_0) - (t - t_0) + f_y(t) && \text{by choice of y}\\ &= L_v(t) - L_y(t) - \mathrm{dist}^{2s-2}(v,y) && \text{Def. of } f_y(t)\end{aligned}$$

This term is positive as the logical clocks run at least as fast as the hardware clock in terms of their rate and the hardware clocks run with rate at least one.

$$\begin{aligned}L_v(t) - L_v(t_0) - (t - t_0) + \max_{x\in V}\{f_x(t)\} &> L_v(t) - L_v(t_0) - (t - t_0) && \text{as } \max > 0\\ &\geq H_v(t) - H_v(t_0) - (t - t_0) && \text{as } \frac{dL_v}{dt} \geq \frac{dH_v}{dt}\\ &\geq 0 && \text{as } \frac{dH_v}{dt} \geq 1\end{aligned}$$

Hence, by Lemma 3, $y$ must be in fast mode as it satisfies FT and, as it satisfies FT it cannot fulfil ST at the same time by Lemma 1.

We also know that by Definition 12, where the parameter $\mu$ is defined, that $f_y$ can show a difference of at least:

$$\frac{df_y(t)}{dt} = 1 - \frac{dL_y(t)}{dt} = 1 - (1+\mu)\frac{dH_y(t)}{dt} \leq -\mu$$

in its clock rate compared to the values of its logical and hardware clock as the hardware clock runs with a rate of at least 1.

With this, we can assume towards contradiction that $\max_{x\in V}\{f_x(t) > 0\}$ for $\forall t \in [t_0, t_1]$; this would mean that no skew got caught up. Then, by Lemma 5, we have:

$$\begin{aligned}\max_{v\in V}(f_x(t_0)) &> -(\max_{x\in V}\{f_x(t_1)\} - \max_{x\in V}\{f_x(t_0)\}) && \max > 0\\ &\geq \mu(t_1 - t_0) && \text{by Lemma 5}\\ &\geq \mu(t_1 - t_0) && \text{Observation 2}\end{aligned}$$

If $s = 1$, we look at the degree of synchronisation at a global level in the network and we get:

$$\max_{v\in V}(f_x(t_0)) = \max_{v\in V}\{L_v(t_0) - L_x(t_0)\} \leq \mathcal{G}(t_0) \leq \mu(t_1 - t_0)$$

by the preconditions of Lemma 7, contradicting Observation 2.
If we look at a finer grained scope in terms of skew, which implies that $s > 1$,

we have:

$$\begin{aligned}\max_{x\in V}(f_x(t_0)) &= \max_{x\in V}\{L_v(t_0) - L_x(t_0)\} - \text{dist}^{2s-2}(v,x) && \text{Def. of } f_x(t_0)\\ &\le \max_{x\in V}\{L_v(t_0) - L_x(t_0)\} - \text{dist}^{2s-3}(v,x) && \text{dist}^{2s-3}(v,x) \ge 0\end{aligned}$$

Hence, we get:

$$\begin{aligned}\max_{x\in V}\{L_v(t_0) - L_x(t_0)\} - \text{dist}^{2s-3}(v,x) &\le \Psi^{s-1}(t_0)\\ &\le \Psi^{s-1}(t_0) && \text{Def. of } \Psi^{s-1}\\ &\le \mu(t_1 - t_0) && \text{Precond. Lemma 7}\end{aligned}$$

This contradicts Observation 2. As both cases yield a contraction, the lemma statement holds true. □

This implies that $\Psi^s$ cannot grow at a rate faster than $\vartheta - 1$. As result, whenever the potential function is positive i.e some skew build up, we can catch up with the nodes in slow mode after the skew described by the global skew $\mathcal{G}$ in the base case where $s = 1$ or after the skew described by $\Psi^{s-1}(t_0)$ in the general case got corrected.

#### 6.2.4 Bounds on the Local Skew

To be able to describe any meaningful upper bound on the local skew achieved by Algorithm 4, we first need to clearly define the starting conditions of the algorithm, i.e, what preconditions apply in the first analysis interval $I_1$ of the sequence $\mathcal{S}$ of executions, that we defined in Section 6.1.2. Most importantly, the structure of the algorithm enforces that, at boot-up time, there has to be some initial degree of synchronisation [39, Lemma 8.26].

This is required as, both in analysis and in practical implementation, we would like to ensure the convergence of the algorithm. However, Algorithm 4 cannot converge out of any arbitrary starting conditions. Hence an initial bound on the skew, implying an initial bound on the potential function, is required. We will thus assume that any implementation of the GCS algorithm satisfies the following definition at boot-up time:

Definition 33. Initial synchronisation [39, adapted from Lemma 8.26]
We assume that at boot-up i.e time $t \in \mathbb{R}_{\geq 0} = 0$, the hardware clocks of any two nodes $\{v, w\} \in E$ connected by an edge in the network are synchronised to a factor of less than:

$$\forall v, w \in V : H_v(0) - H_w(0) \le \sum_{e\in P(v,w)} \kappa_e$$

With this requirement on the initial skew bound, we enforce bounded initial skew as well as an initial skew matching the fast and slow triggers as well as the requirements for the potential function used in the analysis of the algorithm, yielding the desired convergence.

With the precondition defined in Definition 33 in mind, we can show that at boot-up time, the network does not have any skew.

Lemma 8. No skew lemma [39, adapted from Lemma 8.26]
If $H_v(0) - H_w(0) \leq \kappa_e$ for all edges $(v, w) \in E$, then $\Psi^s(0) = 0$ for all skew levels $s \in \mathbb{N}_{>0}$.

Proof. The skew at level $s \in \mathbb{R}_{\geq 0}$ is defined by the level potential function for $t = 0$, i.e:

$$\Psi^s(0) = \max_{v,w \in V} \{L_w(0) - L_v(0) - \text{dist}^{2s-1}(v, w)\}$$

Since for any $s \in \mathbb{N}_{\geq 0}$ it holds that $\sum_{e \in P(v,w)} \kappa_e \geq 0$, we have that:

$$\Psi^s(0) \leq \max_{v,w \in V} \{L_w(0) - L_v(0) - \text{dist}^1(v, w)\}$$

As for $t = 0$ the logical clock values correspond to the hardware clock values by Definition 33, we have:

$$\Psi^s(0) = \max_{v,w \in V} \{H_w(0) - H_w(0) - \text{dist}^1(v, w)\}$$

As by Definition 33, $\forall v, w \in V : H_v(0) - H_w(0) \leq \sum_{e \in P(v,w)} \kappa_e$, we have that $H_v(0) - H_w(0) \leq \text{dist}^1(v, w)$ and hence:

$$\Psi^s(0) = \max_{v,w \in V} \{H_w(0) - H_w(0) - \text{dist}^1(v, w)\} = 0$$

Thus, the difference in skew between $H_v$ and $H_w$ at time $t = 0$ is zero between any two nodes. □

With this initial bound on the skew, we can show an exponential decrease with base $\sigma = \frac{\mu}{\vartheta - 1}$ in the local skew given an upper bound on the global skew: in each computational cycle, the nodes in fast mode catch up by $\mu$ whereas the worst case drift between any two nodes is $\vartheta - 1$ in $G$, yielding the ratio described by the parameter $\sigma$ above. $\sigma$ is called correction-to-drift coefficient

This follows as, given an initial upper bound on the skew, Lemma 4 and Lemma 7 respectively apply, implying that skew cannot be build up quicker at fast running node than it can be caught up by the correction applied at the slow running nodes, as detailed in Section 6.2.2 and Section 6.2.3.

Lemma 9. Global skew bounds local skew [39, adapted from Lemma 8.27]
If $H_v(0) - H_w(0) \leq \kappa_e$ for all edges $(v,w) \in E$ as in definition 30 for all skew levels $s \in \mathbb{N}_{>0}$ then

$$\Psi^s(t) \leq \frac{\mathcal{G}}{\sigma^s}$$

with $\sigma = \frac{\mu}{\vartheta - 1}$, the correction-to-drift coefficient.

Proof. By contradiction.

Suppose that the statement of Lemma 9 is false. We can observe that the potential function is continuous and that $\Psi^s(0) = 0$ by Lemma 8. Hence, there is a time $t_h$ at which the potential function at node $w$ supersedes the skew level build up in the process by a small factor $\iota > 0$ for a matching minimal skew level $s \in \mathbb{N}_{geq0}$. Thus, there is some $w \in V$ defining this skew level potential function:

$$\Psi^s_w(t_h) = \Psi^s(t_h) = \frac{\mathcal{G}}{\sigma^s} + \iota \qquad \text{Observation 3}$$

Let's consider an interval $t' \in [t_s, t_h]$ where $t_s = \max\{t_h - \mathcal{G}/(\mu\sigma^{s-1}, 0\}$ such as $\Psi^s_w(t_s)$ is minimal whilst still being strictly positive on the considered interval. Especially, this yields that $\Psi^s_w(t_s)$ cannot be 0 as otherwise:

$$\begin{aligned}
\frac{\mathcal{G}}{\sigma^s} + \iota &= \Psi^s_w(t_h) && \text{by Obs. 3} \\
&\leq (\vartheta - 1)(t_h - t') && \text{by Corollary 1} \\
&\leq (\vartheta - 1)(t_h - t_s) && \text{by Def. of t'} \\
&\leq \frac{\vartheta - 1}{\mu} \cdot \frac{\mathcal{G}}{\sigma^{s-1}} && \text{by Def. of } t_s \\
&= \frac{\mathcal{G}}{\sigma^s} && \text{as } \sigma = \frac{\mu}{(\vartheta - 1)}
\end{aligned}$$

resulting in a contradiction.

Beyond the fact that $\Psi^s_w(t_s) > 0$, as $\Psi^s_w$ is continuous and $t' = t_s$, this also implies that $t_s \neq 0$ by Lemma 8. If $s > 1$, we thus have $t_s = t_h - \mathcal{G}/(\mu\sigma^{s-1})$, for which it holds that $\Psi^s(t_s) \leq \mathcal{G}/\sigma^{s-1}$. This yields the following contradiction:

$$\begin{aligned}
\frac{\mathcal{G}}{\sigma^s} + \iota &= \Psi^s_w(t_h) && \text{by Obs. 3} \\
&\leq \Psi^s_w(t_s) + \vartheta(t_h - t_s) - (L_w(t_h) - L_w(t_s)) && \text{by Lemma 4} \\
&\leq (\vartheta - 1)(t_h - t_s) && \text{by Lemma 7} \\
&= \frac{\vartheta - 1}{\mu} \cdot \frac{\mathcal{G}}{\sigma^{s-1}} && \text{by Def. } t_s \\
&= \frac{\mathcal{G}}{\sigma^s} && \text{as } \sigma = \frac{\mu}{(\vartheta - 1)}
\end{aligned}$$

Successively applying Lemma 4, showing that nodes wait up, as well as Lemma 7, showing that nodes catch up, before replacing $t_s$ by its definition, proves the claim as we reach a contraction in this case too. □

This means that we get, by iterating over all skew levels, an exponential decrease of the skew potential function that is bounded by the global skew present at boot-up. The central theorem follows immediately:

Theorem 2. Local skew [39, adapted from Theorem 8.28]
Assume that $\forall \{v, w\} \in E : \kappa_e \geq \delta$ and that Definition 33 holds for all edges of $G$, then any implementation of the GCS algorithm maintains a local skew of at most:

$$\mathcal{L} \leq \ 2 \cdot \kappa_e \left\lceil \log_\sigma \frac{\mathcal{G}}{\kappa_e} \right\rceil$$

with $\sigma = \frac{\mu}{\vartheta - 1}$ as correction-to-drift coefficient.

Proof. Given that the local skew is bound by the global skew by Lemma 9, we set the skew level to $s = \log_\sigma(\frac{\mathcal{G}}{\kappa_e})$. Then, for any edge $(v, w) \in E$ and any time $t \in \mathbb{R}_{\geq 0}$, we have:

$$\begin{aligned}
L_v(t) - L_w(t) - (2s-1)\kappa_e &= L_v(t) - L_w(t) - \text{dist}^{2s-1}(v, w) && \text{by Def. of } \Psi^s_w(t) \\
&\leq \Psi^s(t) && \text{by Def. of } \Psi^s(t) \\
&\leq \frac{\mathcal{G}}{\sigma^s} && \text{by Lemma 9} \\
&\leq \kappa_e && \text{as } s \geq \log_\sigma\left(\frac{\mathcal{G}}{\kappa_e}\right)
\end{aligned}$$

Taking the reverse direction on the bidirectional edge, i.e, the edge $(w, v)$, we get:

$$L_w(t) - L_v(t) - (2s-1)\kappa_e \leq \kappa_e$$

by the same argument as above. Rearranging both inequalities yields a local skew of:

$$\mathcal{L}(t) = \max_{(v,w)\in E}\{|L_v(t) - L_w(t)|\} \leq 2s\ \kappa_e = 2\kappa_e \left\lceil \log_\sigma \frac{\mathcal{G}}{\kappa_e} \right\rceil$$

concluding the proof. □

Note that this bound applies to any algorithm implementing the conditions FC and SC, implying that it covers the impact of $\kappa$ on the performance but omits all other factors, such as for the example the computation time required to compute the estimates or limiting effects in the properties of the physical implementations of measurements.

#### 6.2.5 Bounds on the Global Skew

The global skew is a special case of the local skew in which the two measured nodes are within distance up to $D$ of each other, that is the diameter of the network. In the following, we will show that the edge specific estimation errors as well as the correction-to-drift coefficient will allow us to describe an upper bound on the global skew.

Theorem 3. Global skew [39, adapted from Theorem 8.29]
Assume that $\forall\{v,w\} \in E : \kappa_e \geq \delta$ and that Definition 33 holds for all edges of $G$, then any implementation of the GCS algorithm maintains a global skew of at most:

$$\mathcal{G} \leq \left(1 + \frac{1}{\sigma - 1}\right) \sum_{e \in P(v,w)} \kappa_e$$

where $\sigma = \frac{\mu}{\vartheta - 1}$ is the correction-to-drift coefficient and $e \in P(v,w)$ a shortest path of length at most $D$.

Proof. By contradiction.

Assume towards contradiction that Lemma 9 gives us:

$$\begin{aligned}
\mathcal{G}(t_1) &= \left(1 + \frac{1}{\sigma - 1}\right) \sum_{e \in P(v,w)} \kappa_e + \iota \\
&= \frac{\sigma}{\sigma - 1} \sum_{e \in P(v,w)} \kappa_e + \iota && \text{Observation 4}
\end{aligned}$$

for a small $\iota > 0$, a minimal time $t_1$ and a shortest path $P(v,w)$ of length at most $D$. Since $\mathcal{G}$ is continuous, such a minimal time $t_1$ exists. This implies:

$$\mathcal{G}(0) = \max_{v,w \in V} \{L_v(0) - L_w(0)\}$$

by the definition of $\mathcal{G}(0)$. Furthermore, as the shortest path distance between nodes $v$ and $w$ has a length at most $D$, we have:

$$\begin{aligned}
\mathcal{G}(0) &\leq \max_{v,w \in V} \{L_v(0) - L_w(0)\} + \sum_{e \in P(v,w)} \kappa_e \\
&= \Psi^1(0) + \sum_{e \in P(v,w)} \kappa_e
\end{aligned}$$

Applying the definition of $\Psi^1(0) = 0$ as well as Lemma 8, this results in:

$$\begin{aligned}
\mathcal{G}(0) &= \Psi^1(0) + \sum_{e \in P(v,w)} \kappa_e \\
&= \sum_{e \in P(v,w)} \kappa_e
\end{aligned}$$

by Lemma 8. Accordingly, we set $t_0 = \max(t_1 - \mathcal{G}(t_1)/\mu, 0)$ and choose some $w \in V$ such as $\Psi^1_w(t_1) = \Psi^1(t_1)$ i.e, $w$ maximises the potential function of $v$ to become the skew level potential function.

With this, we have to distinguish two cases:

Case 1: We look at the global skew at boot-up time, i.e, at start time $t_0 = 0$. Then, we have:

$$\begin{aligned}
\Psi^1_w(t_1) &\leq \Psi^1_w(0) + (\vartheta - 1)(t_1 - t_0) && \text{by Corollary 1} \\
&= (\vartheta - 1)(t_1 - t_0) && \text{by Lemma 8} \\
&= (\vartheta - 1)(t_1 - t_0) && \text{Observation 5}
\end{aligned}$$

covering the base case.

Case 2: After boot up, i.e, when $t_0 > 0$, Lemma 7 can be applied to compute how much nodes catch up the build up skew. We get $t_1 - t_0 = \mathcal{G}(t_1)/\mu > \mathcal{G}(t_0)/\mu$ by the minimality of $t_1$ and thus:

$$\begin{aligned}
\Psi^1_w(t_1) &\leq \Psi^1_w(t_0) - (L_w(t_1) - L_w(t_0)) + \vartheta(t_1 - t_0) && \text{by Lemma 4} \\
&= \vartheta(t_1 - t_0) && \text{by Lemma 7} \\
&= \vartheta(t_1 - t_0) && \text{Observation 6}
\end{aligned}$$

in the regular case.

In both cases, this yields the following:

$$\begin{aligned}
\frac{\mathcal{G}(t_1)}{\sigma} &= \frac{1}{(\sigma - 1)} \cdot \sum_{e \in P(v,w)} \kappa_e + \frac{\iota}{\sigma} && \text{by Obs. 4} \\
&< \frac{1}{(\sigma - 1)} \cdot \sum_{e \in P(v,w)} \kappa_e + \iota && \text{as } \iota > 0,\ \mu/(\vartheta - 1) > 1 \\
&= \mathcal{G}(t_1) - \sum_{e \in P(v,w)} \kappa_e && \text{by Obs. 4}
\end{aligned}$$

for a shortest path $P(v, w)$ of length at most $D$. Furthermore, we have:

$$\begin{aligned}
\frac{\mathcal{G}(t_1)}{\sigma} &= \mathcal{G}(t_1) - \sum_{e \in P(v,w)} \kappa_e \\
&\leq \max_{v,w \in V} \{L_v(t_1) - L_w(t_1) - \operatorname{dist}^1(v, w)\} && \text{as } H(v, w) \leq D \\
&= \Psi^1(t_1) && \text{by Def. of } \Psi^1 \\
&= \Psi^1_w(t_1) && \text{by Def. of } \Psi^1_w
\end{aligned}$$

for a matching choice of $w$.

By successively applying Observation 5 and Observation 6 as well as replacing $\sigma = \frac{\mu}{(\vartheta-1)}$ and $t_0$ with its defined value, this results in:

$$\begin{aligned}
\Psi_w^1(t_1) &\leq (\vartheta - 1)(t_1 - t_0) && \text{by Obs. 5} \\
&\leq \frac{(\vartheta - 1)\mathcal{G}(t_1)}{\mu} && \text{by Obs. 6} \\
&= \frac{\mathcal{G}(t_1)}{\sigma} && \text{by Def. of } t_0
\end{aligned}$$

yielding a contradiction, proving the claim. □

## 6.3 Interpretation of resulting skew bounds

### 6.3.1 Properties of local and global skew

The local skew, or more specifically the upper bound on the local skew, varies as a function of multiple factors, amongst them the skew to correction factor $\mu$, as well as the (initial) global skew and the edge specific estimation error $\kappa_e$. Assuming that all other factors stay constant, a larger skew-to-correction factor $\sigma$ implies a larger logarithmic base, i.e, a faster exponential decrease of the skew whilst an increase of the initial global skew induces the local skew to grow logarithmically in $\log_\sigma$. Last but not least, larger edge specific weights $\kappa_e$ have the highest impact on the local skew, yielding a linear growth of the local skew as a function of increasing values of $\kappa_e$. This means that the edge specific weights, and thereby the estimation error induced by delays, uncertainties and asymmetries have a big impact on the performance and should be kept as small as possible in practical implementations.

The global skew, in turn, grows linear in the estimation error on the chosen path for a fixed value of $\sigma$. This means that, in a large network, the expected performance in terms of global skew decreases with the path lengths but also with growing uncertainties and asymmetries. Furthermore, the gradient property inherent to GCS accentuates this effect for longer paths, implying that small values of $\kappa$ are desirable in practice. The same applies to the skew-to-correction factor $\sigma$, as the skew grows as inverse proportional function in the value of $\sigma$.

### 6.3.2 Interpretation of results

Asymptotically, we can conclude that the local skew expected for nodes running the GCS algorithm lives in the same range, for both the weighted and the unweighted setting, as in the simple GCS algorithm presented in [39] - and that is $O(d)$. This is due to the fact that the presented variant of GCS is structurally identical to the basic algorithm and hence results in the same expected skew performance.

However, in comparison, our model fundamentally alters the scope and meaning of parameter $d$. Instead of considering an estimation of the actual path delay, as being the delay modelled by the parameter $d$ in [39], the two way measurements presented in Section 4.6 allow us to precisely determine then deduct the actual path delay, reducing the global uncertainty per hop. The indeterminable error term of the system thus just consist of the path asymmetry coefficient $\epsilon_d$ and the measurement uncertainty $\epsilon_m$, which both are multiple orders of magnitude smaller that the actual path delay and uncertainty, yet do depend on the later. This implies that the resulting values of $\kappa$ for a fixed maximum values of the drift $\vartheta$ as well as the skew correction factor $\mu$ - which are usually parameters dictated by the type of frequency reference used - determine the performance of the algorithm. The concrete performance gap between a GCS instance based on prior models and the model suggested in this thesis depends on the use case.

### 6.3.3 Comparative performance evaluation

For networks on chips, the values of $\kappa$ heavily depend on the design and the chosen architecture, resulting a large variance in terms of performance over this parameter. This is due to the fact that cross-talking and other spurious noise inducing signals mentioned in Section 4.3.2 mainly depend on the layout of the wires on the chip. We hence conjecture an expected variance of $\kappa$ estimated to be in the order of 10% to 20% already with in-situ measurements, as the path delays are also very variable in this setup. Hence, in-situ measurements should clearly be preferred over ante-hoc measurements of path length and measurement error to guarantee a decent performance of our GCS variant.

In wired networks, the variance of $\kappa$ mostly depends on the choice of the measurement scheme deployed. When running the algorithm with wire delays pre-defined by an ante-hoc measurement, the variations during runtime cannot be corrected with a high degree of precision, yielding a conjectured variation of the estimated error per edge in the order of 1-5%. With in-situ measurements, which recalibrate the error induced by the delays variations of the wires regularly, we conjecture that the variance can be reduced to be in the per mill range. This implies that the wired setup highly benefits from a model allowing for in-situ measurements in terms of performance of our variant of the GCS algorithm.

In the wireless setup, in-situ measurement are the only viable option to achieve decent performance. This is due to the fact that radio conditions rapidly fluctuate, making ante-hoc measurements too unreliable for predictions, as they would yield linear variance in the path delay. With in-situ measurements, similarly low variance in the per mill range as in the wired setup can be achieved. However, longer transmission distances and thus delays have a higher impact than in the wired setting, resulting in a very high performance only on paths of medium long delays. This is due to fact that for short delays and paths, the measurement noise dominates very early on as described in Section 5.4.2. Given that the measurement noise does not dominate, we conjecture the variance to be sublinear, delivering good performance even in unstable radio environments.

# 7 Conclusion

## 7.1 Results

In this thesis, we developed an implementation-near model for GCS, tailored to support the three flagship use cases specified in literature, namely wired and wireless networks as well as networks on chip. In Section 4, we started by modelling the behaviour of crystal and LC oscillators. In Section 4.4.1, we defined a formal model of physical properties of these hardware clocks and stated that they can indeed be described by a differentiable function. We also formally defined the clock drift and its particular meaning in TCS related publications in Section 4.4.2. Finally, in Section 4.4.3, we formally defined the logical clocks used in the GCS algorithm as a function of the hardware clock and its rate changes, jogged by the rate corrections initiated by GCS, creating an implementation-near clock model. In Section 4.4.4, we argued that this model is indeed realistic as exceptions to the behaviour described are negligible in practice. We hence covered all desirata formulated in Section 3.2.2.

Furthermore, we developed a refined model replacing the one-way measurement based communication model paradigm from prior work by a protocol mirroring the behaviour of physical measurements, which are usually based on the two-way measurement paradigm. We presented Algorithm 1, Algorithm 2, and Algorithm 3 allowing to precisely determine the delay $d$ between sending and arrival time of the two messages exchanged in the course of a two-way measurement as specified in Definition 3. Furthermore, we provided a new metric for the uncertainty in the system. As the delay can be computed in our model instead of being estimated as in prior work, there is no uncertainty on the duration of the delay. We hence redefined the notion of the uncertainty, which only covers the potential difference in duration between the delay on the forward path to a node and the delay on the backward path. As this difference is usually very small, we thereby reduced the contribution of the uncertainty by a significant factor: instead of being in the order of the delay itself, the uncertainty in our model is conjectured to be in range, depending on the use case, 0,1% to 10% of the edge specific delay. Beyond reducing the contribution of the delay to the uncertainty, we also introduced a distinction between short term and long term components of the uncertainty and presented a method to distinguish between the short and longterm fluctuation of the short and long term components. As we detailed in Section 5.4.2, this distinction does not yield any major improvements as it is only beneficial for measurements of very short durations. Together, the enumerated points provide a framework providing detailed information on how to determine all the parameters of the GCS model. Along the way, we detailed many pitfalls and effects, achieving the goal of providing a clear roadmap to a possible implementation of the setup we described in theory. With this, we meet all the requirements formulated in Section 3.2.3.

Finally, in Sections 6.2.4 and 6.2.5, we proved that, while the assumptions made in our model indeed yield tighter upper bounds on local and global skew, structurally, the local and global skew can still be shown analogously to the bounds shown in prior work [39]. This means that our variant of the GCS algorithm does not behave fundamentally different than its original counterpart described in [39]. With this, all additional requirements formulated in Section 3.2.4 hold.

## 7.2 Open questions & further work

In Section 6.3.3, we claimed that the expected range of the edge specific estimation error $\kappa$ is in range of 20% to 0,1% of the edge specific delay, depending on the use case. To verify whether these claims are indeed correct, a physical implementation of our model and GCS variant is required. We leave this for further work.

As mentioned in Section 5.42, our model as well as all models in prior art, assume the worst case clock drift $\vartheta$ in all bounds specified as there is no simple method with low computational overhead known to determine the actual drift in a more precise way. However, this assumption is limiting the performance of GCS. In practice, oscillators often stay below the worst case bound on the clock drift. While no simple predictive model of their drift behaviour exists, their stability over time and under different environmental conditions still can be studied by performing measurements against a high stability reference. By studying this type of data and building a refined model out of the observed fluctuations, a heuristic to predict the stability of oscillators in the context of the estimates could be developed and incorporated into the GCS algorithm. The exact effectuation of this is an open and highly-non trivial question that is, again, left for further work.